\PassOptionsToPackage{sort}{natbib}

\documentclass{article} 

\usepackage{iclr2027_conference,times}
\usepackage{url}
\usepackage[inline]{enumitem}
\usepackage{graphicx}
\usepackage{amsmath}
\usepackage{booktabs}
\usepackage{acronym}
\usepackage{multirow}
\usepackage{graphicx}
\usepackage{wrapfig}

\usepackage[colorlinks=true,
            linkcolor=blue,
            citecolor=blue,
            filecolor=blue,
            urlcolor=blue]{hyperref}
\usepackage{cleveref}

\newcommand{\header}[1]{\vspace{1mm}\noindent\textbf{#1}.}

\newif\ifshowcomments
\showcommentstrue   

\ifshowcomments
    \newcommand{\todo}[1]{\textcolor{purple}{TODO: #1}}
    \newcommand{\todossc}[1]{\textcolor{magenta}{[SSC: #1]}}
    \newcommand{\yt}[1]{\textcolor{blue}{[YT: #1]}}
    \newcommand{\pierre}[1]{\textcolor{teal}{[Pierre: #1]}}
\else
    \newcommand{\todo}[1]{}
    \newcommand{\todossc}[1]{}
    \newcommand{\yt}[1]{}
    \newcommand{\pierre}[1]{}
\fi

\acrodef{GenRec}{generative recommendation}
\acrodef{MU}{machine unlearning}
\acrodef{NAU}{Neighborhood-Aware Unlearning}

\acrodef{SID}{semantic identifier}

\crefname{equation}{Eq.}{Eqs.}
\Crefname{equation}{Eq.}{Eqs.}

\usepackage{amsmath,amsfonts,bm}

\def\eqref#1{Eq.~\ref{#1}}

\def\1{\bm{1}}

\DeclareMathAlphabet{\mathsfit}{\encodingdefault}{\sfdefault}{m}{sl}
\SetMathAlphabet{\mathsfit}{bold}{\encodingdefault}{\sfdefault}{bx}{n}

\title{Neighborhood-Aware Unlearning for Generative Recommendation}

\author{%
Pierre Lubitzsch$^{1}$ \quad Yubao Tang$^{2}$ \quad Maarten de Rijke$^{2}$ \quad Sebastian Schelter$^{1}$ \\
\normalfont\normalsize
$^{1}$BIFOLD \& TU Berlin, Germany \quad
$^{2}$University of Amsterdam, The Netherlands \\
\normalfont\normalsize
\texttt{\{lubitzsch, schelter\}@tu-berlin.de}, \texttt{\{y.tang3, m.derijke\}@uva.nl}
}

\iclrfinalcopy 
\begin{document}

\maketitle

\acresetall


\begin{abstract}
\Acl{GenRec} predicts items by autoregressively generating \acp{SID}, whose tokens are shared across semantically related items. With this paradigm gaining adoption, the question of how to support unlearning in it has become pressing. If training interactions must later be removed, such as injected spam or interactions with items a user no longer wants, this sharing lets an unlearning update unintentionally affect the predictions for the interactions' neighbors. We show that this effect is locally structured and hidden by global metrics: unlearning can leave recommendation utility on the whole test set nearly unchanged while substantially degrading it on neighborhoods of the forgotten interactions' \ac{SID}. We propose Neighborhood-Aware Unlearning (NAU), which augments interaction-level unlearning with explicit control over the probability mass assigned to the forgotten items' neighborhoods. Across three datasets and four \ac{SID} tokenizers, NAU restores spam-induced exposure toward the retrained reference while retaining more than 95\% of retrained utility, and recovers the neighborhood utility that forget-and-repair unlearning loses at no global cost. We identify local probability redistribution as an important failure mode of unlearning in \ac{SID}-based generative recommenders and show that explicitly controlling this redistribution can mitigate its effects.
\end{abstract}

\acresetall

\section{Introduction}

\Ac{GenRec} predicts the next item from a user's interaction history by autoregressively generating its \ac{SID}~\citep{rajput_2023_tiger}.
\Acp{SID} represent items as sequences of discrete tokens derived from item representations, such that semantically related items can share parts of their IDs.
As the paradigm is rapidly being adopted across diverse real-world recommendation settings~\citep{damico2026deployingsemanticidbasedgenerative,li2026genrecllmbackedrecommendationranker,collins2026llmbasedgenerativeretrievalsnapchat,tutorialGenIRtang}, it becomes urgent to consider how unlearning can be achieved within it~\citep{chen-2022-recommendation,schelter2023forget}.
Malicious interactions such as spam, fake reviews, or shilling attacks can manipulate recommendations to promote target items~\citep{wu2021trialattack,NEURIPS2022_c1bb0e3b}.
Users may request the removal of interactions involving unwanted items from their recommendation history and future predictions~\citep{schelter2023forget}.
Retraining without the deleted interactions is the natural reference, but it is often too expensive for frequent deletion requests~\citep{chen2024cure4rec,lubitzsch2026erase}, which motivates efficient approximate unlearning.

\header{Challenges of unlearning for \ac{GenRec}}
In a conventional dense recommender, each item has its own output embedding, so an update that lowers one item's score directly affects that item's embedding and shared model parameters, but not the output embeddings of other items~\citep{li2023_scif,kang2018selfattentivesequentialrecommendation}.
In \ac{GenRec}, candidate items are ranked by the probability of generating their \acp{SID}. The item SID probability is the product of the conditional probabilities of the \ac{SID} tokens. Items that share an \ac{SID} prefix therefore share the corresponding prefix-level factors, so updates to these factors can affect multiple items under the same history, with the directly coupled items identifiable from the \ac{SID} structure.
Straightforward adaptations of existing deletion and machine-unlearning strategies do not fully address this shared-token coupling in \ac{GenRec}:
\begin{enumerate*}[label=(\roman*)]
    \item inference-time filtering can remove a target from the recommendation list, while the model may still prefer to generate its \ac{SID};
    \item directly unlearning the forgotten interactions can also affect other items that share \ac{SID} tokens with the target; and
    \item retain-set repair~\citep{jia23model_sparsity_unlearning} can recover overall recommendation quality but does not explicitly control such spillover.
\end{enumerate*}
\emph{The central challenge is therefore to remove the influence of a target item without unnecessarily propagating the unlearning effect to other items that share \ac{SID} tokens with the target.}

\begin{figure}
    \vspace{-15pt}  
    \centering
    \includegraphics[width=0.8\textwidth]{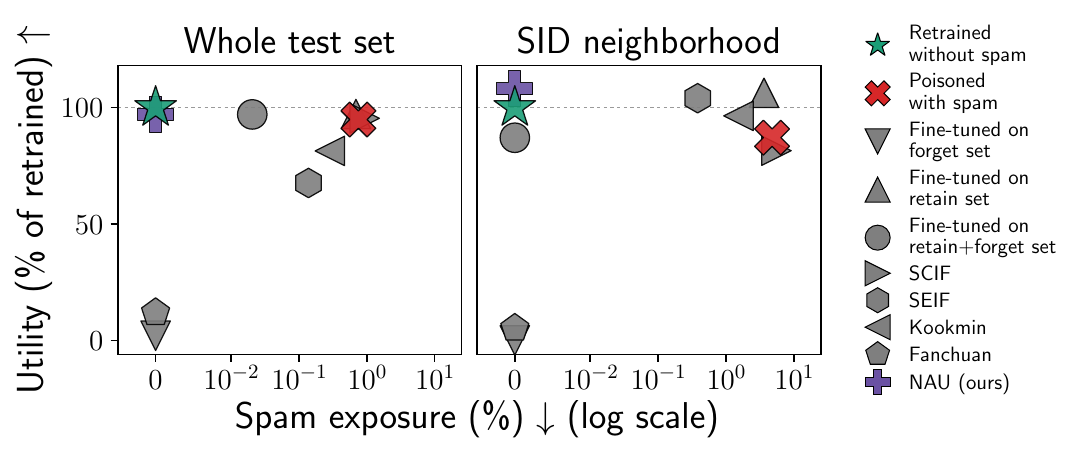}
    \vspace{-10pt}
    \caption{Utility $\uparrow$ ($\mathrm{NDCG}@10$ as a percentage of the retrained model's) versus spam exposure after spam removal, on the whole test set (left) and on the test cases whose next item shares the forgotten target's first \ac{SID} token (right).
    Only NAU removes spam exposure while preserving model utility on both the whole test set and the neighborhood of the forgotten interactions.
    }
    \vspace{-15pt} 
    \label{fig:motivation-locality}
\end{figure}

\header{Diagnosing \ac{SID}-induced unlearning spillover}
We call the items that share \ac{SID} tokens with a target its \emph{\ac{SID} neighborhood}.
Our key observation is that unlearning effects concentrate there: global utility can stay close to the retrained reference while utility on the neighborhood changes considerably (\Cref{fig:motivation-locality}), and methods that ignore the neighborhood cannot both remove influence of interactions to forget
and preserve local and global utility.
This motivates using neighborhoods not only to diagnose unlearning effects, but also to control them.

\header{Our approach}
We propose Neighborhood-Aware Unlearning (NAU) for \ac{GenRec} to remove the influence of forgotten interactions while explicitly controlling how unlearning affects items that share \ac{SID} tokens with the corresponding target items.
Because \ac{GenRec} is trained through token-level \ac{SID} generation whereas deletion requests concern interactions and items, NAU defines an \ac{SID}-aware item-level objective over the generated identifier.
We further design neighborhood mass control to regulate the total generation probability assigned to the target item's neighborhood.
This allows NAU to preserve legitimate neighboring items in spam removal, or suppress neighborhood exposure when nearby items are also relevant to the removal request.
NAU keeps the model architecture and \acp{SID} unchanged and requires no second-order information.
On three Amazon recommendation datasets (Beauty, Sports, and Toys) and four \ac{SID} or tokenization settings, NAU achieves removal behavior close to retraining while retaining over 95\% of its recommendation utility, at three orders of magnitude lower runtime than retraining.
For unwanted item removal, the same neighborhood-control principle further reduces unwanted exposure while preserving recommendation utility.

\header{Contributions}
In summary, our main contributions are:
\begin{itemize}[leftmargin=*,nosep]
    \item We formulate unlearning for \ac{GenRec} under two practical deletion scenarios: spam removal and unwanted item removal (\Cref{sec:problem-definition}).

    \item We identify and analyze \ac{SID}-induced unlearning spillover, showing that unlearning effects are concentrated around items neighboring the target and motivating explicit neighborhood-level control (\Cref{sec:results-spillover}).

    \item We propose neighborhood-aware unlearning (NAU), which combines an \ac{SID}-aware unlearning objective with  neighborhood mass control (\Cref{sec:methodology}).
    
    \item We provide a systematic evaluation of representative general-purpose and recommendation-specific unlearning methods on three Amazon recommendation datasets and four \ac{SID} or tokenization settings, covering both spam removal and unwanted item removal.
    Our results reveal distinct forgetting--utility failure modes of existing methods and show that NAU achieves effective forgetting while preserving over 95\% of retrained utility (\Cref{sec:experimental-setup,sec:results}).

\end{itemize}
We release our implementation, experimental logs, and results at \textcolor{blue}{\url{https://github.com/pierre-lubitzsch/neighborhood-aware-unlearning}}.

\section{{Preliminaries and Problem Definition}}
\label{sec:preliminaries}
\label{sec:problem-definition}

\textbf{Generative recommendation with semantic identifiers.}
A \ac{SID}-based \ac{GenRec} system consists of two main components: an item tokenizer and a generative recommender.
Let $\mathcal{I}$ be the item set.
The tokenizer maps each item $i \in \mathcal{I}$ to a semantic identifier
$\operatorname{SID}(i) = (c_i^1,\ldots,c_i^L)$,
where $c_i^l$ is the semantic token at level $l$ and $L$ is the identifier length.
A common construction obtains these tokens as discrete quantization codes from item representations~\citep{rajput_2023_tiger,luo_2025_rqkmeans,tang2024recentECIR}.
Semantically similar items may share one or more \ac{SID} tokens.
Given these \acp{SID}, the recommender predicts the next item from a user's interaction history by autoregressively generating its identifier.
For a training example $(\bm{x}, i)$, the generation probability of item $i$ is
$
    \mathbb{P}_{\bm{\theta}}(i \mid \bm{x})
    =
    \prod\nolimits_{\scriptscriptstyle l=1}^{\scriptscriptstyle L}
    \mathbb{P}_{\bm{\theta}}
    \bigl(c_i^l \mid c_i^{<l}, \bm{x}\bigr).
$
At inference time, valid generated \acp{SID} are mapped back to items and ranked by their generation probabilities.
Because different items may share \ac{SID} tokens, updates involving shared tokens can affect the generation probabilities of multiple items.

\header{Item neighborhoods}
We use an item neighborhood to capture items close to a target under a chosen proximity criterion.
For an item $i$, let
$
    \mathcal{N}(i) \subseteq \mathcal{I}\setminus\{i\}
$
denote its neighborhood.
It may contain items sharing an \ac{SID} prefix with $i$, or the top-$k$ items under an embedding-based similarity function,
$
    \mathcal{N}_{k}(i)
    :=
    \operatorname*{top-}k_{j \in \mathcal{I}\setminus\{i\}}
    \mathrm{sim}(i,j),
$
We treat neighborhood construction as modular and only assume access to a similarity function.

\header{Unlearning interactions in \ac{GenRec}}
We study interaction-level unlearning in \ac{GenRec}, where deletion requests are specified over individual training interactions.
A training example is represented as $(\bm{x}, i)$, where $\bm{x}$ is the interaction history and $i$ is the next item.
Let $D$ be the original training set, $D_f \subset D$ the forget set, and $D_r = D \setminus D_f$ the retain set.
Let $\bm{\theta}_o$ denote the original model trained on $D$, $\bm{\theta}_r$ the reference model retrained from scratch on $D_r$, and $\bm{\theta}_u$ the model produced by an unlearning algorithm from $\bm{\theta}_o$.
The goal is to remove the influence of $D_f$ while preserving recommendation behavior learned from $D_r$, at substantially lower cost than retraining.

\header{Unlearning scenarios}
We consider two practical deletion scenarios motivated by prior  work~\citep{wu2021trialattack,NEURIPS2022_c1bb0e3b,schelter2023forget,chen2024cure4rec,lubitzsch2026erase}.
They differ in the desired effect on the target item and its neighborhood.
\begin{enumerate}[leftmargin=*,label=(\arabic*)]
\item \emph{Spam removal.}
Fake or poisoned interactions promote a set of target items $\mathcal{I}_{\mathrm{p}} \subset \mathcal{I}$, and $D_f$ contains these injected interactions.
The goal is to remove the resulting promotion effect without globally suppressing target items that may otherwise remain legitimate.

\item \emph{Unwanted item removal.}
A subset of users $\mathcal{U}_{\mathrm{u}} \subset \mathcal{U}$ request removal of interactions involving unwanted items $\mathcal{I}_{\mathrm{u}} \subset \mathcal{I}$, and $D_f$ contains the corresponding interactions.
Unlike spam removal, the goal here is behavioral: requesting users should stop being recommended $\mathcal{I}_{\mathrm{u}}$; we therefore judge removal directly (using $\mathrm{UHF}@10$, see Section~\ref{sec:experimental-setup}) and use the retrained model only as the reference for other users (using $\mathrm{UHR}@10$, see Section~\ref{sec:experimental-setup}).
When nearby items share the semantics motivating the request, the desired reduction in exposure may also extend to their neighborhood.
\end{enumerate}

\header{Evaluation criteria}
We evaluate unlearning along three dimensions:
\begin{enumerate*}[label=(\roman*)]
\item effectiveness, measuring whether the influence associated with $D_f$ is removed;
\item utility, measuring whether recommendation quality on retained data is preserved; and
\item efficiency, the computational cost relative to retraining.
\end{enumerate*}
For spam removal, effectiveness is measured by whether promoted-item exposure returns toward the retrained reference.
For unwanted item removal, it is measured by whether unwanted-item exposure decreases for requesting users.
Because unlearning may affect neighboring items differently from the overall item space, we evaluate both global recommendation quality and local neighborhood behavior.
\section{Methodology}
\label{sec:methodology}

\textbf{Overview.}
We propose Neighborhood-Aware Unlearning (NAU) for \ac{GenRec}.
NAU consists of three main components:
\begin{enumerate*}[label=(\roman*)]
    \item an item-level loss defined over generated \acp{SID};
    \item forget-and-retain optimization that removes the influence of forgotten interactions while preserving retained behavior; and
    \item neighborhood mass control that explicitly regulates how unlearning affects neighboring items.
\end{enumerate*}
Unlike standard forget-and-retain optimization~\citep{kurmanji2023unbounded,triantafillou2024makingprogressunlearningfindings}, NAU does not treat the target of each forgotten interaction in isolation, but explicitly controls its surrounding item neighborhood.

\header{Item-level loss over \acp{SID}}
\Ac{GenRec} is trained by predicting \ac{SID} tokens, whereas unlearning requests concern interactions and their target items.
We therefore define an item-level loss using the likelihood of the complete target \ac{SID}.
For a training pair $(\bm{x}, i)$,
\begin{equation}
    \label{eq:sid-aware-item-loss}
    \ell(i \mid \bm{x}; \bm{\theta})
    =
    -
    \sum\nolimits_{l = 1}^L
    \log
    \mathbb{P}_{\bm{\theta}}
    \left(
        c_i^l
        \mid
        c_i^{<l}, \bm{x}
    \right),
\end{equation}
i.e., the negative log-likelihood of the complete target \ac{SID}.

\header{Forget-and-retain optimization}
We optimize over both forgotten and retained interactions to remove the requested influence while maintaining useful recommendation behavior.

\emph{Retain repair.}
We preserve recommendation behavior on retained interactions with
$
    \mathcal{L}_{\mathrm{ret}}
    =
    \mathbb{E}_{(\bm{x}_r,i_r)\sim D_r}
    \left[
        \ell(i_r \mid \bm{x}_r; \bm{\theta})
    \right].
$

\emph{Forget removal.}
We remove the influence of forgotten interactions by maximizing their item-level loss,
$
    \mathcal{L}_{\mathrm{for}}
    =
    -
    \mathbb{E}_{(\bm{x}_f,i_f)\sim D_f}
    \left[
        \ell(i_f \mid \bm{x}_f; \bm{\theta})
    \right].
$
Minimizing $\mathcal{L}_{\mathrm{for}}$ therefore performs gradient ascent on the forgotten examples and reduces the generation probability of their targets under the corresponding histories.

\emph{Target separation.}
Forget removal suppresses a target under its forgotten histories, but does not directly constrain its ranking under other histories.
We therefore introduce a contrastive separation term over retained examples.
Let
$
    \phi(i \mid \bm{x}; \bm{\theta})
    =
    \log
    \mathbb{P}_{\bm{\theta}}(i \mid \bm{x})
$
denote the item score induced by its generated \ac{SID} probability, and let
$
    \mathcal{I}_f
    =
    \{i : (\bm{x},i)\in D_f\}
$
denote the set of target items associated with forgotten interactions.
We define
\begin{equation}
    \label{eq:separation-loss}
    \mathcal{L}_{\mathrm{sep}}
    =
    \mathbb{E}_{(\bm{x}_r,i_r)\sim D_r}
    \left[
      -
      \log
      \frac{\exp \left(\phi(i_r \mid \bm{x}_r; \bm{\theta})/\tau \right)}
           {\exp \left(\phi(i_r \mid \bm{x}_r; \bm{\theta})/\tau \right)
            + \sum_{i_f \in \mathcal{I}_f \setminus \{ i_r \}}
              \exp \left(\phi(i_f \mid \bm{x}_r; \bm{\theta})/\tau \right)}
    \right],
\end{equation}
where $\tau$ is a temperature.
This term discourages forget-associated target items from remaining highly ranked under retained histories.

\header{Neighborhood mass control}
The preceding terms operate on exact target items but do not explicitly control how unlearning affects their neighbors.
For a forgotten interaction $(\bm{x}_f,i_f)$, let $\mathcal{N}(i_f)$ denote the neighborhood of $i_f$ defined in \Cref{sec:preliminaries}.
We control the total generation probability assigned to this neighborhood through
\begin{equation}
    \label{eq:neighborhood-loss}
    \mathcal{L}_{\mathrm{nbr}}
    =
    \mathbb{E}_{(\bm{x}_f,i_f)\sim D_f}
    \left[
        -
        \log
        \sum\nolimits_{j \in \mathcal{N}(i_f)}
        \mathbb{P}_{\bm{\theta}}
        \left(
            j \mid \bm{x}_f
        \right)
    \right].
\end{equation}

\emph{Full objective.}
Combining the three components, our final unlearning objective NAU is
\begin{equation}
    \label{eq:full-unlearning-objective}
    \mathcal{L}
    =
    \mathcal{L}_{\mathrm{ret}}
    +
    \lambda_f \mathcal{L}_{\mathrm{for}}
    +
    \lambda_s \mathcal{L}_{\mathrm{sep}}
    +
    \lambda_n \mathcal{L}_{\mathrm{nbr}},
\end{equation}
where $\lambda_f$, $\lambda_s$, and $\lambda_n$ control the strengths of the forget, separation, and neighborhood terms, respectively.
The sign of $\lambda_n$ determines how the neighborhood is treated.
When $\lambda_n>0$, minimizing \eqref{eq:full-unlearning-objective} increases the probability assigned to neighboring items, which is useful in spam removal when these items remain legitimate alternatives.
When $\lambda_n<0$, it suppresses neighborhood probability, which is useful in unwanted item removal when nearby items may share the semantics motivating the deletion request.
\section{Experimental Setup}
\label{sec:experimental-setup}

\textbf{Research questions.}
Our experiments are organized around the following research questions:
\begin{enumerate}[leftmargin=*,label=\textbf{(RQ\arabic*)}]
    \item 
    How well does NAU remove scenario-specific target exposure
    while preserving recommendation utility compared to existing unlearning baselines?
    \item 
    How do the retain, forget, separation, and neighborhood mass terms contribute to the final unlearning behavior?
    \item 
    Can the same neighborhood-aware framework handle both the spam removal and unwanted-item removal scenario?
    \item 
    How does unlearning affect the target item's \ac{SID} neighborhood compared with the overall item space?
\end{enumerate}

\textbf{Datasets and backbone.}
We evaluate interaction-level unlearning for \ac{GenRec} on three sequential recommendation datasets from Amazon: Beauty, Sports, and Toys~\citep{mcauley2015image,he2016ups}, using the five-core preprocessing released by \citet{geng2022p5}.
The Amazon subsets are standard evaluation datasets for \ac{GenRec} methods~\citep{rajput_2023_tiger,wang_2024_letter,fu_2026_diffable_sids,kim_2026_mviger,wang_2024_eager,liu_2026_genrec_scaling,cheng2026capsidsoftroutedvariablelengthsemantic,acharya2025glossgenerativelanguagemodels,ju_2025_genrecsemids}.
Dataset statistics are in \Cref{sec:appendix-dataset}.
Our backbone is TIGER~\citep{rajput_2023_tiger}, which quantizes item representations into \acp{SID} and autoregressively generates the next item's identifier.
Our main experiments construct the \acp{SID} with RQ-KMeans~\citep{luo_2025_rqkmeans}; we additionally evaluate RQ-VAE~\citep{rajput_2023_tiger,zeghidour_2021_rqvae} and LETTER's collaboratively regularized tokenizer~\citep{wang_2024_letter} in \Cref{sec:appendix-spam}.

\header{Unlearning scenarios}
For spam removal, we inject a 1\% bandwagon attack promoting one target item, drawn from the unpopular (bottom 20\% by interaction count), mid-popularity (30-70\%), or popular (top 5\%) items, and unlearn the injected fake users; for unwanted-item removal, 12 to 14 users per seed request deletion of their interactions with a sensitive category (\Cref{sec:appendix-scenarios}).

\header{Baselines}
We compare against fine-tuning, forget-only gradient ascent, forget+repair, SCIF ~\citep{li2023_scif}, and the SEIF, Kookmin, and Fanchuan baselines of the NeurIPS 2023 unlearning competition~\citep{triantafillou2024makingprogressunlearningfindings}; TRACER~\citep{chen2026tracertokenreassignmentconcept}, a concept-unlearning method for \ac{GenRec} that reassigns \acp{SID} associated with a target concept, is included where its tokenizer setup is applicable.
Descriptions are in \Cref{sec:appendix-baselines}.

\header{Neighborhood construction}
NAU's training neighborhood $\mathcal{N}(i)$ is the top-$k$ ($k=8$) cosine neighbors of $i$ in the pre-quantization embedding space.
For evaluation we use the tokenizer's own structure instead: the \emph{\ac{SID} neighborhood} of a target is the set of test cases whose next item shares the target's first \ac{SID} token, and the target's own test cases are excluded from the neighborhood scope.

\header{Evaluation metrics}
We evaluate effectiveness, utility, and efficiency, using the retrained model $\bm{\theta}_r$ as the reference where appropriate.
\begin{itemize}[leftmargin=*]
\item For spam removal, $\mathrm{SH}@K$ is the hit rate of promoted target items in the top-$K$ list over the test histories $\mathcal{Q}$:
$
    \mathrm{SH}@K
    =
    \frac{1}{|\mathcal{Q}|}
    \sum_{\bm{x}\in\mathcal{Q}}
    \mathbb{I}
    \left[
        \operatorname{TopK}_{\bm{\theta}}(\bm{x})
        \cap
        \mathcal{I}_{\mathrm{p}}
        \neq
        \emptyset
    \right].
$
Because the retrained model still recommends the promoted items to some users, $\mathrm{SH}@10=0$ is not the target.
We therefore report $\Delta^{\mathrm{retr}}_{\mathrm{SH}} = |\mathrm{SH}_U@10-\mathrm{SH}_R@10|$, averaged over (seed, target) cells, where lower is better and over-suppression counts as much as residual exposure.
Unlike inference-time filtering, all reported unlearning methods modify the model itself; no items are removed from the generated candidate set at evaluation time.
\item Utility is NDCG@10 on the retain test set, where higher is better; 
$\tau_P = \mathrm{NDCG}_U@10/\mathrm{NDCG}_R@10$ measures utility relative to retraining and defines the $95\%$ utility threshold used for model selection.
\item For unwanted-item removal, the unwanted-item hit rates $\mathrm{UHF}@K$ and $\mathrm{UHR}@K$ are computed like $\mathrm{SH}@K$ with $\mathcal{I}_{\mathrm{u}}$ in place of $\mathcal{I}_{\mathrm{p}}$, over the requesting and the remaining users; lower $\mathrm{UHF}@10$ is better, and $\Delta^{\mathrm{retr}}_{\mathrm{UHR}} = |\mathrm{UHR}_U@10-\mathrm{UHR}_R@10|$ measures how far the update moves the remaining users away from retraining.
\item Efficiency $\mathcal{E}$ is the median wall-clock time per request on one H200 GPU.
\end{itemize}
Neighborhood comparisons use paired bootstrap intervals over test cases, since all methods are scored on the same cases (\Cref{sec:appendix-bootstrap}).
$\mathrm{SH}@K$, NDCG@K, and $\tau_P$ are percentages, reported as mean $\pm$ s.d. across five data seeds.

\header{Implementation details}
For every baseline we report the configuration with the lowest validation $\Delta^{\mathrm{retr}}_{\mathrm{SH}}$ (for unwanted-item removal, the lowest $\mathrm{UHF}@10$) among those keeping validation $\tau_P \geq 95\%$, selected separately per dataset (\Cref{tab:appendix-valsel,tab:appendix-valsens}).
NAU's weights ($\lambda_f=0.1$, $\lambda_s=0.01$, $\lambda_n=0.01$) were selected once on the Beauty validation split with RQ-KMeans \acp{SID}: among settings of the full objective whose exposure is not resolvably different from the lowest, the one with the highest mean of whole-set and neighborhood $\tau_P$ (\Cref{tab:appendix-nau-valsel}); they are used unchanged for all datasets and tokenizers, and unwanted-item removal uses $\lambda_s=0$, $\lambda_n=-1$, with $\lambda_n$ selected the same way on the Beauty validation split (\Cref{tab:appendix-valsens}).
\section{Experimental Results}
\label{sec:results}

\textbf{Overall performance for unlearning spam.}
\label{sec:results-spam}
To answer \textbf{RQ1}, we compare NAU with existing unlearning baselines in the spam-removal scenario.
\Cref{tab:bandwagon-unlearn-all-tiger-rkmeans} reports results on Beauty, Sports, and Toys with TIGER and RQ-KMeans \acp{SID}.

\begin{table*}[t]
\centering
\small
\setlength{\tabcolsep}{3pt}
\caption{
Spam removal on Beauty, Sports, and Toys with TIGER + RQ-KMeans \acp{SID} under a 1\% bandwagon attack; mean $\pm$ s.d. over five data seeds $\times$ three targets per dataset.
$\Delta^{\mathrm{retr}}_{\mathrm{SH}} = |\mathrm{SH}_U@10 - \mathrm{SH}_R@10|$ is the distance of the promoted item's exposure from the retrained model's, so leftover exposure and over-suppression both count.
In the Retrain row, the italic value in parentheses is the retrained model's own $\mathrm{SH}@10$, the reference for that column. 
For each baseline, we report the configuration with the smallest validation $\Delta^{\mathrm{retr}}_{\mathrm{SH}}$ among those keeping validation $\tau_P \geq 95\%$ (ties broken by higher $\tau_P$), or, when no configuration reaches $95\%$, the smallest $\Delta^{\mathrm{retr}}_{\mathrm{SH}}$ overall. 
NAU is reported at one operating point for all datasets ($\lambda_f=0.1$, $\lambda_s=0.01$, $\lambda_n=0.01$); the indented row below it uses a stronger separation weight ($\lambda_s=0.1$) and is not ranked.
All values are percentages. Bold and underline mark the best and second-best value per column among methods keeping $\tau_P \geq 95\%$.
TRACER has no Beauty run under RQ-KMeans; cells marked \textsuperscript{\dag} are its Beauty result under its own tokenizer, where $\tau_P$ is measured against that tokenizer's retrained reference.
}
\label{tab:bandwagon-unlearn-all-tiger-rkmeans}
\resizebox{\textwidth}{!}{%
\begin{tabular}{lrrrrrrrrr}
\toprule
& \multicolumn{3}{c}{Beauty} & \multicolumn{3}{c}{Sports} & \multicolumn{3}{c}{Toys} \\
\cmidrule(lr){2-4} \cmidrule(lr){5-7} \cmidrule(lr){8-10}
Method & $\Delta^{\mathrm{retr}}_{\mathrm{SH}}{\downarrow}$ & $\mathrm{NDCG}@10{\uparrow}$ & $\tau_P{\uparrow}$ & $\Delta^{\mathrm{retr}}_{\mathrm{SH}}{\downarrow}$ & $\mathrm{NDCG}@10\uparrow$ & $\tau_P{\uparrow}$ & $\Delta^{\mathrm{retr}}_{\mathrm{SH}}{\downarrow}$ & $\mathrm{NDCG}@10\uparrow$ & $\tau_P{\uparrow}$ \\
\midrule
\textit{Retrained} & \textit{(0.072)} & 3.766\,{\scriptsize$\pm$0.13} & 100.0 & \textit{(0.139)} & 2.033\,{\scriptsize$\pm$0.07} & 100.0 & \textit{(0.073)} & 3.465\,{\scriptsize$\pm$0.11} & 100.0 \\
\textit{Poisoned} & 0.902\,{\scriptsize$\pm$0.59} & 3.589\,{\scriptsize$\pm$0.03} & 95.4\,{\scriptsize$\pm$3.32} & 0.168\,{\scriptsize$\pm$0.17} & 2.031\,{\scriptsize$\pm$0.06} & 100.0\,{\scriptsize$\pm$2.96} & 0.221\,{\scriptsize$\pm$0.14} & 3.387\,{\scriptsize$\pm$0.06} & 97.8\,{\scriptsize$\pm$3.29} \\
\midrule
\multicolumn{10}{l}{\textit{Baselines}} \\
Finetune & 0.767\,{\scriptsize$\pm$0.67} & \textbf{3.626}\,{\scriptsize$\pm$0.04} & \textbf{96.4}\,{\scriptsize$\pm$2.80} & 0.281\,{\scriptsize$\pm$0.23} & \textbf{2.082}\,{\scriptsize$\pm$0.04} & \textbf{102.5}\,{\scriptsize$\pm$3.52} & 0.823\,{\scriptsize$\pm$0.32} & \underline{3.350}\,{\scriptsize$\pm$0.06} & \underline{96.8}\,{\scriptsize$\pm$4.03} \\
Forget only & 0.072\,{\scriptsize$\pm$0.04} & 0.086\,{\scriptsize$\pm$0.14} & 2.2\,{\scriptsize$\pm$3.57} & 0.139\,{\scriptsize$\pm$0.21} & 0.014\,{\scriptsize$\pm$0.02} & 0.7\,{\scriptsize$\pm$1.20} & 0.073\,{\scriptsize$\pm$0.06} & 0.012\,{\scriptsize$\pm$0.01} & 0.3\,{\scriptsize$\pm$0.24} \\
Forget + Repair & \underline{0.106}\,{\scriptsize$\pm$0.12} & 3.607\,{\scriptsize$\pm$0.07} & 95.9\,{\scriptsize$\pm$3.29} & \textbf{0.137}\,{\scriptsize$\pm$0.20} & \underline{2.068}\,{\scriptsize$\pm$0.04} & \underline{101.8}\,{\scriptsize$\pm$3.83} & \underline{0.195}\,{\scriptsize$\pm$0.14} & \textbf{3.355}\,{\scriptsize$\pm$0.05} & \textbf{96.9}\,{\scriptsize$\pm$3.65} \\
SCIF & 1.309\,{\scriptsize$\pm$0.66} & 3.563\,{\scriptsize$\pm$0.05} & 94.7\,{\scriptsize$\pm$2.95} & 0.467\,{\scriptsize$\pm$0.37} & 2.027\,{\scriptsize$\pm$0.05} & 99.8\,{\scriptsize$\pm$3.88} & 1.232\,{\scriptsize$\pm$0.81} & 3.318\,{\scriptsize$\pm$0.06} & 95.8\,{\scriptsize$\pm$3.59} \\
SEIF & 0.243\,{\scriptsize$\pm$0.28} & 2.506\,{\scriptsize$\pm$0.08} & 66.6\,{\scriptsize$\pm$3.57} & 0.139\,{\scriptsize$\pm$0.19} & 1.218\,{\scriptsize$\pm$0.05} & 59.9\,{\scriptsize$\pm$2.09} & 0.206\,{\scriptsize$\pm$0.15} & 2.185\,{\scriptsize$\pm$0.09} & 63.1\,{\scriptsize$\pm$4.29} \\
Kookmin & 0.476\,{\scriptsize$\pm$0.32} & 3.046\,{\scriptsize$\pm$0.09} & 81.0\,{\scriptsize$\pm$3.41} & 0.148\,{\scriptsize$\pm$0.09} & 1.626\,{\scriptsize$\pm$0.07} & 80.0\,{\scriptsize$\pm$3.94} & 0.558\,{\scriptsize$\pm$0.23} & 2.664\,{\scriptsize$\pm$0.03} & 77.0\,{\scriptsize$\pm$2.47} \\
Fanchuan & 0.072\,{\scriptsize$\pm$0.04} & 1.225\,{\scriptsize$\pm$0.09} & 32.6\,{\scriptsize$\pm$3.01} & 0.139\,{\scriptsize$\pm$0.21} & 0.643\,{\scriptsize$\pm$0.03} & 31.6\,{\scriptsize$\pm$0.96} & 0.073\,{\scriptsize$\pm$0.06} & 0.805\,{\scriptsize$\pm$0.10} & 23.3\,{\scriptsize$\pm$3.13} \\
TRACER & 1.063\,{\scriptsize$\pm$0.67}\rlap{\textsuperscript{\dag}} & 3.566\,{\scriptsize$\pm$0.07}\rlap{\textsuperscript{\dag}} & 99.1\,{\scriptsize$\pm$1.98}\rlap{\textsuperscript{\dag}} & 0.410\,{\scriptsize$\pm$0.34} & 2.028\,{\scriptsize$\pm$0.05} & 99.8\,{\scriptsize$\pm$4.03} & 1.079\,{\scriptsize$\pm$0.78} & 3.314\,{\scriptsize$\pm$0.07} & 95.7\,{\scriptsize$\pm$4.08} \\
\midrule
\multicolumn{10}{l}{\textit{Ours}} \\
NAU \textit{(ours)} & \textbf{0.081}\,{\scriptsize$\pm$0.04} & \underline{3.621}\,{\scriptsize$\pm$0.04} & \underline{96.2}\,{\scriptsize$\pm$3.01} & \underline{0.138}\,{\scriptsize$\pm$0.21} & 2.067\,{\scriptsize$\pm$0.04} & 101.8\,{\scriptsize$\pm$3.84} & \textbf{0.074}\,{\scriptsize$\pm$0.06} & 3.349\,{\scriptsize$\pm$0.06} & 96.8\,{\scriptsize$\pm$4.22} \\
\quad NAU, $\lambda_s{=}0.1$ & 0.072\,{\scriptsize$\pm$0.04} & 3.617\,{\scriptsize$\pm$0.04} & 96.1\,{\scriptsize$\pm$3.14} & 0.139\,{\scriptsize$\pm$0.21} & 2.067\,{\scriptsize$\pm$0.04} & 101.8\,{\scriptsize$\pm$3.86} & 0.072\,{\scriptsize$\pm$0.06} & 3.350\,{\scriptsize$\pm$0.06} & 96.8\,{\scriptsize$\pm$4.13} \\
\bottomrule
\end{tabular}}
\vspace*{-10pt}
\end{table*}

Among methods retaining at least $95\%$ of retrained NDCG@10, NAU achieves the smallest
$\Delta_{\mathrm{SH}}^{\mathrm{retr}}$ on Beauty and Toys ($0.081\%$ and $0.074\%$), and is within $0.001$ points of the best result on Sports ($0.138\%$ versus $0.137\%$), while retaining $96.2$ to $101.8\%$ of retrained utility.
The baselines exhibit three main failure modes:
\begin{enumerate*}[label=(\roman*)]
    \item Methods such as Finetune, SCIF, and TRACER preserve utility but remain far from the retrained exposure, on Beauty and Toys often farther than the poisoned model itself;
    \item Forget only, Fanchuan, SEIF, and Kookmin can achieve smaller exposure distances only with substantial utility degradation; and
    \item Forget+Repair provides the strongest baseline trade-off, but still remains farther from retraining than NAU on Beauty and Toys at comparable utility.
\end{enumerate*}
For example, on Toys, Finetune, SCIF, and TRACER have
$\Delta_{\mathrm{SH}}^{\mathrm{retr}}$ values of $0.823\%$, $1.232\%$, and $1.079\%$, respectively, compared with $0.074\%$ for NAU.
Forget only and Fanchuan keep at most $2.2\%$ and $32.6\%$ of retrained utility, while SEIF and Kookmin retain only $59.9$ to $81.0\%$.
Compared with Forget+Repair, NAU reduces $\Delta_{\mathrm{SH}}^{\mathrm{retr}}$ from $0.106\%$ to $0.081\%$ on Beauty and from $0.195\%$ to $0.074\%$ on Toys; on Sports, the two methods are nearly identical ($0.137\%$ versus $0.138\%$).
A stronger separation weight ($\lambda_s=0.1$, indented row) is slightly closer to retrained exposure on Beauty and Toys ($0.072\%$ on both) but preserves less neighborhood utility (\Cref{sec:results-spillover}).
As analyzed next, similar whole-set utility can still mask substantially different behavior around the target's \ac{SID} neighborhood.
Results with RQ-VAE, LETTER, and TRACER identifiers are consistent (\Cref{sec:appendix-spam}).

\begin{table}[t]
\centering
\small
\setlength{\tabcolsep}{2.5pt}
\renewcommand{\arraystretch}{1.05}
\caption{
Ablation of NAU on Beauty with TIGER + RQ-KMeans over five seeds.
Indented rows are alternative additions to
$\mathcal{L}_{\mathrm{ret}}+\mathcal{L}_{\mathrm{for}}$
(siblings, not a ladder).
Since retrained exposure is $0\%$ in both query sets,
$\mathrm{SH}@10$ also equals the absolute distance to retrained exposure.
$\tau_P$ denotes $\mathrm{NDCG}@10$ as a percentage of the retrained model's utility within each query set.
}
\label{tab:scoped-ablation-compact-beauty}

\begin{tabular}{lcccccc}
\toprule
\multirow{2}{*}[-3pt]{\textbf{Objective}}
& \multicolumn{3}{c}{\textbf{Whole test set}}
& \multicolumn{3}{c}{\textbf{SID neighborhood}} \\
\cmidrule(lr){2-4} \cmidrule(lr){5-7}
& $\Delta^{\mathrm{retr}}_{\mathrm{SH}}{\scriptstyle\,(\%)}{\downarrow}$
& $\mathrm{NDCG}@10{\uparrow}$
& $\tau_P{\scriptstyle\,(\%)}{\uparrow}$
& $\Delta^{\mathrm{retr}}_{\mathrm{SH}}{\scriptstyle\,(\%)}{\downarrow}$
& $\mathrm{NDCG}@10{\uparrow}$
& $\tau_P{\scriptstyle\,(\%)}{\uparrow}$ \\
\midrule

\textit{Retrained}
& \textit{0.000\,{\scriptsize$\pm$0.00}}
& \textit{3.766\,{\scriptsize$\pm$0.13}}
& \textit{100.0}
& \textit{0.000\,{\scriptsize$\pm$0.00}}
& \textit{2.779\,{\scriptsize$\pm$1.63}}
& \textit{100.0} \\

\textit{Poisoned}
& \textit{0.753\,{\scriptsize$\pm$0.79}}
& \textit{3.571\,{\scriptsize$\pm$0.09}}
& \textit{\phantom{0}94.8}
& \textit{4.780\,{\scriptsize$\pm$11.03}}
& \textit{2.420\,{\scriptsize$\pm$1.23}}
& \textit{\phantom{0}87.1} \\

\midrule

$\mathcal{L}_{\mathrm{ret}}$ only
& 0.692\,{\scriptsize$\pm$0.77}
& 3.659\,{\scriptsize$\pm$0.06}
& \phantom{0}97.2
& 3.633\,{\scriptsize$\pm$4.84}
& 2.954\,{\scriptsize$\pm$1.76}
& 106.3 \\

$+\ \mathcal{L}_{\mathrm{for}}$
& 0.021\,{\scriptsize$\pm$0.04}
& 3.654\,{\scriptsize$\pm$0.07}
& \phantom{0}97.0
& 0.000\,{\scriptsize$\pm$0.00}
& 2.419\,{\scriptsize$\pm$1.44}
& \phantom{0}87.0 \\

\quad $+\ \mathcal{L}_{\mathrm{sep}}$ ($\lambda_s=.01$)
& 0.001\,{\scriptsize$\pm$0.00}
& 3.648\,{\scriptsize$\pm$0.05}
& \phantom{0}96.9
& 0.000\,{\scriptsize$\pm$0.00}
& 2.428\,{\scriptsize$\pm$1.42}
& \phantom{0}87.4 \\

\quad $+\ \mathcal{L}_{\mathrm{sep}}$ ($\lambda_s=.1$)
& 0.000\,{\scriptsize$\pm$0.00}
& 3.644\,{\scriptsize$\pm$0.04}
& \phantom{0}96.8
& 0.000\,{\scriptsize$\pm$0.00}
& 2.021\,{\scriptsize$\pm$1.50}
& \phantom{0}72.7 \\

\quad $+\ \mathcal{L}_{\mathrm{nbr}}$ ($\lambda_n=.01$)
& 0.004\,{\scriptsize$\pm$0.01}
& 3.651\,{\scriptsize$\pm$0.07}
& \phantom{0}97.0
& 0.000\,{\scriptsize$\pm$0.00}
& 3.036\,{\scriptsize$\pm$1.74}
& 109.3 \\

\quad $+\ \mathcal{L}_{\mathrm{nbr}}$ ($\lambda_n=.1$)
& 0.000\,{\scriptsize$\pm$0.00}
& 3.548\,{\scriptsize$\pm$0.08}
& \phantom{0}94.2
& 0.000\,{\scriptsize$\pm$0.00}
& 4.641\,{\scriptsize$\pm$3.64}
& 167.0 \\

\midrule

NAU ($\lambda_s=.01,\lambda_n=.01$)
& 0.000\,{\scriptsize$\pm$0.00}
& 3.653\,{\scriptsize$\pm$0.05}
& \phantom{0}97.0
& 0.000\,{\scriptsize$\pm$0.00}
& 3.003\,{\scriptsize$\pm$1.70}
& 108.1 \\

NAU ($\lambda_s=.1,\lambda_n=.01$)
& 0.000\,{\scriptsize$\pm$0.00}
& 3.652\,{\scriptsize$\pm$0.05}
& \phantom{0}97.0
& 0.000\,{\scriptsize$\pm$0.00}
& 2.659\,{\scriptsize$\pm$1.54}
& \phantom{0}95.7 \\

NAU ($\lambda_s=.1,\lambda_n=.1$)
& 0.000\,{\scriptsize$\pm$0.00}
& 3.557\,{\scriptsize$\pm$0.05}
& \phantom{0}94.5
& 0.000\,{\scriptsize$\pm$0.00}
& 3.094\,{\scriptsize$\pm$2.00}
& 111.3 \\
\bottomrule
\end{tabular}
\vspace*{-10pt}
\end{table}

\header{Ablation of the unlearning objective}
\label{sec:results-ablation}
We ablate the NAU objective on Beauty to answer \textbf{RQ2}.
\Cref{tab:scoped-ablation-compact-beauty} reports performance on both the whole test set and the target-item \ac{SID} neighborhood.

\emph{Role of each objective term.}
The four terms play complementary roles:
\begin{enumerate*}[label=(\roman*)]
    \item $\mathcal{L}_{\mathrm{ret}}$ preserves recommendation utility but removes little spam exposure on its own;
    \item adding $\mathcal{L}_{\mathrm{for}}$ provides the main removal effect, reducing whole-set $\Delta_{\mathrm{SH}}^{\mathrm{retr}}$ from $0.692\%$ to $0.021\%$ while retaining $97.0\%$ of retrained utility;
    \item $\mathcal{L}_{\mathrm{sep}}$ further suppresses target exposure to near zero; and
    \item $\mathcal{L}_{\mathrm{nbr}}$ primarily regulates how the update affects neighboring items.
\end{enumerate*}

The neighborhood results reveal changes that are largely hidden by whole-set utility.
Adding $\mathcal{L}_{\mathrm{for}}$ keeps whole-set utility at $97.0\%$ of retraining, but neighborhood utility drops to $87.0\%$.
The separation term at $\lambda_s=0.1$, which removes the promoted item completely, reduces neighborhood utility further to $72.7\%$ at an unchanged whole-set utility of $96.8\%$, whereas the weaker $\lambda_s=0.01$ leaves both nearly unchanged ($96.9\%$ and $87.4\%$).
In contrast, the variant with $\mathcal{L}_{\mathrm{nbr}}$ at $\lambda_n=0.01$ yields $109.3\%$ neighborhood utility while maintaining $97.0\%$ whole-set utility.
The full objective ($\lambda_s=0.01,\lambda_n=0.01$) combines both: it matches the retrained exposure and reaches $108.1\%$ of retrained neighborhood utility at $97.0\%$ whole-set utility; with the stronger separation ($\lambda_s=0.1$), the neighborhood term still recovers neighborhood utility from $72.7\%$ to $95.7\%$.

Overall, the retain term preserves utility, the forget and separation terms control target exposure, and neighborhood mass control regulates the local effect of unlearning.
Additional ablations on Sports and Toys show the same qualitative pattern and are reported in \Cref{sec:appendix-ablation}.

\header{Unwanted item removal}
\label{sec:results-unwanted}
To answer \textbf{RQ3}, we evaluate whether the same neighborhood-aware framework can also handle unwanted item removal.
Unlike spam removal, unwanted item removal may require suppressing nearby items that share similar unwanted semantics.
We therefore reverse the direction of neighborhood control by using a negative neighborhood weight.

\begin{table}[t]
\centering
\small
\setlength{\tabcolsep}{4pt}
\renewcommand{\arraystretch}{1.05}
\caption{
Unwanted item removal on Beauty (Hair Loss Products) with TIGER + RQ-KMeans, five data seeds.
$\mathrm{UHF}@10$ and $\mathrm{UHR}@10$ are the unwanted item hit rates of the requesting and the remaining users; $\Delta^{\mathrm{retr}}_{\mathrm{UHR}}$ is the mean over seeds of $|\mathrm{UHR}_U@10-\mathrm{UHR}_R@10|$, the distance to the retrained model of the same seed.
Each method uses the configuration selected on the validation split (\Cref{tab:appendix-valsens}; $^{\dagger}$: $\tau_P<95\%$, never ranked); NAU uses $\lambda_n=-1$, $\lambda_s=0$.
Retraining does not lower $\mathrm{UHF}@10$ below the original model.
}
\label{tab:paper-sensitive-beauty-tiger-rkmeans}
\resizebox{\textwidth}{!}{%
\begin{tabular}{lcccccc}
\toprule
\textbf{Method} & $\mathrm{UHF}@10{\scriptstyle\,(\%)}{\downarrow}$ & $\mathrm{UHR}@10{\scriptstyle\,(\%)}$ & $\Delta^{\mathrm{retr}}_{\mathrm{UHR}}{\scriptstyle\,(\%)}{\downarrow}$ & $\mathrm{NDCG}@10{\uparrow}$ & $\tau_P$ (\%)${\uparrow}$ & $\mathcal{E}{\scriptstyle\,(s)}{\downarrow}$ \\
\midrule
Finetune & 14.424\,{\scriptsize$\pm$13.461} & 2.948\,{\scriptsize$\pm$0.550} & \underline{0.612\,{\scriptsize$\pm$0.299}} & 3.510\,{\scriptsize$\pm$0.162} & 96.3\,{\scriptsize$\pm$5.0} & 0.3 \\
Forget only & \textbf{9.419\,{\scriptsize$\pm$6.109}} & 1.990\,{\scriptsize$\pm$1.176} & 1.111\,{\scriptsize$\pm$0.684} & 3.516\,{\scriptsize$\pm$0.183} & 96.4\,{\scriptsize$\pm$5.8} & 0.3 \\
Forget+Repair & \underline{12.606\,{\scriptsize$\pm$14.899}} & 2.543\,{\scriptsize$\pm$0.840} & 0.734\,{\scriptsize$\pm$0.337} & \underline{3.522\,{\scriptsize$\pm$0.168}} & \underline{96.6\,{\scriptsize$\pm$5.2}} & 0.5 \\
SCIF & 15.981\,{\scriptsize$\pm$8.902} & 3.303\,{\scriptsize$\pm$0.424} & \textbf{0.481\,{\scriptsize$\pm$0.379}} & \textbf{3.682\,{\scriptsize$\pm$0.139}} & \textbf{101.0\,{\scriptsize$\pm$4.6}} & 33.9 \\
SEIF$^{\dagger}$ & 0.000\,{\scriptsize$\pm$0.000} & 0.000\,{\scriptsize$\pm$0.000} & 2.823\,{\scriptsize$\pm$0.474} & 0.030\,{\scriptsize$\pm$0.032} & 0.8\,{\scriptsize$\pm$0.9} & 0.3 \\
Kookmin$^{\dagger}$ & 1.818\,{\scriptsize$\pm$4.066} & 1.230\,{\scriptsize$\pm$0.939} & 1.699\,{\scriptsize$\pm$1.180} & 2.788\,{\scriptsize$\pm$0.179} & 76.5\,{\scriptsize$\pm$5.6} & 0.3 \\
Fanchuan$^{\dagger}$ & 0.000\,{\scriptsize$\pm$0.000} & 0.000\,{\scriptsize$\pm$0.000} & 2.823\,{\scriptsize$\pm$0.474} & 0.164\,{\scriptsize$\pm$0.052} & 4.5\,{\scriptsize$\pm$1.4} & 1.2 \\
\cmidrule(lr){1-7}
NAU \textit{(Ours)} & \textbf{9.419\,{\scriptsize$\pm$6.109}} & 2.277\,{\scriptsize$\pm$0.589} & 0.666\,{\scriptsize$\pm$0.530} & 3.484\,{\scriptsize$\pm$0.191} & 95.5\,{\scriptsize$\pm$5.6} & 3.7 \\
\midrule
\textit{Original (before deletion)} & 14.850\,{\scriptsize$\pm$7.051} & 3.449\,{\scriptsize$\pm$0.843} & 0.702\,{\scriptsize$\pm$0.460} & 3.765\,{\scriptsize$\pm$0.134} & 103.3\,{\scriptsize$\pm$4.4} & -- \\
\textit{Retrain (reference)} & 16.237\,{\scriptsize$\pm$9.987} & 2.823\,{\scriptsize$\pm$0.474} & 0.000 & 3.648\,{\scriptsize$\pm$0.038} & 100.0 & \textit{17,399} \\
\bottomrule
\end{tabular}}
\vspace*{-10pt}
\end{table}


\begin{table}[t]
\centering
\small
\setlength{\tabcolsep}{3.5pt}
\caption{
Effect of the neighborhood term in unwanted item removal across datasets and tokenizers: NAU with $\lambda_n=-1$ against the same objective without the term ($\lambda_n=0$; $\lambda_f=0.1$, $\lambda_s=0$), paired over five data seeds.
Each cell gives $\lambda_n=0 \rightarrow -1$ (mean over seeds) and, in parentheses, on how many of the five seeds the value decreases ($\mathrm{UHF}$, $\mathrm{UHR}$) or increases ($\tau_P$).
Retrained $\mathrm{UHF}@10$ and $\mathrm{UHR}@10$ are given for reference.
}
\label{tab:ablation-unified-sensitive-lambda-n-all}
\begin{tabular}{llccccc}
\toprule
Dataset & \acp{SID} & $\mathrm{UHF}@10{\scriptstyle\,(\%)}{\downarrow}$ & $\mathrm{UHR}@10{\scriptstyle\,(\%)}$ & $\tau_P{\scriptstyle\,(\%)}{\uparrow}$ & Retrained $\mathrm{UHF}$ / $\mathrm{UHR}$ \\
\midrule
Beauty & RQ-KMeans & 12.2 $\rightarrow$ 9.4 (1/5)\phantom{0} & 2.3 $\rightarrow$ 2.3 (2/5) & 93.3 $\rightarrow$ 95.5 (5/5) & 16.2 / 2.82 \\
Beauty & LETTER & 15.4 $\rightarrow$ 10.1 (3/5) & 4.1 $\rightarrow$ 4.4 (3/5) & 89.4 $\rightarrow$ 89.7 (4/5) & 15.0 / 6.66 \\
Sports & RQ-KMeans & 27.6 $\rightarrow$ 11.1 (5/5) & 1.4 $\rightarrow$ 0.5 (5/5) & 91.1 $\rightarrow$ 92.4 (3/5) & 29.3 / 1.42 \\
Sports & LETTER & 22.0 $\rightarrow$ 12.1 (5/5) & 1.4 $\rightarrow$ 0.6 (5/5) & 94.2 $\rightarrow$ 95.0 (3/5) & 37.1 / 1.91 \\
Toys & RQ-KMeans & \phantom{0}5.0 $\rightarrow$ 9.7 (0/5)\phantom{0} & 2.3 $\rightarrow$ 2.4 (3/5) & 92.1 $\rightarrow$ 95.1 (5/5) & 18.1 / 3.51 \\
Toys & LETTER & 14.8 $\rightarrow$ 9.7 (3/5)\phantom{0} & 3.4 $\rightarrow$ 2.9 (5/5) & 90.5 $\rightarrow$ 93.5 (5/5) & \phantom{0}6.6 / 3.44 \\
\bottomrule
\end{tabular}
\vspace*{-10pt}
\end{table}

\emph{Comparison with baselines.}
Table~\ref{tab:paper-sensitive-beauty-tiger-rkmeans} compares NAU with existing unlearning methods on Beauty.
Among methods retaining at least $95\%$ of retrained utility,
\begin{enumerate*}[label=(\roman*)]
    \item NAU and Forget only reach the lowest $\mathrm{UHF}@10$ ($9.419\%$, against $14.850\%$ before deletion);
    \item methods achieving substantially lower exposure, such as SEIF and Fanchuan, incur severe utility degradation; and
    \item at the same $\mathrm{UHF}@10$, NAU moves the remaining users' exposure less than Forget only ($\Delta^{\mathrm{retr}}_{\mathrm{UHR}}$ of $0.666$ versus $1.111$), keeping it about as close to retraining as before deletion ($0.702$), at $95.5\%$ of retrained utility.
\end{enumerate*}

\emph{Effect of neighborhood control.}
In \Cref{tab:ablation-unified-sensitive-lambda-n-all}, $\lambda_n=-1$ lowers $\mathrm{UHF}@10$ on Sports on every seed with both tokenizers ($27.6\%\rightarrow11.1\%$ and $22.0\%\rightarrow12.1\%$) at unchanged utility, while also lowering the remaining users' exposure ($\mathrm{UHR}@10$ $1.4\%\rightarrow0.5\%$), since it acts on the category rather than on specific users.
On Beauty and Toys, where only 12 to 14 users request deletion, it mainly raises $\tau_P$ (on every seed in three of four cells).
The same mechanism thus suppresses rather than preserves probability mass around unwanted items.

\emph{Across datasets.}
In the full comparisons (\Cref{sec:appendix-unwanted}), NAU has the second-lowest $\mathrm{UHF}@10$ on Toys among methods keeping $95\%$ utility ($9.7\%$, behind Finetune's $6.5\%$), and the lowest of any utility-preserving method on Sports ($11.1\%$, from $49.8\%$), at $92.4\%$ utility; there, even Finetune, without a forget term, stays below the threshold ($93.0\%$).

\begin{figure}[t]
    \centering
    \includegraphics[width=\textwidth]{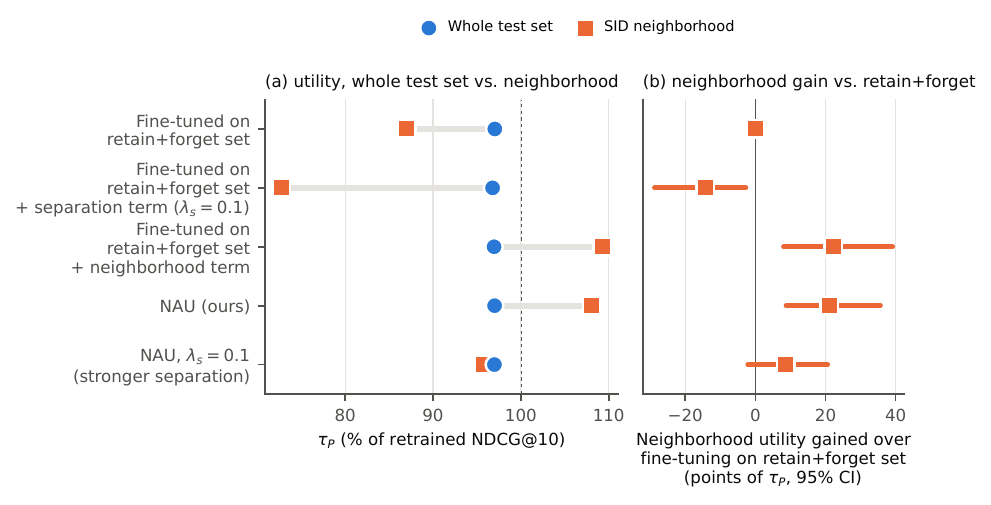}
    \vspace*{-15pt}
    \caption{Effect of each objective term on the neighborhood (spam removal on Beauty, TIGER + RQ-KMeans, mid-popularity target, five seeds). (a) Utility as a percentage of the retrained model's, on the whole test set and on the target's
    \ac{SID} neighborhood (523 test cases whose next item shares the target's first \ac{SID} token). (b) Neighborhood utility gained over fine-tuning on the retain and forget sets, in percentage points, with 95\% paired bootstrap
    intervals (\Cref{sec:appendix-bootstrap}); intervals excluding 0 mark reliable differences. The last row uses a stronger separation weight ($\lambda_s=0.1$) and is not selected (\Cref{tab:appendix-nau-valsel}). Baselines are
    compared in \Cref{fig:motivation-locality} and \Cref{tab:bandwagon-unlearn-all-tiger-rkmeans}.
    }
    \label{fig:scope-locality}
    \vspace*{-15pt}    
\end{figure}

\header{Neighborhood spillover}
\label{sec:results-spillover}
To answer \textbf{RQ4}, \Cref{fig:scope-locality} compares, for the terms of the NAU objective, utility on the whole test set with utility on the forgotten target's \ac{SID} neighborhood; \Cref{fig:motivation-locality} makes the same comparison for the baselines.
On the whole test set, all objectives shown lie within 0.2 points of each other (96.8 to 97.0\% of retrained NDCG@10); on the neighborhood, they spread over 35 points.
Fine-tuning on the retain and forget sets (Forget+Repair) drops to 87.0\% on the neighborhood, and adding the separation term at $\lambda_s=0.1$, which removes the promoted item completely, costs a further 14.3 points (95\% interval $[-28.7, -2.8]$) with no whole-set change.
NAU both removes the promoted item and gains 21.0 points on the neighborhood over Forget+Repair ($[+8.8, +35.6]$) at identical whole-set utility.
With the stronger separation weight ($\lambda_s=0.1$), the neighborhood term still recovers most of the separation loss, 23.0 points over the separation term alone ($[+7.6, +44.0]$), but ends at $95.7\%$ of retrained neighborhood utility; $\lambda_s$ thus trades strict removal of the promoted item against neighborhood preservation.
Across datasets and target popularities, NAU's neighborhood gain over Forget+Repair is positive in eight of nine settings (\Cref{fig:scope-grid}).

\section{Related Work}

\textbf{\Ac{GenRec} with \acp{SID}.} \ac{GenRec} formulates recommendation as generating an item identifier rather than scoring all items~\citep{geng2022p5,rajput_2023_tiger,tang2024recentWWW}, commonly an \ac{SID} obtained by quantizing item representations~\citep{hou_2023_vqrec,rajput_2023_tiger,tang2023recent}, as in TIGER~\citep{rajput_2023_tiger} and LETTER~\citep{wang_2024_letter}.
Recent work studies \ac{SID} construction~\citep{fu_2026_diffable_sids}, generative retrieval objectives~\citep{penha_2025_sids_joint_search_and_rec}, scalability~\citep{liu_2026_genrec_scaling}, long \acp{SID}~\citep{xia2026unleashpotentiallongsemantic}, and deployment~\citep{damico2026deployingsemanticidbasedgenerative}, all under a fixed training set.

\header{Machine unlearning for recommendation}
Machine unlearning removes the influence of selected training data without retraining from scratch~\citep{cao_2015_unlearn}.
Recommender unlearning has been studied for score-based, matrix-factorization, graph-based, and neural recommenders~\citep{li2023_scif,kun_wu_2023,Wu_2023,dong_2024}, including exact unlearning for specific architectures~\citep{schelter2023forget}, typically via influence estimation, gradient updates, retain-set repair, or model-specific approximations.
These methods are not designed for \ac{SID}-based \ac{GenRec}, where each interaction supervises an autoregressive token sequence, and do not model how forgetting one item moves probability mass over items that share its tokens.
TRACER~\citep{chen2026tracertokenreassignmentconcept} unlearns concepts in \ac{GenRec} by reassigning the \acp{SID} of the concept's items; we instead delete interactions while keeping users, items, and \acp{SID} fixed, and control how probability is redistributed over the existing neighborhood.

\section{Conclusion}
\label{sec:conclusion}

We studied interaction-level unlearning for \ac{GenRec}, where items are generated through \acp{SID} and semantically related items may share identifier tokens.
This makes it challenging to remove the influence of forgotten interactions without unnecessarily affecting neighboring items.
We proposed NAU, a neighborhood-aware unlearning framework that combines \ac{SID}-aware item-level unlearning with neighborhood mass control to regulate 
how unlearning affects nearby items.
Across spam removal and unwanted item removal scenarios, NAU 
achieves effective forgetting while preserving recommendation utility, with the model architecture and \acp{SID} kept fixed.

\header{Limitations and future directions}
Our study focuses on practical approximate unlearning for \ac{SID}-based \ac{GenRec}.
One limitation is that we evaluate a representative set of deletion scenarios, while real systems may face more diverse and repeated deletion requests.
Another limitation is that neighborhood construction is treated as a configurable design choice; future work could study adaptive neighborhoods that depend on the tokenizer, item representation space, and deletion goal.
These directions can further strengthen unlearning methods that treat forgetting not only as exact-target removal, but as controlled local change in the item space.

\clearpage
\section*{AI disclosure statement}

In this work, we used generative AI tools for support in implementing methods, assisting with translation, and support in qualitative and thematic data analysis. We have not used generative AI tools for generating synthetic datasets, helping develop theoretical models or conceptual frameworks, formulating mathematical claims, proposing or refining hypotheses, designing or providing feedback on research methodology or experiments, cleaning and reforming the dataset, interpreting results, and providing critical ingredients for proving mathematical claims, assisting in writing proofs are not applicable to this work. Additionally, we used generative AI tools for brainstorming, sourcing/searching for information, and editing a research paper to improve readability. We have reviewed all AI-assisted work.
We take responsibility for the final content of this work, including text, claims or artifacts produced with the aid of generative AI.

\section*{Acknowledgements}
This research was (partially) supported by the Dutch Research Council (NWO), under project numbers 024.004.022, NWA.1389.20.\-183, and KICH3.LTP.20.006, and the European Union under grant agreement No. 101201510 (UNITE).
The authors acknowledge the Scientific Computing of the IT Division at the Charité - Universitätsmedizin Berlin for providing computational resources that have contributed to the research results reported in this paper.
Views and opinions expressed are those of the author(s) only and do not necessarily reflect those of their respective employers, funders and/or granting authorities.

\bibliography{iclr2027_conference}

@article{chen2024cure4rec,
  title={{CURE4Rec}: A benchmark for recommendation unlearning with deeper influence},
  author={Chen, Chaochao and Zhang, Jiaming and Zhang, Yizhao and Zhang, Li and Lyu, Lingjuan and Li, Yuyuan and Gong, Biao and Yan, Chenggang},
  journal={Advances in Neural Information Processing Systems},
  volume={37},
  pages={99128--99144},
  year={2024}
}

@inproceedings{tang2024recentECIR,
  title={Recent Advances in Generative Information Retrieval},
  author={Tang, Yubao and Zhang, Ruqing and Ren, Zhaochun and Guo, Jiafeng and de Rijke, Maarten},
  booktitle={European Conference on Information Retrieval},
  pages={363--368},
  year={2024},
  organization={Springer}
}

@inproceedings{tang2024recentWWW,
  title={Recent Advances in Generative Information Retrieval},
  author={Tang, Yubao and Zhang, Ruqing and Sun, Weiwei and Guo, Jiafeng and De Rijke, Maarten},
  booktitle={Companion Proceedings of the ACM Web Conference 2024},
  pages={1238--1241},
  year={2024}
}

@inproceedings{tang2023recent,
  title={Recent advances in generative information retrieval},
  author={Tang, Yubao and Zhang, Ruqing and Guo, Jiafeng and de Rijke, Maarten},
  booktitle={Proceedings of the Annual International ACM SIGIR Conference on Research and Development in Information Retrieval in the Asia Pacific Region},
  pages={294--297},
  year={2023}
}

@inproceedings{tutorialGenIRtang,
author = {Tang, Yubao and Zhang, Ruqing and Ren, Zhaochun and Guo, Jiafeng and de Rijke, Maarten},
title = {Recent Advances in Generative Information Retrieval},
year = {2024},
isbn = {9798400704314},
publisher = {Association for Computing Machinery},
address = {New York, NY, USA},
url = {https://doi.org/10.1145/3626772.3661379},
doi = {10.1145/3626772.3661379},
booktitle = {Proceedings of the 47th International ACM SIGIR Conference on Research and Development in Information Retrieval},
pages = {3005–3008},
numpages = {4},
location = {Washington DC, USA},
series = {SIGIR '24}
}

@inproceedings{lubitzsch2026erase,
author = {Lubitzsch, Pierre Sicco and de Rijke, Maarten and Schelter, Sebastian},
booktitle = {SIGIR 2026: 49th international ACM SIGIR Conference on Research and Development in Information Retrieval},
month = {July},
pages = {3310--3318},
publisher = {ACM},
title = {{ERASE} -- A Real-World Aligned Benchmark for Unlearning in Recommender Systems},
year = {2026},
url = {https://doi.org/10.1145/3805712.3808596},
}

@inproceedings{schelter2023forget,
  title={Forget me now: Fast and exact unlearning in neighborhood-based recommendation},
  author={Schelter, Sebastian and Ariannezhad, Mozhdeh and de Rijke, Maarten},
  booktitle={Proceedings of the 46th international ACM SIGIR conference on research and development in information retrieval},
  pages={2011--2015},
  year={2023},
url = {https://doi.org/10.1145/3539618.3591989},  
}

@misc{damico2026deployingsemanticidbasedgenerative,
      title={Deploying Semantic {ID}-based Generative Retrieval for Large-Scale Podcast Discovery at {Spotify}}, 
      author={Edoardo D'Amico and Marco De Nadai and Praveen Chandar and Divita Vohra and Shawn Lin and Max Lefarov and Paul Gigioli and Gustavo Penha and Ilya Kopysitsky and Ivo Joel Senese and Darren Mei and Francesco Fabbri and Oguz Semerci and Yu Zhao and Vincent Tang and Brian St. Thomas and Alexandra Ranieri and Matthew N. K. Smith and Aaron Bernkopf and Bryan Leung and Ghazal Fazelnia and Mark VanMiddlesworth and Timothy Christopher Heath and Petter Pehrson Skiden and Alice Y. Wang and Doug J. Cole and Andreas Damianou and Maya Hristakeva and Reid Wilbur and Tarun Chillara and Vladan Radosavljevic and Pooja Chitkara and Sainath Adapa and Juan Elenter and Bernd Huber and Jacqueline Wood and Saaketh Vedantam and Jan Stypka and Sandeep Ghael and Martin D. Gould and David Murgatroyd and Yves Raimond and Mounia Lalmas and Paul N. Bennett},
      year={2026},
      eprint={2603.17540},
      archivePrefix={arXiv},
      primaryClass={cs.IR},
      url={https://arxiv.org/abs/2603.17540}, 
}

@misc{li2026genrecllmbackedrecommendationranker,
      title={{GenRec}: An {LLM}-Backed Recommendation Ranker at {Netflix}}, 
      author={Ying Li and Shradha Sehgal and Arjun Rao and Rein Houthooft and Yaochen Zhu and Ashish Rastogi},
      year={2026},
      eprint={2608.10257},
      archivePrefix={arXiv},
      primaryClass={cs.IR},
      url={https://arxiv.org/abs/2608.10257}, 
}

@misc{collins2026llmbasedgenerativeretrievalsnapchat,
      title={LLM-Based Generative Retrieval for Snapchat Content Recommendation}, 
      author={Liam Collins and Jiwen Ren and Donald Loveland and Bhuvesh Kumar and Clark Mingxuan Ju and Xuan Guo and Mo Li and Alvin Hou and Yi Cui and Peng Yang and Jian Wang and Saud Afzal Shafi and Nga Than and Ruiming Lu and Wenfeng Zhuo and Dongheng Li and Lili Zhang and Mingtao Zhang and Jinchao Ye and Vincent Xue and Chunhui Zhu and Neil Shah},
      year={2026},
      eprint={2607.28895},
      archivePrefix={arXiv},
      primaryClass={cs.IR},
      url={https://arxiv.org/abs/2607.28895}, 
}

@inproceedings{wu2021trialattack,
author = {Wu, Chenwang and Lian, Defu and Ge, Yong and Zhu, Zhihao and Chen, Enhong},
title = {Triple Adversarial Learning for Influence based Poisoning Attack in Recommender Systems},
year = {2021},
publisher = {Association for Computing Machinery},
url = {https://doi.org/10.1145/3447548.3467335},
booktitle = {Proceedings of the 27th ACM SIGKDD Conference on Knowledge Discovery \& Data Mining},
pages = {1830–1840},
}

@inproceedings{NEURIPS2022_c1bb0e3b,
 author = {Li, Haoyang and Di, Shimin and Chen, Lei },
 booktitle = {Advances in Neural Information Processing Systems},
 doi = {10.52202/068431-2174},
 editor = {S. Koyejo and S. Mohamed and A. Agarwal and D. Belgrave and K. Cho and A. Oh},
 pages = {29989--30002},
 publisher = {Curran Associates, Inc.},
 title = {Revisiting Injective Attacks on Recommender Systems},
 volume = {35},
 year = {2022}
}

@inproceedings{jia23model_sparsity_unlearning,
 author = {Jia, Jinghan and Liu, Jiancheng and Ram, Parikshit and Yao, Yuguang and Liu, Gaowen and Liu, Yang and Sharma, Pranay and Liu, Sijia},
 booktitle = {Advances in Neural Information Processing Systems},
 doi = {10.52202/075280-2246},
 editor = {A. Oh and T. Naumann and A. Globerson and K. Saenko and M. Hardt and S. Levine},
 pages = {51584--51605},
 publisher = {Curran Associates, Inc.},
 title = {Model Sparsity Can Simplify Machine Unlearning},
 volume = {36},
 year = {2023}
}

@inproceedings{kurmanji2023unbounded,
  title     = {Towards Unbounded Machine Unlearning},
  author    = {Kurmanji, Meghdad and Triantafillou, Peter and Hayes, Jamie
               and Triantafillou, Eleni},
  booktitle = {Advances in Neural Information Processing Systems},
  volume    = {36},
  year      = {2023},
}

@misc{triantafillou2024makingprogressunlearningfindings,
      title={Are We Making Progress in Unlearning? {Findings} from the First {NeurIPS} Unlearning Competition}, 
      author={Eleni Triantafillou and Peter Kairouz and Fabian Pedregosa and Jamie Hayes and Meghdad Kurmanji and Kairan Zhao and Vincent Dumoulin and Julio Jacques Junior and Ioannis Mitliagkas and Jun Wan and Lisheng Sun Hosoya and Sergio Escalera and Gintare Karolina Dziugaite and Peter Triantafillou and Isabelle Guyon},
      year={2024},
      eprint={2406.09073},
      archivePrefix={arXiv},
      primaryClass={cs.LG},
      url={https://arxiv.org/abs/2406.09073}, 
}

@article{li2023_scif,
author = {Li, Yuyuan and Chen, Chaochao and Zheng, Xiaolin and Zhang, Yizhao and Gong, Biao and Wang, Jun and Chen, Linxun},
title = {Selective and collaborative influence function for efficient recommendation unlearning},
year = {2023},
issue_date = {Dec 2023},
publisher = {Pergamon Press, Inc.},
address = {USA},
volume = {234},
number = {C},
url = {https://doi.org/10.1016/j.eswa.2023.121025},
journal = {Expert Syst. Appl.},
month = dec,
numpages = {11}
}

@misc{chen2026tracertokenreassignmentconcept,
      title={{TRACER}: Token ReAssignment for Concept ERasure in Generative Recommendation}, 
      author={Ziheng Chen and Jiali Cheng and Zezhong Fan and Hadi Amiri and Diyuan Wu and Gabriele Tolomei and Yang Zhang},
      year={2026},
      eprint={2606.07688},
      archivePrefix={arXiv},
      primaryClass={cs.IR},
      url={https://arxiv.org/abs/2606.07688}, 
}

@inproceedings{rajput_2023_tiger,
 author = {Rajput, Shashank and Mehta, Nikhil and Singh, Anima and Hulikal Keshavan, Raghunandan and Vu, Trung and Heldt, Lukasz and Hong, Lichan and Tay, Yi and Tran, Vinh and Samost, Jonah and Kula, Maciej and Chi, Ed and Sathiamoorthy, Maheswaran},
 booktitle = {Advances in Neural Information Processing Systems},
 doi = {10.52202/075280-0452},
 editor = {A. Oh and T. Naumann and A. Globerson and K. Saenko and M. Hardt and S. Levine},
 pages = {10299--10315},
 publisher = {Curran Associates, Inc.},
 title = {Recommender Systems with Generative Retrieval},
 volume = {36},
 year = {2023}
}

@article{zeghidour_2021_rqvae,
author = {Zeghidour, Neil and Luebs, Alejandro and Omran, Ahmed and Skoglund, Jan and Tagliasacchi, Marco},
title = {{SoundStream}: An End-to-End Neural Audio Codec},
year = {2021},
issue_date = {2022},
publisher = {IEEE Press},
volume = {30},
url = {https://doi.org/10.1109/TASLP.2021.3129994},
journal = {IEEE/ACM Trans. Audio, Speech and Lang. Proc.},
month = nov,
pages = {495–507},
numpages = {13}
}

@inproceedings{luo_2025_rqkmeans,
author = {Luo, Xinchen and Cao, Jiangxia and Sun, Tianyu and Yu, Jinkai and Huang, Rui and Yuan, Wei and Lin, Hezheng and Zheng, Yichen and Wang, Shiyao and Hu, Qigen and Qiu, Changqing and Zhang, Jiaqi and Zhang, Xu and Yan, Zhiheng and Zhang, Jingming and Zhang, Simin and Wen, Mingxing and Liu, Zhaojie and Zhou, Guorui},
title = {QARM: Quantitative Alignment Multi-Modal Recommendation at Kuaishou},
year = {2025},
publisher = {Association for Computing Machinery},
address = {New York, NY, USA},
url = {https://doi.org/10.1145/3746252.3761502},
booktitle = {Proceedings of the 34th ACM International Conference on Information and Knowledge Management},
pages = {5915–5922},
}

@inproceedings{cao_2015_unlearn,
author = {Cao, Yinzhi and Yang, Junfeng},
title = {Towards Making Systems Forget with Machine Unlearning},
year = {2015},
isbn = {9781467369497},
publisher = {IEEE Computer Society},
address = {USA},
url = {https://doi.org/10.1109/SP.2015.35},
doi = {10.1109/SP.2015.35},
booktitle = {Proceedings of the 2015 IEEE Symposium on Security and Privacy},
pages = {463–480},
numpages = {18},
series = {SP '15}
}

@inproceedings{wang_2024_letter,
author = {Wang, Wenjie and Bao, Honghui and Lin, Xinyu and Zhang, Jizhi and Li, Yongqi and Feng, Fuli and Ng, See-Kiong and Chua, Tat-Seng},
title = {Learnable Item Tokenization for Generative Recommendation},
year = {2024},
publisher = {Association for Computing Machinery},
url = {https://doi.org/10.1145/3627673.3679569},
booktitle = {Proceedings of the 33rd ACM International Conference on Information and Knowledge Management},
pages = {2400–2409},
}

@article{kang2018selfattentivesequentialrecommendation,
      title={Self-Attentive Sequential Recommendation}, 
      author={Wang-Cheng Kang and Julian McAuley},
      year={2018},
      journal={arXiv preprint arXiv:1808.09781},
      url={https://arxiv.org/abs/1808.09781}, 
}

@inproceedings{mcauley2015image,
title     = {Image-Based Recommendations on Styles and Substitutes},
author    = {McAuley, Julian and Targett, Christopher and Shi, Qinfeng and
             van den Hengel, Anton},
booktitle = {Proceedings of the 38th International ACM SIGIR Conference on
             Research and Development in Information Retrieval},
pages     = {43--52},
year      = {2015},
doi       = {10.1145/2766462.2767755}
}

@inproceedings{he2016ups,
title     = {Ups and Downs: Modeling the Visual Evolution of Fashion Trends
             with One-Class Collaborative Filtering},
author    = {He, Ruining and McAuley, Julian},
booktitle = {Proceedings of the 25th International Conference on World Wide Web},
pages     = {507--517},
year      = {2016},
doi       = {10.1145/2872427.2883037}
}

@inproceedings{geng2022p5,
title     = {Recommendation as Language Processing ({RLP}): A Unified Pretrain,
             Personalized Prompt \& Predict Paradigm ({P5})},
author    = {Geng, Shijie and Liu, Shuchang and Fu, Zuohui and Ge, Yingqiang
             and Zhang, Yongfeng},
booktitle = {Proceedings of the 16th ACM Conference on Recommender Systems},
pages     = {299--315},
year      = {2022},
doi       = {10.1145/3523227.3546767}
}

@inproceedings{hou_2023_vqrec,
author = {Hou, Yupeng and He, Zhankui and McAuley, Julian and Zhao, Wayne Xin},
title = {Learning Vector-Quantized Item Representation for Transferable Sequential Recommenders},
year = {2023},
publisher = {Association for Computing Machinery},
url = {https://doi.org/10.1145/3543507.3583434},
booktitle = {Proceedings of the ACM Web Conference 2023},
pages = {1162–1171},
numpages = {10},
location = {Austin, TX, USA},
series = {WWW '23}
}

@inproceedings{fu_2026_diffable_sids,
author = {Fu, Junchen and Ge, Xuri and Karatzoglou, Alexandros and Arapakis, Ioannis and Verberne, Suzan and Jose, Joemon M. and Ren, Zhaochun},
title = {Differentiable Semantic {ID} for Generative Recommendation},
year = {2026},
publisher = {Association for Computing Machinery},
url = {https://doi.org/10.1145/3805712.3809641},
booktitle = {Proceedings of the 49th International ACM SIGIR Conference on Research and Development in Information Retrieval},
pages = {369–379},
}

@inproceedings{penha_2025_sids_joint_search_and_rec,
author = {Penha, Gustavo and D'Amico, Edoardo and De Nadai, Marco and Palumbo, Enrico and Tamborrino, Alexandre and Vardasbi, Ali and Lefarov, Max and Lin, Shawn and Heath, Timothy and Fabbri, Francesco and Bouchard, Hugues},
title = {Semantic {IDs} for Joint Generative Search and Recommendation},
year = {2025},
publisher = {Association for Computing Machinery},
url = {https://doi.org/10.1145/3705328.3759300},
booktitle = {Proceedings of the Nineteenth ACM Conference on Recommender Systems},
pages = {1296–1301},
}

@inproceedings{liu_2026_genrec_scaling,
author = {Liu, Jingzhe and Collins, Liam and Tang, Jiliang and Zhao, Tong and Shah, Neil and Ju, Mingxuan},
title = {Understanding Generative Recommendation with Semantic IDs from a Model-scaling View},
year = {2026},
publisher = {Association for Computing Machinery},
url = {https://doi.org/10.1145/3770855.3817976},
booktitle = {Proceedings of the 32nd ACM SIGKDD Conference on Knowledge Discovery and Data Mining V.2},
pages = {3234–3245},
}

@article{xia2026unleashpotentiallongsemantic,
      title={Unleash the Potential of Long Semantic {IDs} for Generative Recommendation}, 
      author={Ming Xia and Guoxin Ma and Zhiqin Zhou and Dongmin Huang},
      year={2026},
      journal={arXiv preprint arXiv:2602.13573},
      url={https://arxiv.org/abs/2602.13573}, 
}

@inproceedings{kun_wu_2023,
author = {Wu, Kun and Shen, Jie and Ning, Yue and Wang, Ting and Wang, Wendy Hui},
title = {Certified Edge Unlearning for Graph Neural Networks},
year = {2023},
publisher = {Association for Computing Machinery},
url = {https://doi.org/10.1145/3580305.3599271},
booktitle = {Proceedings of the 29th ACM SIGKDD Conference on Knowledge Discovery and Data Mining},
pages = {2606–2617},
}

@inproceedings{Wu_2023, series={WWW ’23},
   title={{GIF}: A General Graph Unlearning Strategy via Influence Function},
   url={http://dx.doi.org/10.1145/3543507.3583521},
   booktitle={Proceedings of the ACM Web Conference 2023},
   publisher={ACM},
   author={Wu, Jiancan and Yang, Yi and Qian, Yuchun and Sui, Yongduo and Wang, Xiang and He, Xiangnan},
   year={2023},
   month=apr, pages={651–661},
}

@inproceedings{dong_2024,
author = {Dong, Yushun and Zhang, Binchi and Lei, Zhenyu and Zou, Na and Li, Jundong},
title = {{IDEA}: A Flexible Framework of Certified Unlearning for Graph Neural Networks},
year = {2024},
publisher = {Association for Computing Machinery},
url = {https://doi.org/10.1145/3637528.3671744},
booktitle = {Proceedings of the 30th ACM SIGKDD Conference on Knowledge Discovery and Data Mining},
pages = {621–630},
}

@inproceedings{kim_2026_mviger,
author = {Kim, Tongyoung and Yoon, SooJin and Kang, SeongKu and Yeo, Jinyoung and Lee, Dongha},
title = {{MVIGER}: Multi-View Variational Integration of Complementary Knowledge for Generative Recommender},
year = {2026},
publisher = {Association for Computing Machinery},
url = {https://doi.org/10.1145/3805712.3809591},
booktitle = {Proceedings of the 49th International ACM SIGIR Conference on Research and Development in Information Retrieval},
pages = {834–844},
}

@inproceedings{wang_2024_eager,
author = {Wang, Ye and Xun, Jiahao and Hong, Minjie and Zhu, Jieming and Jin, Tao and Lin, Wang and Li, Haoyuan and Li, Linjun and Xia, Yan and Zhao, Zhou and Dong, Zhenhua},
title = {{EAGER}: Two-Stream Generative Recommender with Behavior-Semantic Collaboration},
year = {2024},
publisher = {Association for Computing Machinery},
url = {https://doi.org/10.1145/3637528.3671775},
booktitle = {Proceedings of the 30th ACM SIGKDD Conference on Knowledge Discovery and Data Mining},
pages = {3245–3254},
}

@misc{cheng2026capsidsoftroutedvariablelengthsemantic,
      title={{CapsID}: Soft-Routed Variable-Length Semantic IDs for Generative Recommendation}, 
      author={Wenzhuo Cheng and Menghang Gong and Qixin Guo and Hang Zheng and Zhaobin Yang and Jianguo Lou and Zhengwei Zheng},
      year={2026},
      eprint={2605.05096},
      archivePrefix={arXiv},
      primaryClass={cs.IR},
      url={https://arxiv.org/abs/2605.05096}, 
}

@misc{acharya2025glossgenerativelanguagemodels,
      title={{GLoSS}: Generative Language Models with Semantic Search for Sequential Recommendation}, 
      author={Krishna Acharya and Aleksandr V. Petrov and Juba Ziani},
      year={2025},
      eprint={2506.01910},
      archivePrefix={arXiv},
      primaryClass={cs.IR},
      url={https://arxiv.org/abs/2506.01910}, 
}

@inproceedings{ju_2025_genrecsemids,
author = {Ju, Clark Mingxuan and Collins, Liam and Neves, Leonardo and Kumar, Bhuvesh and Wang, Louis Yufeng and Zhao, Tong and Shah, Neil},
title = {Generative Recommendation with Semantic {IDs}: A Practitioner's Handbook},
year = {2025},
publisher = {Association for Computing Machinery},
url = {https://doi.org/10.1145/3746252.3761612},
booktitle = {Proceedings of the 34th ACM International Conference on Information and Knowledge Management},
pages = {6420–6425},
numpages = {6},
location = {Seoul, Republic of Korea},
series = {CIKM '25}
}

@inproceedings{chen-2022-recommendation,
author = {Chen, Chong and Sun, Fei and Zhang, Min and Ding, Bolin},
title = {Recommendation Unlearning},
year = {2022},
publisher = {Association for Computing Machinery},
booktitle = {Proceedings of the ACM Web Conference 2022},
pages = {2768–2777},
}
\bibliographystyle{iclr2027_conference}

\clearpage
\appendix

\section*{Appendix Overview}
\label[appendix]{sec:appendix-overview}

\begin{itemize}[leftmargin=*,nosep]
    \item \Cref{sec:appendix-dataset}: Dataset statistics and training set construction.
    \item \Cref{sec:appendix-baselines}: Baselines.
    \item \Cref{sec:appendix-scenarios}: Deletion scenario construction.
    \item \Cref{sec:appendix-spam}: Additional spam-removal results.
    \item \Cref{sec:appendix-ablation}: Additional ablation results.
    \item \Cref{sec:appendix-unwanted}: Additional unwanted-item results.
    \item \Cref{sec:appendix-bootstrap}: Intervals for neighborhood comparisons.
    \item \Cref{sec:appendix-diagnostics}: \ac{SID} and neighborhood diagnostics.
\end{itemize}

\section{Dataset Statistics}
\label[appendix]{sec:appendix-dataset}

\Cref{tab:dataset-statistics} summarizes the dataset statistics after preprocessing. Following TIGER, a user's history is capped at 20 items during training. The statistics are computed before this cap.
For each dataset, we follow the next-item prediction setting, where each training example is represented as $(\bm{x}, i)$, with $\bm{x}$ denoting the user history and $i$ the next item.
Interactions are split into training, validation, and test sets in chronological order.
The backbone \ac{GenRec} model is trained to autoregressively generate the \ac{SID} of the next item.
All unlearning methods start from the same original model $\bm{\theta}_o$.
During evaluation, valid generated \acp{SID} are mapped back to items and ranked by generation probability.

\begin{table}[h]
\centering
\caption{Dataset statistics after preprocessing.}
\label{tab:dataset-statistics}
\small
\begin{tabular}{lcccc}
\toprule
Dataset & \#Users & \#Items & \#Interactions & Avg. length \\
\midrule
Beauty & 22,363 & 12,101 & 198,502 & 8.88 \\
Sports & 35,598 & 18,357 & 296,337 & 8.32 \\
Toys   & 19,412 & 11,924 & 167,597 & 8.63 \\
\bottomrule
\end{tabular}
\end{table}

\section{Baselines}
\label[appendix]{sec:appendix-baselines}

We compare with representative removal and unlearning baselines:
\begin{enumerate*}[label=(\roman*)]
    \item \emph{Original}, the model $\bm{\theta}_o$ before unlearning;
    \item \emph{Retrain}, the reference $\bm{\theta}_r$ trained from scratch on $D_r$;
    \item \emph{Finetune} (retain only), which fine-tunes $\bm{\theta}_o$ on $D_r$;
    \item \emph{Forget only}, gradient ascent on $D_f$;
    \item \emph{Forget+Repair}, gradient ascent on $D_f$ combined with fine-tuning on $D_r$;
    \item \emph{SCIF}~\citep{li2023_scif}, which applies an influence-function update to a subset of the parameters to remove the forget set while keeping retain performance;
    \item \emph{SEIF}, which adds noise to $\bm{\theta}_o$ and then fine-tunes on $D_r$;
    \item \emph{Kookmin}, which reinitializes a subset of the parameters and fine-tunes on $D_r$;
    \item \emph{Fanchuan}, which first pushes the predictions on forgotten examples toward a uniform distribution and then alternates a contrastive forget objective with fine-tuning on $D_r$ (SEIF, Kookmin, and Fanchuan are from the NeurIPS 2023 unlearning competition~\citep{triantafillou2024makingprogressunlearningfindings}); and
    \item \emph{TRACER}~\citep{chen2026tracertokenreassignmentconcept}, which reassigns the \acp{SID} of the items to be forgotten and regularizes the recommender toward the original identifiers.
\end{enumerate*}



\header{Selection of NAU's weights}
\Cref{tab:appendix-nau-valsel} reports the validation grid behind NAU's operating point.
Whole-set exposure barely separates the settings: among settings of the full objective ($\lambda_s, \lambda_n > 0$) that keep $\tau_P \geq 95\%$, $\lambda_s=0.1$ has the lowest $\Delta^{\mathrm{retr}}_{\mathrm{SH}}$ ($0.079\%$), but $\lambda_s=0.01$ ($0.086\%$) is identical to it on 12 of 15 cells and their paired difference is not resolved.
Among such tied settings we select the highest combined utility, the mean of whole-set and neighborhood $\tau_P$: $\lambda_s=0.01$ reaches $100.2\%$ against $95.8\%$ ($103.6\%$ against $95.0\%$ on the neighborhood) and is ahead on both parts, so the choice does not depend on their weighting.
Settings without the separation term ($\lambda_s=0$) are ablations of NAU rather than candidates; with $\lambda_n=0.01$ such a setting reaches a similar combined utility ($101.7\%$).

\begin{table}[t]
\centering
\small
\setlength{\tabcolsep}{4pt}
\caption{
Selection of NAU's weights on the validation split (Beauty, TIGER + RQ-KMeans, $\lambda_f=0.1$, three targets $\times$ five data seeds).
Whole set: $\Delta^{\mathrm{retr}}_{\mathrm{SH}}$ and $\tau_P$ on all validation cases; neighborhood: the same on the validation cases whose next item shares the target's first \ac{SID} token (the target's own cases excluded), with $\tau_P$ as a ratio of pooled $\mathrm{NDCG}@10$.
Selection is among settings of the full objective ($\lambda_s, \lambda_n > 0$) that keep whole-set $\tau_P \geq 95\%$: settings whose whole-set $\Delta^{\mathrm{retr}}_{\mathrm{SH}}$ is not resolvably different from the lowest one (paired 95\% interval over the five seeds includes 0) are treated as tied, and among them the highest combined utility, the mean of whole-set and neighborhood $\tau_P$, is selected (\checkmark); the choice is the same for any weighting of the two.
Settings below $95\%$ whole-set utility are shown in gray; -- marks settings not scored on the neighborhood.}
\label{tab:appendix-nau-valsel}
\begin{tabular}{llccccc c}
\toprule
 & & \multicolumn{2}{c}{Whole set} & \multicolumn{2}{c}{SID neighborhood} & Combined & \\
\cmidrule(lr){3-4} \cmidrule(lr){5-6} \cmidrule(lr){7-7}
$\lambda_s$ & $\lambda_n$ & $\Delta^{\mathrm{retr}}_{\mathrm{SH}}{\scriptstyle\,(\%)}\downarrow$ & $\tau_P{\scriptstyle\,(\%)}\uparrow$ & $\Delta^{\mathrm{retr}}_{\mathrm{SH}}{\scriptstyle\,(\%)}\downarrow$ & $\tau_P{\scriptstyle\,(\%)}\uparrow$ & $\tau_P{\scriptstyle\,(\%)}\uparrow$ & Sel. \\
\midrule
0.0 & 0.0 & 0.112 & 96.9 & 0.678 & 86.5 & 91.7 &  \\
0.0 & 0.01 & 0.089 & 96.7 & 0.863 & 106.6 & 101.7 &  \\
\textcolor{gray}{0.0} & \textcolor{gray}{0.1} & \textcolor{gray}{0.094} & \textcolor{gray}{93.7} & \textcolor{gray}{0.874} & \textcolor{gray}{157.9} & \textcolor{gray}{125.8} &  \\
0.01 & 0.0 & 0.086 & 97.0 & 0.879 & 87.6 & 92.3 &  \\
0.01 & 0.01 & 0.086 & 96.9 & 0.874 & 103.6 & 100.2 & \checkmark \\
\textcolor{gray}{0.01} & \textcolor{gray}{0.1} & \textcolor{gray}{0.072} & \textcolor{gray}{93.9} & \textcolor{gray}{--} & \textcolor{gray}{--} & \textcolor{gray}{--} &  \\
0.1 & 0.0 & 0.079 & 96.8 & 0.874 & 80.2 & 88.5 &  \\
0.1 & 0.01 & 0.079 & 96.6 & 0.874 & 95.0 & 95.8 &  \\
\textcolor{gray}{0.1} & \textcolor{gray}{0.1} & \textcolor{gray}{0.079} & \textcolor{gray}{94.2} & \textcolor{gray}{0.874} & \textcolor{gray}{147.4} & \textcolor{gray}{120.8} &  \\
\bottomrule
\end{tabular}
\end{table}

The forget weight was checked in the same way: without the separation and neighborhood terms, $\lambda_f=0.01$ barely removes the promoted item (validation $\Delta^{\mathrm{retr}}_{\mathrm{SH}}$ of $0.832\%$, close to the poisoned model) and $\lambda_f=1.0$ removes it but keeps only $76.3\%$ of retrained utility, while $\lambda_f=0.1$ reaches $0.112\%$ at $96.9\%$.

\header{Selection of baseline configurations}
\Cref{tab:appendix-valsel} lists every baseline configuration run on all five seeds and three targets, with validation and test metrics; the selected configurations are those in \Cref{tab:bandwagon-unlearn-all-tiger-rkmeans}, and validation and test agree closely.
\begin{table}
\centering
\small
\setlength{\tabcolsep}{3pt}
\caption{
Hyperparameter selection on the validation split (TIGER + RQ-KMeans; TRACER on Beauty with its own tokenizer). For every candidate configuration, pooled over five data seeds and three targets: distance to the retrained model's exposure $\Delta^{\mathrm{retr}}_{\mathrm{SH}}$ and $\tau_P$, on the validation split (used for selection) and on the test split (reported in \Cref{tab:bandwagon-unlearn-all-tiger-rkmeans}). The selected configuration (\checkmark) has the smallest validation $\Delta^{\mathrm{retr}}_{\mathrm{SH}}$ among those with validation $\tau_P \geq 95\%$, ties broken by higher $\tau_P$; if none reaches $95\%$, the smallest validation $\Delta^{\mathrm{retr}}_{\mathrm{SH}}$ overall. Finetune and SCIF have a single configuration; NAU's selection is shown in \Cref{tab:appendix-nau-valsel}.}
\label{tab:appendix-valsel}
\resizebox{\columnwidth}{!}{%
\begin{tabular}{lllrrrrc}
\toprule
 & & & \multicolumn{2}{c}{Validation} & \multicolumn{2}{c}{Test} & \\
\cmidrule(lr){4-5} \cmidrule(lr){6-7}
Dataset & Method & Configuration & $\Delta^{\mathrm{retr}}_{\mathrm{SH}}\downarrow$ & $\tau_P\uparrow$ & $\Delta^{\mathrm{retr}}_{\mathrm{SH}}\downarrow$ & $\tau_P\uparrow$ & Sel. \\
\midrule
Beauty & Fanchuan & fanchuan\_t0.007 & 0.079 & 11.9 & 0.072 & 12.6 &  \\
Beauty & Fanchuan & fanchuan\_t0.07 & 0.079 & 32.0 & 0.072 & 32.4 & \checkmark \\
Beauty & Forget only & unified\_abl\_lf0.1\_ls0.0\_ln0.0\_lr0.0 & 0.079 & 2.1 & 0.072 & 1.9 &  \\
Beauty & Forget only & unified\_abl\_lf1.0\_ls0.0\_ln0.0\_lr0.0 & 0.079 & 2.7 & 0.072 & 2.5 & \checkmark \\
Beauty & Forget+Repair & unified\_abl\_lf1.0\_ls0.0\_ln0.0 & 0.079 & 75.8 & 0.072 & 72.8 &  \\
Beauty & Forget+Repair & unified\_abl\_lf0.1\_ls0.0\_ln0.0 & 0.108 & 96.4 & 0.106 & 95.8 & \checkmark \\
Beauty & Kookmin & kookmin\_ir0.0001 & 0.643 & 82.0 & 0.575 & 81.0 &  \\
Beauty & Kookmin & kookmin\_ir0.001 & 0.550 & 81.9 & 0.493 & 80.9 & \checkmark \\
Beauty & Kookmin & kookmin\_ir0.01 & 0.846 & 74.8 & 0.775 & 75.0 &  \\
Beauty & SEIF & seif\_std0.06 & 0.258 & 66.1 & 0.258 & 66.4 & \checkmark \\
Beauty & SEIF & seif\_std0.006 & 0.278 & 68.0 & 0.256 & 68.6 &  \\
Beauty & TRACER & tracer\_lf1.0\_lc0.1\_plr0.001\_ep4 & 1.185 & 98.2 & 1.063 & 99.1 & \checkmark \\
Beauty & TRACER & tracer\_lf1.0\_lc1.0\_plr0.001\_ep4 & 4.262 & 98.5 & 3.546 & 99.3 &  \\
Beauty & TRACER & tracer\_lf0.1\_lc1.0\_plr0.001\_ep4 & 5.821 & 98.6 & 4.848 & 99.5 &  \\
Beauty & TRACER & tracer\_lf0.1\_lc0.1\_plr0.001\_ep4 & 1.520 & 98.5 & 1.323 & 99.5 &  \\
\midrule
Sports & Fanchuan & fanchuan\_t0.07 & 0.146 & 30.5 & 0.139 & 31.7 & \checkmark \\
Sports & Fanchuan & fanchuan\_t0.007 & 0.146 & 11.7 & 0.139 & 12.8 &  \\
Sports & Forget only & unified\_abl\_lf0.1\_ls0.0\_ln0.0\_lr0.0 & 0.146 & 0.7 & 0.139 & 0.5 &  \\
Sports & Forget only & unified\_abl\_lf1.0\_ls0.0\_ln0.0\_lr0.0 & 0.146 & 0.8 & 0.139 & 0.7 & \checkmark \\
Sports & Forget+Repair & unified\_abl\_lf0.1\_ls0.0\_ln0.0 & 0.142 & 99.8 & 0.139 & 102.1 & \checkmark \\
Sports & Forget+Repair & unified\_abl\_lf1.0\_ls0.0\_ln0.0 & 0.146 & 68.3 & 0.139 & 69.5 &  \\
Sports & Kookmin & kookmin\_ir0.001 & 0.177 & 79.3 & 0.165 & 80.3 & \checkmark \\
Sports & Kookmin & kookmin\_ir0.01 & 0.252 & 73.3 & 0.224 & 72.5 &  \\
Sports & Kookmin & kookmin\_ir0.0001 & 0.242 & 79.8 & 0.203 & 80.8 &  \\
Sports & SEIF & seif\_std0.006 & 0.157 & 60.5 & 0.144 & 61.4 &  \\
Sports & SEIF & seif\_std0.06 & 0.122 & 58.7 & 0.112 & 59.5 & \checkmark \\
Sports & TRACER & tracer\_lf1.0\_lc0.1\_plr0.001\_ep4 & 0.462 & 97.0 & 0.410 & 99.8 & \checkmark \\
Sports & TRACER & tracer\_lf0.1\_lc0.1\_plr0.001\_ep4 & 0.694 & 97.0 & 0.602 & 99.8 &  \\
Sports & TRACER & tracer\_lf0.1\_lc1.0\_plr0.001\_ep4 & 3.877 & 96.9 & 3.488 & 99.3 &  \\
Sports & TRACER & tracer\_lf1.0\_lc1.0\_plr0.001\_ep4 & 3.312 & 96.6 & 2.949 & 99.2 &  \\
\midrule
Toys & Fanchuan & fanchuan\_t0.07 & 0.073 & 21.4 & 0.074 & 23.4 & \checkmark \\
Toys & Fanchuan & fanchuan\_t0.007 & 0.073 & 3.2 & 0.074 & 3.5 &  \\
Toys & Forget only & unified\_abl\_lf0.1\_ls0.0\_ln0.0\_lr0.0 & 0.073 & 0.3 & 0.074 & 0.4 &  \\
Toys & Forget only & unified\_abl\_lf1.0\_ls0.0\_ln0.0\_lr0.0 & 0.073 & 0.5 & 0.074 & 0.4 & \checkmark \\
Toys & Forget+Repair & unified\_abl\_lf0.1\_ls0.0\_ln0.0 & 0.164 & 97.0 & 0.185 & 96.7 & \checkmark \\
Toys & Forget+Repair & unified\_abl\_lf1.0\_ls0.0\_ln0.0 & 0.073 & 69.6 & 0.074 & 69.8 &  \\
Toys & Kookmin & kookmin\_ir0.001 & 0.403 & 75.5 & 0.397 & 74.7 &  \\
Toys & Kookmin & kookmin\_ir0.01 & 0.592 & 68.2 & 0.573 & 68.9 &  \\
Toys & Kookmin & kookmin\_ir0.0001 & 0.384 & 76.8 & 0.385 & 76.1 & \checkmark \\
Toys & SEIF & seif\_std0.06 & 0.206 & 60.4 & 0.216 & 61.3 &  \\
Toys & SEIF & seif\_std0.006 & 0.183 & 62.6 & 0.206 & 63.1 & \checkmark \\
Toys & TRACER & tracer\_lf1.0\_lc1.0\_plr0.001\_ep4 & 2.555 & 96.7 & 2.293 & 96.2 &  \\
Toys & TRACER & tracer\_lf1.0\_lc0.1\_plr0.001\_ep4 & 1.150 & 96.6 & 1.079 & 95.7 & \checkmark \\
Toys & TRACER & tracer\_lf0.1\_lc1.0\_plr0.001\_ep4 & 3.327 & 97.1 & 2.999 & 96.6 &  \\
Toys & TRACER & tracer\_lf0.1\_lc0.1\_plr0.001\_ep4 & 1.471 & 97.1 & 1.377 & 96.2 &  \\
\bottomrule
\end{tabular}}
\end{table}

For unwanted-item removal, \Cref{tab:appendix-valsens} applies the same procedure with $\mathrm{UHF}@10$ in place of $\Delta^{\mathrm{retr}}_{\mathrm{SH}}$. NAU's $\lambda_n$ is chosen on Beauty: $\lambda_n=-1$ has the lowest validation $\mathrm{UHF}@10$ ($9.9\%$, against $11.4\%$ at $\lambda_n=-10$ and $13.0\%$ at $\lambda_n=0$). No setting of NAU or of SEIF, Kookmin and Fanchuan keeps validation $\tau_P \geq 95\%$ (NAU reaches $93.3\%$), so the lowest $\mathrm{UHF}@10$ decides.
\begin{table}
\centering
\footnotesize
\setlength{\tabcolsep}{4pt}
\renewcommand{\arraystretch}{0.92}
\caption{
Hyperparameter selection for unwanted-item removal on the validation split (TIGER + RQ-KMeans). For every candidate configuration, averaged over five data seeds: $\mathrm{UHF}@10$ and $\tau_P$ on the validation split (used for selection) and on the test split (reported in \Cref{sec:appendix-unwanted}). The selected configuration (\checkmark) has the lowest validation $\mathrm{UHF}@10$ among those with validation $\tau_P \geq 95\%$; if none reaches $95\%$, the lowest validation $\mathrm{UHF}@10$ overall. NAU ($\lambda_f=0.1$, $\lambda_s=0$) selects $\lambda_n$ once on the Beauty validation split with the same rule and uses it on all datasets. Finetune, Forget only and Forget+Repair have a single configuration.}
\label{tab:appendix-valsens}
\begin{tabular}{lllrrrrc}
\toprule
 & & & \multicolumn{2}{c}{Validation} & \multicolumn{2}{c}{Test} & \\
\cmidrule(lr){4-5} \cmidrule(lr){6-7}
Dataset & Method & Configuration & $\mathrm{UHF}@10\downarrow$ & $\tau_P\uparrow$ & $\mathrm{UHF}@10\downarrow$ & $\tau_P\uparrow$ & Sel. \\
\midrule
Beauty & SCIF & scif\_d0.001 & 24.99 & 99.6 & 15.98 & 101.0 &  \\
Beauty & SCIF & scif\_d0.01 & 24.99 & 99.6 & 15.98 & 101.0 & \checkmark \\
Beauty & SCIF & scif\_d0.1 & 26.53 & 99.6 & 15.98 & 100.9 &  \\
Beauty & SEIF & seif\_std0.006 & 15.50 & 60.4 & 10.06 & 63.9 &  \\
Beauty & SEIF & seif\_std0.06 & 17.04 & 59.2 & 8.25 & 62.0 &  \\
Beauty & SEIF & seif\_std0.6 & 0.00 & 0.8 & 0.00 & 0.9 & \checkmark \\
Beauty & Kookmin & kookmin\_ir0.0001 & 19.81 & 81.9 & 7.99 & 85.5 &  \\
Beauty & Kookmin & kookmin\_ir0.001 & 9.62 & 80.5 & 4.90 & 84.6 &  \\
Beauty & Kookmin & kookmin\_ir0.01 & 8.23 & 73.1 & 6.71 & 77.4 & \checkmark \\
Beauty & Fanchuan & fanchuan\_t0.007 & 0.00 & 4.1 & 0.00 & 5.0 & \checkmark \\
Beauty & Fanchuan & fanchuan\_t0.07 & 0.00 & 10.4 & 0.00 & 10.8 &  \\
Beauty & Fanchuan & fanchuan\_t1.15 & 0.00 & 21.6 & 0.00 & 23.0 &  \\
Beauty & NAU & $\lambda_n=-0.1$ & 13.10 & 91.0 & 11.07 & 93.9 &  \\
Beauty & NAU & $\lambda_n=-1$ & 9.90 & 93.3 & 9.42 & 95.8 & \checkmark \\
Beauty & NAU & $\lambda_n=-10$ & 11.43 & 93.6 & 7.75 & 95.8 &  \\
Beauty & NAU & $\lambda_n=0$ & 12.97 & 90.1 & 7.71 & 93.2 &  \\
\midrule
Sports & SCIF & scif\_d0.001 & 44.91 & 100.0 & 43.71 & 98.9 &  \\
Sports & SCIF & scif\_d0.01 & 38.37 & 101.3 & 32.87 & 100.4 &  \\
Sports & SCIF & scif\_d0.1 & 38.29 & 101.3 & 35.73 & 100.5 & \checkmark \\
Sports & SEIF & seif\_std0.006 & 22.19 & 44.8 & 15.32 & 46.2 &  \\
Sports & SEIF & seif\_std0.06 & 17.89 & 44.6 & 19.07 & 45.7 &  \\
Sports & SEIF & seif\_std0.6 & 0.00 & 0.0 & 0.00 & 0.1 & \checkmark \\
Sports & Kookmin & kookmin\_ir0.0001 & 31.85 & 75.2 & 29.90 & 74.7 &  \\
Sports & Kookmin & kookmin\_ir0.001 & 31.61 & 74.2 & 29.90 & 74.4 &  \\
Sports & Kookmin & kookmin\_ir0.01 & 28.74 & 68.8 & 27.05 & 68.7 & \checkmark \\
Sports & Fanchuan & fanchuan\_t0.007 & 0.00 & 7.6 & 0.00 & 7.4 & \checkmark \\
Sports & Fanchuan & fanchuan\_t0.07 & 0.00 & 16.3 & 0.00 & 17.3 &  \\
Sports & Fanchuan & fanchuan\_t1.15 & 0.00 & 21.1 & 0.00 & 22.0 &  \\
Sports & TRACER & tracer\_lf0.1\_lc0.1\_ep4 & 34.32 & 101.4 & 32.87 & 100.6 &  \\
Sports & TRACER & tracer\_lf0.1\_lc1.0\_ep4 & 37.19 & 101.5 & 39.51 & 100.6 &  \\
Sports & TRACER & tracer\_lf1.0\_lc0.1\_ep4 & 34.32 & 101.2 & 32.87 & 100.4 & \checkmark \\
Sports & TRACER & tracer\_lf1.0\_lc1.0\_ep4 & 37.19 & 101.4 & 36.65 & 100.6 &  \\
Sports & NAU & $\lambda_n=-0.1$ & 25.02 & 91.5 & 19.68 & 90.9 &  \\
Sports & NAU & $\lambda_n=-1$ & 15.72 & 93.6 & 12.68 & 92.5 & \checkmark \\
Sports & NAU & $\lambda_n=-10$ & 12.72 & 94.3 & 9.97 & 93.9 &  \\
Sports & NAU & $\lambda_n=0$ & 31.13 & 92.1 & 27.64 & 91.0 &  \\
\midrule
Toys & SCIF & scif\_d0.001 & 23.12 & 102.2 & 11.15 & 103.3 &  \\
Toys & SCIF & scif\_d0.01 & 23.12 & 102.1 & 11.15 & 103.3 & \checkmark \\
Toys & SCIF & scif\_d0.1 & 23.12 & 102.1 & 11.15 & 103.3 &  \\
Toys & SEIF & seif\_std0.006 & 8.36 & 54.1 & 6.54 & 55.0 &  \\
Toys & SEIF & seif\_std0.06 & 10.02 & 53.5 & 6.67 & 54.1 &  \\
Toys & SEIF & seif\_std0.6 & 0.00 & 0.5 & 0.00 & 1.2 & \checkmark \\
Toys & Kookmin & kookmin\_ir0.0001 & 19.79 & 77.9 & 9.74 & 79.0 &  \\
Toys & Kookmin & kookmin\_ir0.001 & 11.69 & 77.1 & 5.00 & 77.8 & \checkmark \\
Toys & Kookmin & kookmin\_ir0.01 & 15.02 & 70.0 & 1.67 & 71.2 &  \\
Toys & Fanchuan & fanchuan\_t0.007 & 0.00 & 4.0 & 0.00 & 5.2 & \checkmark \\
Toys & Fanchuan & fanchuan\_t0.07 & 0.00 & 7.4 & 0.00 & 8.9 &  \\
Toys & Fanchuan & fanchuan\_t1.15 & 0.00 & 17.3 & 1.54 & 18.3 &  \\
Toys & TRACER & tracer\_ep4 & 18.12 & 103.2 & 11.15 & 104.2 &  \\
Toys & TRACER & tracer\_lf0.1\_lc0.1\_ep4 & 21.46 & 102.4 & 11.15 & 103.4 &  \\
Toys & TRACER & tracer\_lf0.1\_lc1.0\_ep4 & 18.12 & 103.3 & 11.15 & 104.4 &  \\
Toys & TRACER & tracer\_lf1.0\_lc0.1\_ep4 & 21.46 & 102.0 & 11.15 & 103.3 &  \\
Toys & TRACER & tracer\_lf1.0\_lc1.0\_ep4 & 18.12 & 103.2 & 11.15 & 104.2 & \checkmark \\
Toys & NAU & $\lambda_n=-0.1$ & 9.87 & 91.6 & 8.21 & 92.7 &  \\
Toys & NAU & $\lambda_n=-1$ & 11.54 & 94.1 & 9.74 & 95.0 & \checkmark \\
Toys & NAU & $\lambda_n=-10$ & 8.33 & 94.5 & 8.21 & 95.6 &  \\
Toys & NAU & $\lambda_n=0$ & 11.69 & 91.5 & 5.00 & 92.5 &  \\
\bottomrule
\end{tabular}
\end{table}

\section{Deletion Scenario Construction}
\label[appendix]{sec:appendix-scenarios}
\label[appendix]{sec:appendix-popularity}

\header{Spam removal}
We inject a bandwagon attack before training.
Items are ranked by interaction count, and one target is drawn per attack from the bottom 20\% (unpopular), the 30th to 70th percentile (mid), or the top 5\% (popular) of this ranking.
The attack adds fake users equal to 1\% of the real users (226 on Beauty, 360 on Sports, 197 on Toys).
Each fake session has a length drawn from a lognormal-Poisson fit to the real session lengths, scaled by 0.8; it contains the target once, at a random position between 20\% and 90\% of the session, and every other position is a filler item drawn uniformly from the 20\% most popular items.
The forget set $D_f$ is all interactions of the fake users.
We build one attack per (dataset, data seed, popularity stratum), with data seeds $\{2,3,5,7,11\}$, and train the poisoned model $\bm{\theta}_o$ and the retrained reference $\bm{\theta}_r$ (on the unattacked data) for each.

\header{Unwanted-item removal}
Following ERASE~\citep{lubitzsch2026erase}, the unwanted set $\mathcal{I}_{\mathrm{u}}$ is a sensitive category: Hair Loss Products on Beauty (taxonomy level 2, 106 items, 0.88\% of the catalog), Guns \& Rifles on Sports (level 3, 202 items, 1.10\%), and toy weapons on Toys (68 items, 0.57\%, matched by the keywords gun, rifle, pistol, weapon, sword, knife, and dagger and their plurals).
Users who interacted with the category are shuffled with the data seed and added one at a time until their interactions with it reach 0.01\% of all interactions, which gives 12 to 14 requesting users per seed.
The forget set is the requesting users' interactions with $\mathcal{I}_{\mathrm{u}}$; their other interactions stay in the retain set.
The retrained reference is trained on the training data with exactly these interactions removed.

\section{Additional Spam-Removal Results}
\label[appendix]{sec:appendix-spam}

All tables follow the selection rule of the main results (the smallest $\Delta^{\mathrm{retr}}_{\mathrm{SH}}$ among configurations with $\tau_P \geq 95\%$); validation runs were made for TIGER + RQ-KMeans, so for the other tokenizers the rule is applied on the test split.

\header{Relative efficiency}
\Cref{tab:efficiency-beauty} reports $\tau_E$, the fraction of retraining cost saved, on Beauty; every selected configuration saves more than 99.7\%.

\begin{table}[ht]
\centering
\small
\caption{
Efficiency of unlearning methods on Beauty (TIGER + RQ-KMeans), each at its configuration in \Cref{tab:bandwagon-unlearn-all-tiger-rkmeans}.
$\mathcal{E}$ is the median wall-clock time per unlearning request on one H200 GPU over three targets $\times$ five data seeds, and $\tau_E = 1 - \mathcal{E}/\mathcal{E}_{\mathrm{retrain}}$ the fraction of retraining cost saved.
\textsuperscript{\dag}\,TRACER under its own tokenizer, as in \Cref{tab:bandwagon-unlearn-all-tiger-rkmeans}.
}
\label{tab:efficiency-beauty}
\begin{tabular}{lrr}
\toprule
Method & $\mathcal{E}$ (s) $\downarrow$ & $\tau_E$ $\uparrow$ \\
\midrule
Finetune & 0.5 & 0.99997 \\
Forget only & 0.3 & 0.99998 \\
Forget + Repair & 0.7 & 0.99995 \\
SCIF & 35.3 & 0.99774 \\
SEIF & 0.5 & 0.99997 \\
Kookmin & 0.4 & 0.99998 \\
Fanchuan & 1.6 & 0.99990 \\
TRACER\textsuperscript{\dag} & 2.7 & 0.99983 \\
\midrule
NAU (ours) & 3.9 & 0.99975 \\
\quad NAU, $\lambda_s=0.1$ & 3.5 & 0.99978 \\
\midrule
\textit{Retrain (reference)} & \textit{15,586} & \textit{0} \\
\bottomrule
\end{tabular}
\end{table}

\begin{table}[t]
\centering
\small
\caption{
Additional NAU variants on Beauty with TIGER and RQ-KMeans \acp{SID} under the bandwagon attack.
The variants differ in the separation score and the neighborhood/separation weight selection strategy.
$\mathrm{SH}@10$, $\mathrm{NDCG}@10$, and $\tau_P$ are reported as percentages.
}
\label{tab:nau-variants-beauty}
\begin{tabular}{llrrrr}
\toprule
Target & Variant & $\mathrm{SH}@10$ (\%) & $\mathrm{NDCG}@10$ (\%) & $\tau_P$ (\%) & $\mathcal{E}$ (s) \\
\midrule
\multirow{4}{*}{\rotatebox[origin=c]{90}{Unpopular}}
 & NAU (gen. sep.) & 0.000$\pm$0.000 & 3.600$\pm$0.119 & 95.8$\pm$6.2 & 3.5 \\
 & NAU (cos. sep.) & 0.000$\pm$0.000 & 3.504$\pm$0.092 & 93.2$\pm$5.4 & 2.9 \\
 & NAU & 0.000$\pm$0.000 & 3.596$\pm$0.115 & 95.6$\pm$6.1 & 3.4 \\
 & NAU (grid) & 0.000$\pm$0.000 & 3.612$\pm$0.106 & 96.1$\pm$5.8 & 3.7 \\
\midrule
\multirow{4}{*}{\rotatebox[origin=c]{90}{Mid}}
 & NAU (gen. sep.) & 0.000$\pm$0.000 & 3.654$\pm$0.041 & 97.1$\pm$2.8 & 3.5 \\
 & NAU (cos. sep.) & 0.000$\pm$0.000 & 3.545$\pm$0.053 & 94.2$\pm$2.3 & 2.9 \\
 & NAU & 0.000$\pm$0.000 & 3.656$\pm$0.044 & 97.2$\pm$3.1 & 1.7 \\
 & NAU (grid) & 0.004$\pm$0.008 & 3.599$\pm$0.041 & 95.7$\pm$3.2 & 1.4 \\
\midrule
\multirow{4}{*}{\rotatebox[origin=c]{90}{Popular}}
 & NAU (gen. sep.) & 0.000$\pm$0.000 & 3.596$\pm$0.148 & 95.5$\pm$2.6 & 3.4 \\
 & NAU (cos. sep.) & 0.107$\pm$0.240 & 3.515$\pm$0.161 & 93.4$\pm$3.1 & 2.9 \\
 & NAU & 0.000$\pm$0.000 & 3.603$\pm$0.145 & 95.7$\pm$2.6 & 1.6 \\
 & NAU (grid) & 0.000$\pm$0.000 & 3.610$\pm$0.163 & 95.9$\pm$2.6 & 3.0 \\
\bottomrule
\end{tabular}
\end{table}

\header{Additional NAU variants}
\Cref{tab:nau-variants-beauty} compares NAU variants on Beauty that differ in the separation score and in how the neighborhood and separation weights are chosen.

\header{TIGER with RQ-VAE \acp{SID}}
\Cref{tab:bandwagon-unlearn-beauty-tiger-rqvae,tab:bandwagon-unlearn-sports-tiger-rqvae,tab:bandwagon-unlearn-toys-tiger-rqvae} report spam-removal results when \acp{SID} are constructed using RQ-VAE.
For unpopular and mid-popularity targets, NAU is closest to retrained exposure in most blocks while keeping $\tau_P \geq 95\%$. For popular targets, which the retrained model still recommends, NAU suppresses the target almost completely and is farther from retraining than the best baselines; the separation term is the likely cause (\Cref{tab:ablation-compact-beauty}), and a target-dependent $\lambda_s$ is a natural extension.
\begin{table*}[t]
\centering
\small
\setlength{\tabcolsep}{4pt}
\renewcommand{\arraystretch}{1.05}
\caption{
Spam-removal results on Beauty with TIGER + RQ-VAE \acp{SID} under the bandwagon attack.
Each baseline uses the configuration selected for TIGER + RQ-KMeans on the validation split of the same dataset (\Cref{tab:appendix-valsel}).
NAU is reported at its operating point ($\lambda_f=0.1$, $\lambda_s=0.01$, $\lambda_n=0.01$, generative separation score); the indented row below it uses a stronger separation weight ($\lambda_s=0.1$) and is not ranked.
$\mathrm{SH}@10$, $\Delta^{\mathrm{retr}}_{\mathrm{SH}}$, $\mathrm{NDCG}@10$, and $\tau_P$ are reported as percentages.
$\Delta^{\mathrm{retr}}_{\mathrm{SH}}$ is the mean over (seed, target) cells of $|\mathrm{SH}@10-\mathrm{SH}@10_{\mathrm{retr}}|$, so it measures closeness to retraining rather than raw suppression; it is the same quantity Table~\ref{tab:bandwagon-unlearn-all-tiger-rkmeans} reports.
SH@10 measures target suppression; closeness to retraining also depends on retrained exposure. Higher NDCG@10 and $\tau_P$ indicate better utility preservation.
$\mathcal{E}$ reports unlearning time per request in seconds.
}
\label{tab:bandwagon-unlearn-beauty-tiger-rqvae}
\resizebox{\textwidth}{!}{%
\begin{tabular}{llrrrrr}
\toprule
Target & Method & $\mathrm{SH}@10$ (\%) & $\Delta^{\mathrm{retr}}_{\mathrm{SH}}$ (\%) $\downarrow$ & $\mathrm{NDCG}@10$ (\%) & $\tau_P$ (\%) & $\mathcal{E}$ (s) \\
\midrule
\multirow{12}{*}{\rotatebox[origin=c]{90}{Unpopular}}
 & \multicolumn{6}{l}{\textit{Baselines}} \\
 & Finetune & 0.365$\pm$0.362 & 0.361$\pm$0.362 & \textbf{3.206$\pm$0.106} & \textbf{98.6$\pm$2.4} & \underline{0.9} \\
 & Forget only & 0.000$\pm$0.000 & 0.004$\pm$0.008 & 0.166$\pm$0.288 & 5.2$\pm$9.1 & 0.4 \\
 & Forget + Repair & \underline{0.016$\pm$0.036} & \underline{0.020$\pm$0.035} & 3.196$\pm$0.104 & 98.3$\pm$2.6 & \textbf{0.8} \\
 & SCIF & 0.729$\pm$1.081 & 0.725$\pm$1.082 & 3.203$\pm$0.097 & 98.5$\pm$1.2 & 35.2 \\
 & SEIF & 0.119$\pm$0.190 & 0.115$\pm$0.182 & 2.267$\pm$0.192 & 69.8$\pm$6.6 & 0.5 \\
 & Kookmin & 0.193$\pm$0.119 & 0.190$\pm$0.114 & 2.811$\pm$0.156 & 86.5$\pm$5.2 & 0.4 \\
 & Fanchuan & 0.000$\pm$0.000 & 0.004$\pm$0.008 & 1.113$\pm$0.291 & 34.4$\pm$9.7 & 1.6 \\
 & TRACER & 0.658$\pm$1.220 & 0.655$\pm$1.221 & 3.173$\pm$0.101 & 97.6$\pm$1.6 & 2.3 \\
\cmidrule(lr){2-7}
 & \multicolumn{6}{l}{\textit{Ours}} \\
 & NAU & \textbf{0.000$\pm$0.000} & \textbf{0.004$\pm$0.008} & \underline{3.206$\pm$0.084} & \underline{98.6$\pm$2.0} & 4.0 \\
 & \quad NAU, $\lambda_s=0.1$ & 0.000$\pm$0.000 & 0.004$\pm$0.008 & 3.205$\pm$0.093 & 98.6$\pm$2.1 & 3.5 \\
\midrule
\multirow{12}{*}{\rotatebox[origin=c]{90}{Mid}}
 & \multicolumn{6}{l}{\textit{Baselines}} \\
 & Finetune & 1.116$\pm$1.561 & 1.116$\pm$1.561 & \textbf{3.224$\pm$0.107} & \textbf{99.2$\pm$3.4} & \textbf{0.5} \\
 & Forget only & 0.000$\pm$0.000 & 0.000$\pm$0.000 & 0.045$\pm$0.055 & 1.4$\pm$1.7 & 0.4 \\
 & Forget + Repair & \underline{0.058$\pm$0.080} & \underline{0.058$\pm$0.080} & \underline{3.213$\pm$0.089} & \underline{98.8$\pm$3.0} & \underline{0.7} \\
 & SCIF & 2.482$\pm$4.410 & 2.482$\pm$4.410 & 3.209$\pm$0.157 & 98.7$\pm$5.3 & 35.2 \\
 & SEIF & 0.061$\pm$0.050 & 0.061$\pm$0.050 & 2.358$\pm$0.112 & 72.6$\pm$3.6 & 0.5 \\
 & Kookmin & 0.391$\pm$0.553 & 0.391$\pm$0.553 & 2.716$\pm$0.077 & 83.5$\pm$2.5 & 0.4 \\
 & Fanchuan & 0.000$\pm$0.000 & 0.000$\pm$0.000 & 1.078$\pm$0.206 & 33.2$\pm$6.8 & 1.6 \\
 & TRACER & 0.794$\pm$1.121 & 0.794$\pm$1.121 & 3.077$\pm$0.118 & 94.7$\pm$3.3 & 2.4 \\
\cmidrule(lr){2-7}
 & \multicolumn{6}{l}{\textit{Ours}} \\
 & NAU & \textbf{0.020$\pm$0.029} & \textbf{0.020$\pm$0.029} & 3.211$\pm$0.090 & 98.8$\pm$3.0 & 4.3 \\
 & \quad NAU, $\lambda_s=0.1$ & 0.000$\pm$0.000 & 0.000$\pm$0.000 & 3.189$\pm$0.103 & 98.1$\pm$3.3 & 3.4 \\
\midrule
\multirow{12}{*}{\rotatebox[origin=c]{90}{Popular}}
 & \multicolumn{6}{l}{\textit{Baselines}} \\
 & Finetune & 0.441$\pm$0.402 & \underline{0.227$\pm$0.226} & \textbf{3.286$\pm$0.071} & \textbf{101.1$\pm$2.9} & \textbf{0.9} \\
 & Forget only & 0.000$\pm$0.000 & 0.436$\pm$0.479 & 0.030$\pm$0.051 & 0.9$\pm$1.5 & 0.3 \\
 & Forget + Repair & \underline{0.002$\pm$0.004} & 0.434$\pm$0.478 & \underline{3.284$\pm$0.080} & \underline{101.0$\pm$2.8} & \underline{1.3} \\
 & SCIF & 0.494$\pm$0.369 & \textbf{0.208$\pm$0.118} & 3.241$\pm$0.070 & 99.7$\pm$2.3 & 35.1 \\
 & SEIF & 0.056$\pm$0.045 & 0.395$\pm$0.424 & 2.277$\pm$0.126 & 70.0$\pm$4.0 & 0.5 \\
 & Kookmin & 0.849$\pm$1.083 & 0.485$\pm$0.555 & 2.760$\pm$0.193 & 84.9$\pm$6.7 & 0.4 \\
 & Fanchuan & 0.000$\pm$0.000 & 0.436$\pm$0.479 & 0.938$\pm$0.086 & 28.9$\pm$2.8 & 1.6 \\
 & TRACER & 0.454$\pm$0.249 & 0.265$\pm$0.251 & 3.184$\pm$0.046 & 97.9$\pm$1.4 & 2.5 \\
\cmidrule(lr){2-7}
 & \multicolumn{6}{l}{\textit{Ours}} \\
 & NAU & \textbf{0.000$\pm$0.000} & 0.436$\pm$0.479 & 3.281$\pm$0.080 & 100.9$\pm$2.9 & 4.2 \\
 & \quad NAU, $\lambda_s=0.1$ & 0.000$\pm$0.000 & 0.436$\pm$0.479 & 3.275$\pm$0.075 & 100.7$\pm$3.0 & 3.5 \\
\bottomrule
\end{tabular}}
\end{table*}

\begin{table*}[t]
\centering
\small
\setlength{\tabcolsep}{4pt}
\renewcommand{\arraystretch}{1.05}
\caption{
Spam-removal results on Sports with TIGER + RQ-VAE \acp{SID} under the bandwagon attack.
Each baseline uses the configuration selected for TIGER + RQ-KMeans on the validation split of the same dataset (\Cref{tab:appendix-valsel}).
NAU is reported at its operating point ($\lambda_f=0.1$, $\lambda_s=0.01$, $\lambda_n=0.01$, generative separation score); the indented row below it uses a stronger separation weight ($\lambda_s=0.1$) and is not ranked.
$\mathrm{SH}@10$, $\Delta^{\mathrm{retr}}_{\mathrm{SH}}$, $\mathrm{NDCG}@10$, and $\tau_P$ are reported as percentages.
$\Delta^{\mathrm{retr}}_{\mathrm{SH}}$ is the mean over (seed, target) cells of $|\mathrm{SH}@10-\mathrm{SH}@10_{\mathrm{retr}}|$, so it measures closeness to retraining rather than raw suppression; it is the same quantity Table~\ref{tab:bandwagon-unlearn-all-tiger-rkmeans} reports.
SH@10 measures target suppression; closeness to retraining also depends on retrained exposure. Higher NDCG@10 and $\tau_P$ indicate better utility preservation.
$\mathcal{E}$ reports unlearning time per request in seconds.
}
\label{tab:bandwagon-unlearn-sports-tiger-rqvae}
\resizebox{\textwidth}{!}{%
\begin{tabular}{llrrrrr}
\toprule
Target & Method & $\mathrm{SH}@10$ (\%) & $\Delta^{\mathrm{retr}}_{\mathrm{SH}}$ (\%) $\downarrow$ & $\mathrm{NDCG}@10$ (\%) & $\tau_P$ (\%) & $\mathcal{E}$ (s) \\
\midrule
\multirow{12}{*}{\rotatebox[origin=c]{90}{Unpopular}}
 & \multicolumn{6}{l}{\textit{Baselines}} \\
 & Finetune & 0.494$\pm$0.690 & 0.494$\pm$0.690 & \textbf{1.719$\pm$0.095} & \textbf{106.0$\pm$8.9} & \underline{1.3} \\
 & Forget only & 0.000$\pm$0.000 & 0.000$\pm$0.000 & 0.085$\pm$0.178 & 6.0$\pm$12.7 & 0.4 \\
 & Forget + Repair & \underline{0.178$\pm$0.395} & \underline{0.178$\pm$0.395} & \underline{1.710$\pm$0.095} & \underline{105.4$\pm$9.0} & \textbf{1.0} \\
 & SCIF & 0.725$\pm$1.004 & 0.725$\pm$1.004 & 1.658$\pm$0.091 & 102.3$\pm$10.5 & 34.9 \\
 & SEIF & 0.011$\pm$0.015 & 0.011$\pm$0.015 & 1.001$\pm$0.073 & 61.8$\pm$6.8 & 0.7 \\
 & Kookmin & 0.147$\pm$0.211 & 0.147$\pm$0.211 & 1.336$\pm$0.092 & 82.4$\pm$8.4 & 0.5 \\
 & Fanchuan & 0.000$\pm$0.000 & 0.000$\pm$0.000 & 0.535$\pm$0.088 & 33.1$\pm$6.8 & 2.1 \\
 & TRACER & 0.725$\pm$1.035 & 0.725$\pm$1.035 & 1.657$\pm$0.091 & 102.4$\pm$11.0 & 4.5 \\
\cmidrule(lr){2-7}
 & \multicolumn{6}{l}{\textit{Ours}} \\
 & NAU & \textbf{0.042$\pm$0.090} & \textbf{0.042$\pm$0.090} & 1.700$\pm$0.098 & 104.8$\pm$8.2 & 8.1 \\
 & \quad NAU, $\lambda_s=0.1$ & 0.000$\pm$0.000 & 0.000$\pm$0.000 & 1.711$\pm$0.097 & 105.4$\pm$8.5 & 4.2 \\
\midrule
\multirow{12}{*}{\rotatebox[origin=c]{90}{Mid}}
 & \multicolumn{6}{l}{\textit{Baselines}} \\
 & Finetune & 1.237$\pm$1.330 & 1.189$\pm$1.268 & \underline{1.690$\pm$0.064} & \underline{104.2$\pm$8.6} & \textbf{1.3} \\
 & Forget only & 0.000$\pm$0.000 & 0.048$\pm$0.107 & 0.000$\pm$0.001 & 0.0$\pm$0.1 & 0.5 \\
 & Forget + Repair & \underline{0.107$\pm$0.154} & \underline{0.059$\pm$0.075} & \textbf{1.696$\pm$0.065} & \textbf{104.7$\pm$9.2} & \underline{1.7} \\
 & SCIF & 1.705$\pm$1.942 & 1.657$\pm$1.903 & 1.664$\pm$0.063 & 102.7$\pm$9.6 & 35.2 \\
 & SEIF & 0.052$\pm$0.056 & 0.054$\pm$0.058 & 0.983$\pm$0.082 & 60.4$\pm$2.8 & 0.7 \\
 & Kookmin & 0.411$\pm$0.380 & 0.364$\pm$0.371 & 1.322$\pm$0.077 & 81.4$\pm$5.3 & 0.5 \\
 & Fanchuan & 0.000$\pm$0.000 & 0.048$\pm$0.107 & 0.545$\pm$0.073 & 33.4$\pm$2.3 & 2.1 \\
 & TRACER & 1.549$\pm$1.736 & 1.501$\pm$1.673 & 1.660$\pm$0.064 & 102.5$\pm$9.9 & 4.7 \\
\cmidrule(lr){2-7}
 & \multicolumn{6}{l}{\textit{Ours}} \\
 & NAU & \textbf{0.077$\pm$0.143} & \textbf{0.029$\pm$0.041} & 1.680$\pm$0.062 & 103.7$\pm$9.4 & 5.1 \\
 & \quad NAU, $\lambda_s=0.1$ & 0.000$\pm$0.000 & 0.048$\pm$0.107 & 1.680$\pm$0.064 & 103.7$\pm$9.1 & 4.2 \\
\midrule
\multirow{12}{*}{\rotatebox[origin=c]{90}{Popular}}
 & \multicolumn{6}{l}{\textit{Baselines}} \\
 & Finetune & 0.406$\pm$0.417 & 0.393$\pm$0.413 & \underline{1.679$\pm$0.058} & \underline{103.7$\pm$9.9} & \textbf{1.4} \\
 & Forget only & 0.000$\pm$0.000 & 0.035$\pm$0.055 & 0.001$\pm$0.001 & 0.0$\pm$0.1 & 0.4 \\
 & Forget + Repair & \textbf{0.000$\pm$0.000} & \textbf{0.035$\pm$0.055} & \textbf{1.686$\pm$0.041} & \textbf{104.2$\pm$10.5} & \underline{1.7} \\
 & SCIF & 0.747$\pm$0.715 & 0.712$\pm$0.738 & 1.618$\pm$0.050 & 100.0$\pm$10.5 & 35.4 \\
 & SEIF & 0.011$\pm$0.009 & 0.030$\pm$0.049 & 1.006$\pm$0.092 & 62.2$\pm$8.5 & 0.7 \\
 & Kookmin & 0.289$\pm$0.482 & 0.303$\pm$0.470 & 1.403$\pm$0.102 & 86.6$\pm$9.3 & 0.5 \\
 & Fanchuan & 0.000$\pm$0.000 & 0.035$\pm$0.055 & 0.555$\pm$0.049 & 34.4$\pm$4.8 & 2.1 \\
 & TRACER & 0.475$\pm$0.420 & 0.469$\pm$0.405 & 1.621$\pm$0.048 & 100.2$\pm$10.6 & 4.5 \\
\cmidrule(lr){2-7}
 & \multicolumn{6}{l}{\textit{Ours}} \\
 & NAU & \underline{0.000$\pm$0.000} & \underline{0.035$\pm$0.055} & 1.663$\pm$0.057 & 102.8$\pm$11.2 & 5.2 \\
 & \quad NAU, $\lambda_s=0.1$ & 0.000$\pm$0.000 & 0.035$\pm$0.055 & 1.672$\pm$0.044 & 103.3$\pm$10.7 & 4.2 \\
\bottomrule
\end{tabular}}
\end{table*}

\begin{table*}[t]
\centering
\small
\setlength{\tabcolsep}{4pt}
\renewcommand{\arraystretch}{1.05}
\caption{
Spam-removal results on Toys with TIGER + RQ-VAE \acp{SID} under the bandwagon attack.
Each baseline uses the configuration selected for TIGER + RQ-KMeans on the validation split of the same dataset (\Cref{tab:appendix-valsel}).
NAU is reported at its operating point ($\lambda_f=0.1$, $\lambda_s=0.01$, $\lambda_n=0.01$, generative separation score); the indented row below it uses a stronger separation weight ($\lambda_s=0.1$) and is not ranked.
$\mathrm{SH}@10$, $\Delta^{\mathrm{retr}}_{\mathrm{SH}}$, $\mathrm{NDCG}@10$, and $\tau_P$ are reported as percentages.
$\Delta^{\mathrm{retr}}_{\mathrm{SH}}$ is the mean over (seed, target) cells of $|\mathrm{SH}@10-\mathrm{SH}@10_{\mathrm{retr}}|$, so it measures closeness to retraining rather than raw suppression; it is the same quantity Table~\ref{tab:bandwagon-unlearn-all-tiger-rkmeans} reports.
SH@10 measures target suppression; closeness to retraining also depends on retrained exposure. Higher NDCG@10 and $\tau_P$ indicate better utility preservation.
$\mathcal{E}$ reports unlearning time per request in seconds.
}
\label{tab:bandwagon-unlearn-toys-tiger-rqvae}
\resizebox{\textwidth}{!}{%
\begin{tabular}{llrrrrr}
\toprule
Target & Method & $\mathrm{SH}@10$ (\%) & $\Delta^{\mathrm{retr}}_{\mathrm{SH}}$ (\%) $\downarrow$ & $\mathrm{NDCG}@10$ (\%) & $\tau_P$ (\%) & $\mathcal{E}$ (s) \\
\midrule
\multirow{12}{*}{\rotatebox[origin=c]{90}{Unpopular}}
 & \multicolumn{6}{l}{\textit{Baselines}} \\
 & Finetune & 0.457$\pm$0.295 & 0.453$\pm$0.294 & \underline{3.009$\pm$0.048} & \underline{98.1$\pm$3.6} & \textbf{0.9} \\
 & Forget only & 0.000$\pm$0.000 & 0.004$\pm$0.009 & 0.048$\pm$0.062 & 1.5$\pm$1.9 & 0.4 \\
 & Forget + Repair & \underline{0.031$\pm$0.043} & \underline{0.033$\pm$0.042} & 3.005$\pm$0.063 & 97.9$\pm$3.7 & \underline{1.3} \\
 & SCIF & 0.651$\pm$0.577 & 0.647$\pm$0.579 & \textbf{3.034$\pm$0.058} & \textbf{98.8$\pm$2.9} & 34.2 \\
 & SEIF & 0.104$\pm$0.091 & 0.100$\pm$0.088 & 1.984$\pm$0.145 & 64.7$\pm$5.9 & 0.5 \\
 & Kookmin & 0.282$\pm$0.252 & 0.278$\pm$0.246 & 2.364$\pm$0.057 & 77.0$\pm$2.6 & 0.4 \\
 & Fanchuan & 0.000$\pm$0.000 & 0.004$\pm$0.009 & 0.710$\pm$0.136 & 23.0$\pm$3.8 & 1.6 \\
 & TRACER & 0.500$\pm$0.423 & 0.496$\pm$0.422 & 2.975$\pm$0.088 & 96.9$\pm$2.8 & 4.0 \\
\cmidrule(lr){2-7}
 & \multicolumn{6}{l}{\textit{Ours}} \\
 & NAU & \textbf{0.011$\pm$0.025} & \textbf{0.015$\pm$0.025} & 3.001$\pm$0.071 & 97.8$\pm$3.7 & 4.1 \\
 & \quad NAU, $\lambda_s=0.1$ & 0.001$\pm$0.002 & 0.005$\pm$0.009 & 2.999$\pm$0.066 & 97.7$\pm$3.5 & 3.4 \\
\midrule
\multirow{12}{*}{\rotatebox[origin=c]{90}{Mid}}
 & \multicolumn{6}{l}{\textit{Baselines}} \\
 & Finetune & 0.511$\pm$0.522 & 0.475$\pm$0.545 & 2.951$\pm$0.156 & 96.0$\pm$2.4 & \textbf{0.9} \\
 & Forget only & 0.000$\pm$0.000 & 0.036$\pm$0.066 & 0.020$\pm$0.033 & 0.6$\pm$1.1 & 0.4 \\
 & Forget + Repair & \underline{0.052$\pm$0.085} & \underline{0.077$\pm$0.089} & \underline{2.957$\pm$0.134} & \underline{96.2$\pm$1.7} & \underline{1.3} \\
 & SCIF & 1.289$\pm$1.985 & 1.253$\pm$2.004 & 2.948$\pm$0.102 & 96.0$\pm$1.7 & 34.4 \\
 & SEIF & 0.040$\pm$0.014 & 0.043$\pm$0.031 & 1.961$\pm$0.127 & 63.9$\pm$5.3 & 0.5 \\
 & Kookmin & 0.175$\pm$0.161 & 0.139$\pm$0.174 & 2.348$\pm$0.112 & 76.4$\pm$2.1 & 0.4 \\
 & Fanchuan & 0.000$\pm$0.000 & 0.036$\pm$0.066 & 0.791$\pm$0.221 & 25.8$\pm$7.4 & 1.6 \\
 & TRACER & 0.656$\pm$0.821 & 0.620$\pm$0.843 & 2.933$\pm$0.093 & 95.5$\pm$1.8 & 3.9 \\
\cmidrule(lr){2-7}
 & \multicolumn{6}{l}{\textit{Ours}} \\
 & NAU & \textbf{0.001$\pm$0.002} & \textbf{0.035$\pm$0.067} & \textbf{2.959$\pm$0.135} & \textbf{96.3$\pm$2.0} & 4.1 \\
 & \quad NAU, $\lambda_s=0.1$ & 0.000$\pm$0.000 & 0.036$\pm$0.066 & 2.955$\pm$0.147 & 96.2$\pm$2.5 & 3.4 \\
\midrule
\multirow{12}{*}{\rotatebox[origin=c]{90}{Popular}}
 & \multicolumn{6}{l}{\textit{Baselines}} \\
 & Finetune & 1.005$\pm$0.589 & 0.787$\pm$0.636 & \textbf{3.005$\pm$0.081} & \textbf{97.9$\pm$4.9} & \textbf{0.9} \\
 & Forget only & 0.000$\pm$0.000 & 0.217$\pm$0.120 & 0.004$\pm$0.006 & 0.1$\pm$0.2 & 0.4 \\
 & Forget + Repair & \underline{0.097$\pm$0.046} & \textbf{0.129$\pm$0.110} & \underline{2.992$\pm$0.078} & \underline{97.5$\pm$5.1} & \underline{1.3} \\
 & SCIF & 9.370$\pm$16.800 & 9.153$\pm$16.800 & 2.981$\pm$0.072 & 97.1$\pm$4.0 & 34.3 \\
 & SEIF & 0.108$\pm$0.104 & 0.140$\pm$0.115 & 2.014$\pm$0.154 & 65.5$\pm$3.7 & 0.5 \\
 & Kookmin & 0.561$\pm$0.387 & 0.353$\pm$0.403 & 2.329$\pm$0.072 & 75.8$\pm$2.4 & 0.4 \\
 & Fanchuan & 0.000$\pm$0.000 & 0.217$\pm$0.120 & 0.685$\pm$0.120 & 22.3$\pm$3.8 & 1.6 \\
 & TRACER & 1.327$\pm$0.744 & 1.110$\pm$0.699 & 2.921$\pm$0.077 & 95.2$\pm$5.2 & 4.4 \\
\cmidrule(lr){2-7}
 & \multicolumn{6}{l}{\textit{Ours}} \\
 & NAU & \textbf{0.019$\pm$0.025} & \underline{0.199$\pm$0.116} & 2.991$\pm$0.087 & 97.5$\pm$5.3 & 3.9 \\
 & \quad NAU, $\lambda_s=0.1$ & 0.000$\pm$0.000 & 0.217$\pm$0.120 & 2.979$\pm$0.074 & 97.1$\pm$5.0 & 3.4 \\
\bottomrule
\end{tabular}}
\end{table*}

\header{LETTER-style identifiers}
\Cref{tab:bandwagon-unlearn-beauty-letter-letter,tab:bandwagon-unlearn-sports-letter-letter,tab:bandwagon-unlearn-toys-letter-letter} report results with LETTER-style identifiers.
These results further test whether the observed behavior depends on a specific \ac{SID} construction.
NAU removes the promoted target in every stratum; Forget+Repair often matches its utility and its distance to retrained exposure, so the advantage on these tokenizers lies in removal rather than utility.
\begin{table*}[t]
\centering
\small
\setlength{\tabcolsep}{4pt}
\renewcommand{\arraystretch}{1.05}
\caption{
Spam-removal results on Beauty with TIGER + LETTER \acp{SID} under the bandwagon attack.
Each baseline uses the configuration selected for TIGER + RQ-KMeans on the validation split of the same dataset (\Cref{tab:appendix-valsel}).
NAU is reported at its operating point ($\lambda_f=0.1$, $\lambda_s=0.01$, $\lambda_n=0.01$, generative separation score); the indented row below it uses a stronger separation weight ($\lambda_s=0.1$) and is not ranked.
$\mathrm{SH}@10$, $\Delta^{\mathrm{retr}}_{\mathrm{SH}}$, $\mathrm{NDCG}@10$, and $\tau_P$ are reported as percentages.
$\Delta^{\mathrm{retr}}_{\mathrm{SH}}$ is the mean over (seed, target) cells of $|\mathrm{SH}@10-\mathrm{SH}@10_{\mathrm{retr}}|$, so it measures closeness to retraining rather than raw suppression; it is the same quantity Table~\ref{tab:bandwagon-unlearn-all-tiger-rkmeans} reports.
SH@10 measures target suppression; closeness to retraining also depends on retrained exposure. Higher NDCG@10 and $\tau_P$ indicate better utility preservation.
$\mathcal{E}$ reports unlearning time per request in seconds.
}
\label{tab:bandwagon-unlearn-beauty-letter-letter}
\resizebox{\textwidth}{!}{%
\begin{tabular}{llrrrrr}
\toprule
Target & Method & $\mathrm{SH}@10$ (\%) & $\Delta^{\mathrm{retr}}_{\mathrm{SH}}$ (\%) $\downarrow$ & $\mathrm{NDCG}@10$ (\%) & $\tau_P$ (\%) & $\mathcal{E}$ (s) \\
\midrule
\multirow{12}{*}{\rotatebox[origin=c]{90}{Unpopular}}
 & \multicolumn{6}{l}{\textit{Baselines}} \\
 & Finetune & 0.194$\pm$0.142 & 0.193$\pm$0.142 & \textbf{3.823$\pm$0.062} & \textbf{101.0$\pm$5.3} & \textbf{0.9} \\
 & Forget only & 0.000$\pm$0.000 & 0.001$\pm$0.002 & 0.011$\pm$0.015 & 0.3$\pm$0.4 & 0.4 \\
 & Forget + Repair & \textbf{0.000$\pm$0.000} & \textbf{0.001$\pm$0.002} & 3.778$\pm$0.082 & 99.8$\pm$5.8 & \underline{1.3} \\
 & SCIF & 0.245$\pm$0.183 & 0.244$\pm$0.182 & 3.764$\pm$0.062 & 99.4$\pm$3.8 & 40.0 \\
 & SEIF & 0.123$\pm$0.157 & 0.123$\pm$0.156 & 2.684$\pm$0.240 & 70.7$\pm$3.3 & 0.6 \\
 & Kookmin & 0.225$\pm$0.124 & 0.224$\pm$0.124 & 3.128$\pm$0.104 & 82.6$\pm$3.7 & 0.6 \\
 & Fanchuan & 0.000$\pm$0.000 & 0.001$\pm$0.002 & 1.244$\pm$0.189 & 32.7$\pm$3.7 & 1.6 \\
 & TRACER & 0.140$\pm$0.166 & 0.140$\pm$0.166 & 3.767$\pm$0.063 & 99.4$\pm$3.7 & 4.1 \\
\cmidrule(lr){2-7}
 & \multicolumn{6}{l}{\textit{Ours}} \\
 & NAU & \underline{0.000$\pm$0.000} & \underline{0.001$\pm$0.002} & \underline{3.797$\pm$0.090} & \underline{100.3$\pm$5.8} & 3.5 \\
 & \quad NAU, $\lambda_s=0.1$ & 0.000$\pm$0.000 & 0.001$\pm$0.002 & 3.791$\pm$0.089 & 100.2$\pm$5.8 & 3.5 \\
\midrule
\multirow{12}{*}{\rotatebox[origin=c]{90}{Mid}}
 & \multicolumn{6}{l}{\textit{Baselines}} \\
 & Finetune & 0.276$\pm$0.411 & 0.276$\pm$0.411 & \textbf{3.799$\pm$0.052} & \textbf{100.3$\pm$4.6} & \textbf{0.5} \\
 & Forget only & 0.000$\pm$0.000 & 0.000$\pm$0.000 & 0.015$\pm$0.023 & 0.4$\pm$0.6 & 0.4 \\
 & Forget + Repair & \underline{0.001$\pm$0.002} & \underline{0.001$\pm$0.002} & \underline{3.785$\pm$0.083} & \underline{100.0$\pm$4.6} & \underline{1.3} \\
 & SCIF & 0.523$\pm$0.731 & 0.523$\pm$0.731 & 3.781$\pm$0.083 & 99.9$\pm$5.0 & 56.7 \\
 & SEIF & 0.013$\pm$0.012 & 0.013$\pm$0.012 & 2.746$\pm$0.160 & 72.4$\pm$3.5 & 0.6 \\
 & Kookmin & 0.127$\pm$0.144 & 0.127$\pm$0.144 & 3.232$\pm$0.155 & 85.4$\pm$6.1 & 0.6 \\
 & Fanchuan & 0.000$\pm$0.000 & 0.000$\pm$0.000 & 1.359$\pm$0.227 & 36.0$\pm$6.6 & 1.8 \\
 & TRACER & 0.275$\pm$0.410 & 0.275$\pm$0.410 & 3.778$\pm$0.090 & 99.8$\pm$4.9 & 3.9 \\
\cmidrule(lr){2-7}
 & \multicolumn{6}{l}{\textit{Ours}} \\
 & NAU & \textbf{0.000$\pm$0.000} & \textbf{0.000$\pm$0.000} & 3.766$\pm$0.067 & 99.5$\pm$4.6 & 3.4 \\
 & \quad NAU, $\lambda_s=0.1$ & 0.000$\pm$0.000 & 0.000$\pm$0.000 & 3.785$\pm$0.050 & 100.0$\pm$4.3 & 3.5 \\
\midrule
\multirow{12}{*}{\rotatebox[origin=c]{90}{Popular}}
 & \multicolumn{6}{l}{\textit{Baselines}} \\
 & Finetune & 0.268$\pm$0.205 & \underline{0.198$\pm$0.226} & \textbf{3.789$\pm$0.099} & \textbf{100.1$\pm$5.0} & \textbf{0.6} \\
 & Forget only & 0.000$\pm$0.000 & 0.247$\pm$0.238 & 0.014$\pm$0.031 & 0.4$\pm$0.8 & 0.3 \\
 & Forget + Repair & \underline{0.001$\pm$0.002} & 0.246$\pm$0.238 & 3.744$\pm$0.063 & \underline{98.9$\pm$4.1} & \underline{0.7} \\
 & SCIF & 0.296$\pm$0.239 & 0.248$\pm$0.249 & 3.721$\pm$0.131 & 98.2$\pm$4.8 & 69.2 \\
 & SEIF & 0.080$\pm$0.102 & 0.210$\pm$0.234 & 2.627$\pm$0.066 & 69.4$\pm$4.3 & 1.0 \\
 & Kookmin & 0.354$\pm$0.424 & 0.326$\pm$0.438 & 3.136$\pm$0.148 & 82.9$\pm$7.2 & 0.7 \\
 & Fanchuan & 0.000$\pm$0.000 & 0.247$\pm$0.238 & 1.135$\pm$0.071 & 29.9$\pm$1.4 & 1.9 \\
 & TRACER & 0.141$\pm$0.122 & \textbf{0.195$\pm$0.215} & 3.722$\pm$0.127 & 98.3$\pm$4.8 & 2.7 \\
\cmidrule(lr){2-7}
 & \multicolumn{6}{l}{\textit{Ours}} \\
 & NAU & \textbf{0.000$\pm$0.000} & 0.247$\pm$0.238 & \underline{3.747$\pm$0.062} & 98.9$\pm$3.9 & 3.9 \\
 & \quad NAU, $\lambda_s=0.1$ & 0.000$\pm$0.000 & 0.247$\pm$0.238 & 3.751$\pm$0.063 & 99.1$\pm$4.0 & 3.5 \\
\bottomrule
\end{tabular}}
\end{table*}

\begin{table*}[t]
\centering
\small
\setlength{\tabcolsep}{4pt}
\renewcommand{\arraystretch}{1.05}
\caption{
Spam-removal results on Sports with TIGER + LETTER \acp{SID} under the bandwagon attack.
Each baseline uses the configuration selected for TIGER + RQ-KMeans on the validation split of the same dataset (\Cref{tab:appendix-valsel}).
NAU is reported at its operating point ($\lambda_f=0.1$, $\lambda_s=0.01$, $\lambda_n=0.01$, generative separation score); the indented row below it uses a stronger separation weight ($\lambda_s=0.1$) and is not ranked.
$\mathrm{SH}@10$, $\Delta^{\mathrm{retr}}_{\mathrm{SH}}$, $\mathrm{NDCG}@10$, and $\tau_P$ are reported as percentages.
$\Delta^{\mathrm{retr}}_{\mathrm{SH}}$ is the mean over (seed, target) cells of $|\mathrm{SH}@10-\mathrm{SH}@10_{\mathrm{retr}}|$, so it measures closeness to retraining rather than raw suppression; it is the same quantity Table~\ref{tab:bandwagon-unlearn-all-tiger-rkmeans} reports.
SH@10 measures target suppression; closeness to retraining also depends on retrained exposure. Higher NDCG@10 and $\tau_P$ indicate better utility preservation.
$\mathcal{E}$ reports unlearning time per request in seconds.
}
\label{tab:bandwagon-unlearn-sports-letter-letter}
\resizebox{\textwidth}{!}{%
\begin{tabular}{llrrrrr}
\toprule
Target & Method & $\mathrm{SH}@10$ (\%) & $\Delta^{\mathrm{retr}}_{\mathrm{SH}}$ (\%) $\downarrow$ & $\mathrm{NDCG}@10$ (\%) & $\tau_P$ (\%) & $\mathcal{E}$ (s) \\
\midrule
\multirow{12}{*}{\rotatebox[origin=c]{90}{Unpopular}}
 & \multicolumn{6}{l}{\textit{Baselines}} \\
 & Finetune & 0.240$\pm$0.231 & 0.167$\pm$0.204 & \textbf{1.992$\pm$0.053} & \textbf{105.8$\pm$7.0} & \underline{1.3} \\
 & Forget only & 0.000$\pm$0.000 & 0.078$\pm$0.105 & 0.011$\pm$0.021 & 0.6$\pm$1.1 & 0.3 \\
 & Forget + Repair & \underline{0.030$\pm$0.066} & \underline{0.097$\pm$0.100} & 1.976$\pm$0.055 & 104.9$\pm$7.2 & \textbf{1.0} \\
 & SCIF & 0.334$\pm$0.352 & 0.256$\pm$0.345 & 1.946$\pm$0.060 & 103.3$\pm$6.4 & 34.8 \\
 & SEIF & 0.157$\pm$0.210 & 0.079$\pm$0.106 & 1.206$\pm$0.087 & 64.0$\pm$6.3 & 0.8 \\
 & Kookmin & 0.276$\pm$0.325 & 0.198$\pm$0.231 & 1.555$\pm$0.148 & 82.7$\pm$10.7 & 0.5 \\
 & Fanchuan & 0.000$\pm$0.000 & 0.078$\pm$0.105 & 0.653$\pm$0.038 & 34.7$\pm$3.1 & 2.1 \\
 & TRACER & 0.221$\pm$0.240 & 0.166$\pm$0.227 & 1.954$\pm$0.061 & 103.7$\pm$6.6 & 4.9 \\
\cmidrule(lr){2-7}
 & \multicolumn{6}{l}{\textit{Ours}} \\
 & NAU & \textbf{0.013$\pm$0.030} & \textbf{0.081$\pm$0.103} & \underline{1.979$\pm$0.049} & \underline{105.1$\pm$7.0} & 4.2 \\
 & \quad NAU, $\lambda_s=0.1$ & 0.000$\pm$0.000 & 0.078$\pm$0.105 & 1.982$\pm$0.038 & 105.2$\pm$6.6 & 4.2 \\
\midrule
\multirow{12}{*}{\rotatebox[origin=c]{90}{Mid}}
 & \multicolumn{6}{l}{\textit{Baselines}} \\
 & Finetune & 0.625$\pm$1.247 & 0.625$\pm$1.248 & \textbf{1.972$\pm$0.026} & \textbf{104.6$\pm$4.0} & \underline{1.4} \\
 & Forget only & 0.000$\pm$0.000 & 0.001$\pm$0.001 & 0.000$\pm$0.000 & 0.0$\pm$0.0 & 0.3 \\
 & Forget + Repair & \underline{0.007$\pm$0.016} & \underline{0.008$\pm$0.016} & \underline{1.966$\pm$0.026} & \underline{104.3$\pm$4.2} & \textbf{1.0} \\
 & SCIF & 1.233$\pm$2.583 & 1.233$\pm$2.583 & 1.934$\pm$0.044 & 102.6$\pm$3.8 & 39.5 \\
 & SEIF & 0.020$\pm$0.027 & 0.020$\pm$0.027 & 1.141$\pm$0.074 & 60.5$\pm$3.4 & 0.8 \\
 & Kookmin & 0.240$\pm$0.269 & 0.239$\pm$0.270 & 1.496$\pm$0.069 & 79.4$\pm$5.2 & 0.5 \\
 & Fanchuan & 0.000$\pm$0.000 & 0.001$\pm$0.001 & 0.617$\pm$0.021 & 32.8$\pm$2.3 & 2.3 \\
 & TRACER & 0.951$\pm$2.067 & 0.951$\pm$2.068 & 1.932$\pm$0.046 & 102.4$\pm$3.7 & 4.7 \\
\cmidrule(lr){2-7}
 & \multicolumn{6}{l}{\textit{Ours}} \\
 & NAU & \textbf{0.000$\pm$0.000} & \textbf{0.001$\pm$0.001} & 1.943$\pm$0.025 & 103.0$\pm$3.8 & 4.2 \\
 & \quad NAU, $\lambda_s=0.1$ & 0.000$\pm$0.000 & 0.001$\pm$0.001 & 1.943$\pm$0.031 & 103.0$\pm$3.8 & 4.3 \\
\midrule
\multirow{12}{*}{\rotatebox[origin=c]{90}{Popular}}
 & \multicolumn{6}{l}{\textit{Baselines}} \\
 & Finetune & 0.545$\pm$0.903 & 0.524$\pm$0.915 & \textbf{1.915$\pm$0.047} & \textbf{101.6$\pm$2.8} & \textbf{0.8} \\
 & Forget only & 0.000$\pm$0.000 & 0.021$\pm$0.032 & 0.003$\pm$0.004 & 0.2$\pm$0.2 & 0.5 \\
 & Forget + Repair & \underline{0.002$\pm$0.005} & \underline{0.023$\pm$0.031} & 1.908$\pm$0.056 & 101.1$\pm$2.1 & \underline{0.9} \\
 & SCIF & 2.331$\pm$4.948 & 2.310$\pm$4.960 & 1.886$\pm$0.061 & 100.0$\pm$2.7 & 68.6 \\
 & SEIF & 0.013$\pm$0.011 & 0.022$\pm$0.027 & 1.234$\pm$0.124 & 65.3$\pm$3.8 & 0.8 \\
 & Kookmin & 0.133$\pm$0.094 & 0.137$\pm$0.067 & 1.571$\pm$0.145 & 83.1$\pm$4.9 & 0.5 \\
 & Fanchuan & 0.000$\pm$0.000 & 0.021$\pm$0.032 & 0.692$\pm$0.097 & 36.6$\pm$4.6 & 2.1 \\
 & TRACER & 1.615$\pm$3.515 & 1.595$\pm$3.527 & 1.884$\pm$0.054 & 99.9$\pm$2.9 & 4.7 \\
\cmidrule(lr){2-7}
 & \multicolumn{6}{l}{\textit{Ours}} \\
 & NAU & \textbf{0.000$\pm$0.000} & \textbf{0.021$\pm$0.032} & \underline{1.910$\pm$0.058} & \underline{101.2$\pm$2.2} & 4.2 \\
 & \quad NAU, $\lambda_s=0.1$ & 0.000$\pm$0.000 & 0.021$\pm$0.032 & 1.904$\pm$0.066 & 100.9$\pm$2.2 & 4.2 \\
\bottomrule
\end{tabular}}
\end{table*}

\begin{table*}[t]
\centering
\small
\setlength{\tabcolsep}{4pt}
\renewcommand{\arraystretch}{1.05}
\caption{
Spam-removal results on Toys with TIGER + LETTER \acp{SID} under the bandwagon attack.
Each baseline uses the configuration selected for TIGER + RQ-KMeans on the validation split of the same dataset (\Cref{tab:appendix-valsel}).
NAU is reported at its operating point ($\lambda_f=0.1$, $\lambda_s=0.01$, $\lambda_n=0.01$, generative separation score); the indented row below it uses a stronger separation weight ($\lambda_s=0.1$) and is not ranked.
$\mathrm{SH}@10$, $\Delta^{\mathrm{retr}}_{\mathrm{SH}}$, $\mathrm{NDCG}@10$, and $\tau_P$ are reported as percentages.
$\Delta^{\mathrm{retr}}_{\mathrm{SH}}$ is the mean over (seed, target) cells of $|\mathrm{SH}@10-\mathrm{SH}@10_{\mathrm{retr}}|$, so it measures closeness to retraining rather than raw suppression; it is the same quantity Table~\ref{tab:bandwagon-unlearn-all-tiger-rkmeans} reports.
SH@10 measures target suppression; closeness to retraining also depends on retrained exposure. Higher NDCG@10 and $\tau_P$ indicate better utility preservation.
$\mathcal{E}$ reports unlearning time per request in seconds.
}
\label{tab:bandwagon-unlearn-toys-letter-letter}
\resizebox{\textwidth}{!}{%
\begin{tabular}{llrrrrr}
\toprule
Target & Method & $\mathrm{SH}@10$ (\%) & $\Delta^{\mathrm{retr}}_{\mathrm{SH}}$ (\%) $\downarrow$ & $\mathrm{NDCG}@10$ (\%) & $\tau_P$ (\%) & $\mathcal{E}$ (s) \\
\midrule
\multirow{12}{*}{\rotatebox[origin=c]{90}{Unpopular}}
 & \multicolumn{6}{l}{\textit{Baselines}} \\
 & Finetune & 0.518$\pm$0.398 & 0.501$\pm$0.415 & 3.374$\pm$0.059 & 94.6$\pm$3.3 & 0.5 \\
 & Forget only & 0.000$\pm$0.000 & 0.018$\pm$0.034 & 0.000$\pm$0.000 & 0.0$\pm$0.0 & 0.3 \\
 & Forget + Repair & \textbf{0.012$\pm$0.013} & \textbf{0.026$\pm$0.031} & 3.391$\pm$0.048 & 95.0$\pm$2.6 & \textbf{1.2} \\
 & SCIF & 0.874$\pm$0.741 & 0.856$\pm$0.759 & \textbf{3.411$\pm$0.054} & \textbf{95.6$\pm$2.8} & 39.1 \\
 & SEIF & 0.098$\pm$0.087 & 0.109$\pm$0.072 & 2.124$\pm$0.117 & 59.6$\pm$4.4 & 0.6 \\
 & Kookmin & 0.440$\pm$0.294 & 0.422$\pm$0.280 & 2.469$\pm$0.172 & 69.1$\pm$4.2 & 0.4 \\
 & Fanchuan & 0.000$\pm$0.000 & 0.018$\pm$0.034 & 0.923$\pm$0.157 & 25.9$\pm$4.6 & 1.8 \\
 & TRACER & \underline{0.554$\pm$0.483} & \underline{0.537$\pm$0.501} & \underline{3.403$\pm$0.059} & \underline{95.3$\pm$2.9} & \underline{2.4} \\
\cmidrule(lr){2-7}
 & \multicolumn{6}{l}{\textit{Ours}} \\
 & NAU & 0.002$\pm$0.003 & 0.018$\pm$0.034 & 3.384$\pm$0.037 & 94.8$\pm$2.5 & 3.4 \\
 & \quad NAU, $\lambda_s=0.1$ & 0.000$\pm$0.000 & 0.018$\pm$0.034 & 3.383$\pm$0.038 & 94.8$\pm$2.5 & 3.4 \\
\midrule
\multirow{12}{*}{\rotatebox[origin=c]{90}{Mid}}
 & \multicolumn{6}{l}{\textit{Baselines}} \\
 & Finetune & 0.378$\pm$0.201 & 0.264$\pm$0.206 & 3.525$\pm$0.129 & \underline{98.8$\pm$4.6} & \textbf{0.6} \\
 & Forget only & 0.000$\pm$0.000 & 0.121$\pm$0.200 & 0.000$\pm$0.000 & 0.0$\pm$0.0 & 0.3 \\
 & Forget + Repair & \underline{0.013$\pm$0.019} & \underline{0.130$\pm$0.191} & 3.459$\pm$0.193 & 96.9$\pm$5.7 & \underline{1.3} \\
 & SCIF & 0.778$\pm$0.657 & 0.657$\pm$0.754 & \underline{3.526$\pm$0.098} & 98.8$\pm$3.8 & 66.8 \\
 & SEIF & 0.062$\pm$0.110 & 0.077$\pm$0.089 & 2.334$\pm$0.374 & 65.5$\pm$11.4 & 0.5 \\
 & Kookmin & 0.226$\pm$0.232 & 0.150$\pm$0.085 & 2.681$\pm$0.247 & 75.2$\pm$7.9 & 0.4 \\
 & Fanchuan & 0.000$\pm$0.000 & 0.121$\pm$0.200 & 1.082$\pm$0.341 & 30.4$\pm$9.9 & 1.8 \\
 & TRACER & 0.510$\pm$0.408 & 0.424$\pm$0.469 & \textbf{3.536$\pm$0.087} & \textbf{99.1$\pm$3.6} & 4.1 \\
\cmidrule(lr){2-7}
 & \multicolumn{6}{l}{\textit{Ours}} \\
 & NAU & \textbf{0.004$\pm$0.009} & \textbf{0.123$\pm$0.199} & 3.510$\pm$0.160 & 98.4$\pm$5.1 & 3.4 \\
 & \quad NAU, $\lambda_s=0.1$ & 0.000$\pm$0.000 & 0.121$\pm$0.200 & 3.504$\pm$0.159 & 98.2$\pm$5.1 & 3.4 \\
\midrule
\multirow{12}{*}{\rotatebox[origin=c]{90}{Popular}}
 & \multicolumn{6}{l}{\textit{Baselines}} \\
 & Finetune & 0.522$\pm$0.393 & 0.385$\pm$0.335 & \underline{3.585$\pm$0.066} & \underline{100.4$\pm$3.0} & \textbf{0.5} \\
 & Forget only & 0.000$\pm$0.000 & 0.176$\pm$0.062 & 0.000$\pm$0.000 & 0.0$\pm$0.0 & 0.4 \\
 & Forget + Repair & \underline{0.071$\pm$0.105} & \textbf{0.132$\pm$0.073} & \textbf{3.587$\pm$0.065} & \textbf{100.5$\pm$3.3} & \underline{0.8} \\
 & SCIF & 0.919$\pm$0.698 & 0.768$\pm$0.660 & 3.563$\pm$0.053 & 99.8$\pm$3.3 & 66.8 \\
 & SEIF & 0.123$\pm$0.167 & 0.103$\pm$0.064 & 2.455$\pm$0.279 & 68.8$\pm$7.7 & 0.7 \\
 & Kookmin & 0.352$\pm$0.339 & 0.252$\pm$0.273 & 2.862$\pm$0.168 & 80.2$\pm$5.6 & 0.4 \\
 & Fanchuan & 0.000$\pm$0.000 & 0.176$\pm$0.062 & 1.088$\pm$0.114 & 30.5$\pm$3.1 & 1.6 \\
 & TRACER & 0.652$\pm$0.566 & 0.530$\pm$0.500 & 3.557$\pm$0.056 & 99.7$\pm$3.3 & 4.2 \\
\cmidrule(lr){2-7}
 & \multicolumn{6}{l}{\textit{Ours}} \\
 & NAU & \textbf{0.026$\pm$0.036} & \underline{0.150$\pm$0.071} & 3.570$\pm$0.054 & 100.0$\pm$2.7 & 3.5 \\
 & \quad NAU, $\lambda_s=0.1$ & 0.004$\pm$0.009 & 0.172$\pm$0.062 & 3.562$\pm$0.061 & 99.8$\pm$3.0 & 3.5 \\
\bottomrule
\end{tabular}}
\end{table*}

\header{TRACER-compatible tokenization}
\Cref{tab:bandwagon-unlearn-beauty-tiger-tracerv3} reports results under the TRACER-compatible tokenization setting.
This comparison provides an additional robustness check under an identifier construction aligned with the TRACER baseline.

\begin{table*}[t]
\centering
\small
\setlength{\tabcolsep}{4pt}
\renewcommand{\arraystretch}{1.05}
\caption{
Spam-removal results on Beauty with TIGER + TRACER \acp{SID} under the bandwagon attack.
Each baseline uses the configuration selected for TIGER + RQ-KMeans on the validation split of the same dataset (\Cref{tab:appendix-valsel}).
NAU is reported at its operating point ($\lambda_f=0.1$, $\lambda_s=0.01$, $\lambda_n=0.01$, generative separation score); the indented row below it uses a stronger separation weight ($\lambda_s=0.1$) and is not ranked.
$\mathrm{SH}@10$, $\Delta^{\mathrm{retr}}_{\mathrm{SH}}$, $\mathrm{NDCG}@10$, and $\tau_P$ are reported as percentages.
$\Delta^{\mathrm{retr}}_{\mathrm{SH}}$ is the mean over (seed, target) cells of $|\mathrm{SH}@10-\mathrm{SH}@10_{\mathrm{retr}}|$, so it measures closeness to retraining rather than raw suppression; it is the same quantity Table~\ref{tab:bandwagon-unlearn-all-tiger-rkmeans} reports.
SH@10 measures target suppression; closeness to retraining also depends on retrained exposure. Higher NDCG@10 and $\tau_P$ indicate better utility preservation.
$\mathcal{E}$ reports unlearning time per request in seconds.
}
\label{tab:bandwagon-unlearn-beauty-tiger-tracerv3}
\resizebox{\textwidth}{!}{%
\begin{tabular}{llrrrrr}
\toprule
Target & Method & $\mathrm{SH}@10$ (\%) & $\Delta^{\mathrm{retr}}_{\mathrm{SH}}$ (\%) $\downarrow$ & $\mathrm{NDCG}@10$ (\%) & $\tau_P$ (\%) & $\mathcal{E}$ (s) \\
\midrule
\multirow{12}{*}{\rotatebox[origin=c]{90}{Unpopular}}
 & \multicolumn{6}{l}{\textit{Baselines}} \\
 & Finetune & 1.250$\pm$0.936 & 1.250$\pm$0.936 & 3.700$\pm$0.072 & \underline{102.8$\pm$3.3} & \textbf{0.5} \\
 & Forget only & 0.000$\pm$0.000 & 0.000$\pm$0.000 & 0.020$\pm$0.034 & 0.5$\pm$0.9 & 0.4 \\
 & Forget + Repair & \underline{0.097$\pm$0.129} & \underline{0.097$\pm$0.129} & \textbf{3.701$\pm$0.077} & 102.8$\pm$3.2 & \underline{0.7} \\
 & SCIF & 1.729$\pm$1.660 & 1.729$\pm$1.660 & 3.649$\pm$0.056 & 101.4$\pm$3.0 & 39.9 \\
 & SEIF & 0.535$\pm$0.714 & 0.535$\pm$0.714 & 2.607$\pm$0.193 & 72.5$\pm$5.6 & 0.5 \\
 & Kookmin & 0.917$\pm$0.406 & 0.917$\pm$0.406 & 3.115$\pm$0.181 & 86.6$\pm$6.0 & 0.7 \\
 & Fanchuan & 0.000$\pm$0.000 & 0.000$\pm$0.000 & 1.243$\pm$0.255 & 34.6$\pm$7.7 & 1.8 \\
 & TRACER & 1.403$\pm$1.303 & 1.403$\pm$1.303 & 3.650$\pm$0.046 & 101.5$\pm$3.0 & 2.7 \\
\cmidrule(lr){2-7}
 & \multicolumn{6}{l}{\textit{Ours}} \\
 & NAU & \textbf{0.004$\pm$0.006} & \textbf{0.004$\pm$0.006} & \underline{3.701$\pm$0.070} & \textbf{102.9$\pm$3.4} & 4.1 \\
 & \quad NAU, $\lambda_s=0.1$ & 0.000$\pm$0.000 & 0.000$\pm$0.000 & 3.687$\pm$0.063 & 102.5$\pm$3.6 & 3.4 \\
\midrule
\multirow{12}{*}{\rotatebox[origin=c]{90}{Mid}}
 & \multicolumn{6}{l}{\textit{Baselines}} \\
 & Finetune & 0.636$\pm$0.597 & 0.635$\pm$0.596 & \textbf{3.572$\pm$0.107} & \textbf{99.3$\pm$2.7} & \textbf{0.6} \\
 & Forget only & 0.000$\pm$0.000 & 0.001$\pm$0.002 & 0.020$\pm$0.041 & 0.6$\pm$1.2 & 0.4 \\
 & Forget + Repair & \underline{0.046$\pm$0.090} & \underline{0.045$\pm$0.088} & \underline{3.563$\pm$0.116} & \underline{99.0$\pm$2.8} & \underline{0.7} \\
 & SCIF & 0.960$\pm$0.877 & 0.959$\pm$0.876 & 3.520$\pm$0.157 & 97.8$\pm$3.3 & 40.0 \\
 & SEIF & 0.111$\pm$0.126 & 0.110$\pm$0.125 & 2.516$\pm$0.156 & 70.0$\pm$5.4 & 0.9 \\
 & Kookmin & 0.706$\pm$1.076 & 0.705$\pm$1.076 & 3.104$\pm$0.147 & 86.2$\pm$4.0 & 0.6 \\
 & Fanchuan & 0.000$\pm$0.000 & 0.001$\pm$0.002 & 1.249$\pm$0.237 & 34.8$\pm$7.5 & 1.8 \\
 & TRACER & 0.683$\pm$0.657 & 0.682$\pm$0.655 & 3.505$\pm$0.158 & 97.4$\pm$3.3 & 4.4 \\
\cmidrule(lr){2-7}
 & \multicolumn{6}{l}{\textit{Ours}} \\
 & NAU & \textbf{0.021$\pm$0.048} & \textbf{0.021$\pm$0.046} & 3.558$\pm$0.112 & 98.9$\pm$2.8 & 4.1 \\
 & \quad NAU, $\lambda_s=0.1$ & 0.000$\pm$0.000 & 0.001$\pm$0.002 & 3.554$\pm$0.115 & 98.8$\pm$3.0 & 3.5 \\
\midrule
\multirow{12}{*}{\rotatebox[origin=c]{90}{Popular}}
 & \multicolumn{6}{l}{\textit{Baselines}} \\
 & Finetune & 0.855$\pm$0.705 & \underline{1.039$\pm$1.399} & \textbf{3.633$\pm$0.087} & \textbf{101.0$\pm$4.2} & \textbf{0.5} \\
 & Forget only & 0.000$\pm$0.000 & 1.269$\pm$2.020 & 0.021$\pm$0.035 & 0.6$\pm$1.0 & 0.4 \\
 & Forget + Repair & \underline{0.074$\pm$0.094} & 1.226$\pm$2.044 & 3.616$\pm$0.081 & 100.5$\pm$4.1 & \underline{0.8} \\
 & SCIF & 0.887$\pm$0.726 & \textbf{1.020$\pm$1.311} & 3.550$\pm$0.072 & 98.6$\pm$1.8 & 69.0 \\
 & SEIF & 0.470$\pm$0.603 & 1.333$\pm$1.890 & 2.547$\pm$0.142 & 70.8$\pm$4.0 & 0.6 \\
 & Kookmin & 1.261$\pm$1.460 & 0.905$\pm$0.934 & 3.047$\pm$0.085 & 84.7$\pm$3.5 & 0.6 \\
 & Fanchuan & 0.000$\pm$0.000 & 1.269$\pm$2.020 & 1.123$\pm$0.115 & 31.2$\pm$3.0 & 1.6 \\
 & TRACER & 0.807$\pm$0.653 & 1.103$\pm$1.461 & 3.544$\pm$0.074 & 98.5$\pm$2.1 & 2.7 \\
\cmidrule(lr){2-7}
 & \multicolumn{6}{l}{\textit{Ours}} \\
 & NAU & \textbf{0.000$\pm$0.000} & 1.269$\pm$2.020 & \underline{3.622$\pm$0.086} & \underline{100.7$\pm$4.4} & 4.0 \\
 & \quad NAU, $\lambda_s=0.1$ & 0.000$\pm$0.000 & 1.269$\pm$2.020 & 3.628$\pm$0.088 & 100.9$\pm$4.3 & 3.4 \\
\bottomrule
\end{tabular}}
\end{table*}

\section{Additional Ablation Results}
\label[appendix]{sec:appendix-ablation}

This section provides additional per-term ablations of the NAU objective on Sports and Toys.
\Cref{tab:scoped-ablation-compact-toys} is the Toys counterpart of \Cref{tab:scoped-ablation-compact-beauty}.
The per-term tables for Sports and Toys (\Cref{tab:ablation-unified-per-term-sports,tab:ablation-unified-per-term-toys}) use the earlier protocol (all strata, removal and utility on the whole test set only) with the stronger separation weight ($\lambda_s=0.1$); \Cref{tab:scoped-ablation-compact-toys} includes the reported setting.
Lower $\mathrm{SH}@10$ indicates stronger removal of promoted targets, while higher $\mathrm{NDCG}@10$ and $\tau_P$ indicate better utility preservation.

On the whole test set, the forget term carries most of the removal, separation drives it to zero or near zero, and $\lambda_n=0.01$ leaves utility unchanged while $\lambda_n=0.1$ costs 2 to 7 points of $\tau_P$.

\begin{table}[t]
\centering
\small
\setlength{\tabcolsep}{4pt}
\renewcommand{\arraystretch}{1.05}
\caption{
Ablation of NAU on Toys with TIGER + RQ-KMeans (mid-popularity target, five seeds).
The \ac{SID} neighborhood contains 563 test cases whose next item shares the target's first \ac{SID} token; the target's own test cases are excluded.
Indented rows are alternative additions to
$\mathcal{L}_{\mathrm{ret}}+\mathcal{L}_{\mathrm{for}}$
(siblings, not a ladder).
$\Delta^{\mathrm{retr}}_{\mathrm{SH}}$ is the absolute distance to the retrained model's exposure
(0.067\% on the whole test set and 2.664\% in the neighborhood).
$\tau_P$ denotes $\mathrm{NDCG}@10$ as a percentage of the retrained model's utility within each query set.
}
\label{tab:scoped-ablation-compact-toys}

\begin{tabular}{lcccc}
\toprule
\multicolumn{5}{c}{\textbf{Whole test set}} \\
\cmidrule(lr){1-5}
\textbf{Objective}
& $\mathrm{SH}@10{\scriptstyle\,(\%)}\downarrow$
& $\Delta^{\mathrm{retr}}_{\mathrm{SH}}{\scriptstyle\,(\%)}\downarrow$
& $\mathrm{NDCG}@10\uparrow$
& $\tau_P{\scriptstyle\,(\%)}\uparrow$ \\
\midrule

\textit{Retrained}
& \textit{0.067\,{\scriptsize$\pm$0.12}}
& \textit{0.000}
& \textit{3.461\,{\scriptsize$\pm$0.11}}
& \textit{100.0} \\

\textit{Poisoned}
& \textit{0.297\,{\scriptsize$\pm$0.23}}
& \textit{0.230}
& \textit{3.377\,{\scriptsize$\pm$0.06}}
& \textit{97.6} \\

\midrule

$\mathcal{L}_{\mathrm{ret}}$ only
& 0.656\,{\scriptsize$\pm$0.57}
& 0.589
& 3.348\,{\scriptsize$\pm$0.03}
& 96.7 \\

$+\ \mathcal{L}_{\mathrm{for}}$
& 0.159\,{\scriptsize$\pm$0.23}
& 0.092
& 3.345\,{\scriptsize$\pm$0.03}
& 96.6 \\

\quad $+\ \mathcal{L}_{\mathrm{sep}}$ ($\lambda_s=.01$)
& 0.113\,{\scriptsize$\pm$0.18}
& 0.046
& 3.352\,{\scriptsize$\pm$0.03}
& 96.8 \\

\quad $+\ \mathcal{L}_{\mathrm{sep}}$ ($\lambda_s=.1$)
& 0.018\,{\scriptsize$\pm$0.04}
& 0.049
& 3.340\,{\scriptsize$\pm$0.03}
& 96.5 \\

\quad $+\ \mathcal{L}_{\mathrm{nbr}}$ ($\lambda_n=.01$)
& 0.007\,{\scriptsize$\pm$0.01}
& 0.060
& 3.345\,{\scriptsize$\pm$0.04}
& 96.6 \\

\quad $+\ \mathcal{L}_{\mathrm{nbr}}$ ($\lambda_n=.1$)
& 0.000\,{\scriptsize$\pm$0.00}
& 0.067
& 3.231\,{\scriptsize$\pm$0.05}
& 93.3 \\

\midrule

NAU ($\lambda_s=.01,\lambda_n=.01$, ours)
& 0.001\,{\scriptsize$\pm$0.00}
& 0.066
& 3.361\,{\scriptsize$\pm$0.04}
& 97.1 \\

NAU ($\lambda_s=.1,\lambda_n=.01$)
& 0.000\,{\scriptsize$\pm$0.00}
& 0.067
& 3.358\,{\scriptsize$\pm$0.05}
& 97.0 \\

NAU ($\lambda_s=.1,\lambda_n=.1$)
& 0.000\,{\scriptsize$\pm$0.00}
& 0.067
& 3.236\,{\scriptsize$\pm$0.04}
& 93.5 \\

\addlinespace[5pt]
\midrule

\multicolumn{5}{c}{\textbf{SID neighborhood}} \\
\cmidrule(lr){1-5}
\textbf{Objective}
& $\mathrm{SH}@10{\scriptstyle\,(\%)}\downarrow$
& $\Delta^{\mathrm{retr}}_{\mathrm{SH}}{\scriptstyle\,(\%)}\downarrow$
& $\mathrm{NDCG}@10\uparrow$
& $\tau_P{\scriptstyle\,(\%)}\uparrow$ \\
\midrule

\textit{Retrained}
& \textit{2.664\,{\scriptsize$\pm$3.71}}
& \textit{0.000}
& \textit{5.206\,{\scriptsize$\pm$4.75}}
& \textit{100.0} \\

\textit{Poisoned}
& \textit{3.730\,{\scriptsize$\pm$4.47}}
& \textit{1.066}
& \textit{4.224\,{\scriptsize$\pm$4.34}}
& \textit{81.1} \\

\midrule

$\mathcal{L}_{\mathrm{ret}}$ only
& 8.526\,{\scriptsize$\pm$15.72}
& 5.861
& 4.907\,{\scriptsize$\pm$4.30}
& 94.3 \\

$+\ \mathcal{L}_{\mathrm{for}}$
& 4.440\,{\scriptsize$\pm$9.53}
& 1.776
& 4.416\,{\scriptsize$\pm$4.48}
& 84.8 \\

\quad $+\ \mathcal{L}_{\mathrm{sep}}$ ($\lambda_s=.01$)
& 2.664\,{\scriptsize$\pm$2.74}
& 0.000
& 4.270\,{\scriptsize$\pm$4.32}
& 82.0 \\

\quad $+\ \mathcal{L}_{\mathrm{sep}}$ ($\lambda_s=.1$)
& 0.710\,{\scriptsize$\pm$0.81}
& 1.954
& 3.893\,{\scriptsize$\pm$4.17}
& 74.8 \\

\quad $+\ \mathcal{L}_{\mathrm{nbr}}$ ($\lambda_n=.01$)
& 0.000\,{\scriptsize$\pm$0.00}
& 2.664
& 4.788\,{\scriptsize$\pm$4.14}
& 92.0 \\

\quad $+\ \mathcal{L}_{\mathrm{nbr}}$ ($\lambda_n=.1$)
& 0.000\,{\scriptsize$\pm$0.00}
& 2.664
& 6.182\,{\scriptsize$\pm$3.33}
& 118.8 \\

\midrule

NAU ($\lambda_s=.01,\lambda_n=.01$, ours)
& 0.000\,{\scriptsize$\pm$0.00}
& 2.664
& 4.879\,{\scriptsize$\pm$4.32}
& 93.7 \\

NAU ($\lambda_s=.1,\lambda_n=.01$)
& 0.000\,{\scriptsize$\pm$0.00}
& 2.664
& 4.912\,{\scriptsize$\pm$4.19}
& 94.4 \\

NAU ($\lambda_s=.1,\lambda_n=.1$)
& 0.000\,{\scriptsize$\pm$0.00}
& 2.664
& 5.923\,{\scriptsize$\pm$3.19}
& 113.8 \\

\bottomrule
\end{tabular}
\end{table}

\header{Probability-mass ablation}
\Cref{tab:ablation-compact-beauty} repeats the Beauty ablation over all three popularity strata and five seeds and reports the local effect in log probability mass instead of neighborhood utility; beyond prefix-disjoint controls, the neighborhood's mass relative to retraining grows with $\lambda_n$ ($+0.04$ without the term, $+0.18$ for NAU, $+1.55$ at $\lambda_n=0.1$), ($+0.04$ without the term, $+0.33$ for NAU, $+1.55$ at $\lambda_n=0.1$), and the separation term pushes the target below its retrained mass (about 4 nats for NAU, about 18.5 with the stronger $\lambda_s=0.1$), which is also why NAU can remove popular targets that the retrained model still recommends (\Cref{sec:appendix-spam}).

\begin{table}[t]
\centering
\small
\setlength{\tabcolsep}{4pt}
\caption{
Ablation of NAU on Beauty (TIGER + RQ-KMeans, five seeds $\times$ three targets). Indented rows each add one term to $\mathcal{L}_{\mathrm{ret}}+\mathcal{L}_{\mathrm{for}}$ (siblings, not a ladder).
Local effect: signed change in log probability mass relative to the retrained model of the same seed, on held-out test histories, for the forgotten target ($\Delta_{\mathrm{tgt}}$) and its \ac{SID}-prefix neighbors beyond prefix-disjoint controls ($\Delta_{\mathrm{nbr}}-\Delta_{\mathrm{ctrl}}$); mean $\pm$ s.d. across seeds.
\textsuperscript{\dag}\,Local effect measured on a matched configuration rather than the ablation's own runs.
$\mathcal{L}_{\mathrm{for}}$ does most of the removal, $\mathcal{L}_{\mathrm{nbr}}$ moves mass into the neighborhood in proportion to $\lambda_n$, and a strong separation term ($\lambda_s=0.1$) pushes the target far below its retrained mass.
}
\label{tab:ablation-compact-beauty}
\resizebox{\columnwidth}{!}{%
\begin{tabular}{lcccccc}
\toprule
\multirow{2}{*}[-3pt]{\textbf{Objective}}
& \multicolumn{4}{c}{\textbf{Removal and utility}} & \multicolumn{2}{c}{\textbf{Local effect} (nats)} \\
\cmidrule(lr){2-5} \cmidrule(lr){6-7}
 & $\mathrm{SH}@10 {\scriptstyle\,(\%)}$ & $\Delta^{\mathrm{retr}}_{\mathrm{SH}}{\scriptstyle\,(\%)}\downarrow$ & $\mathrm{NDCG}@10\uparrow$ & $\tau_P{\scriptstyle\,(\%)}\uparrow$ & $\Delta_{\mathrm{tgt}}$ & $\Delta_{\mathrm{nbr}}-\Delta_{\mathrm{ctrl}}$ \\
\midrule
$\mathcal{L}_{\mathrm{ret}}$ only & 0.986\,{\scriptsize$\pm$0.947} & 0.914 & 3.630\,{\scriptsize$\pm$0.109} & 96.5\,{\scriptsize$\pm$3.7} & $+5.07$\,{\scriptsize$\pm$1.69} & $-0.14$\,{\scriptsize$\pm$0.21} \\
$+\ \mathcal{L}_{\mathrm{for}}$ & 0.110\,{\scriptsize$\pm$0.211} & 0.108 & 3.622\,{\scriptsize$\pm$0.118} & 96.3\,{\scriptsize$\pm$4.0} & $+1.67$\,{\scriptsize$\pm$4.17} & $+0.04$\,{\scriptsize$\pm$0.31} \\
\quad $+\ \mathcal{L}_{\mathrm{sep}}$ ($\lambda_s=0.1$) & 0.000\,{\scriptsize$\pm$0.000} & 0.072 & 3.615\,{\scriptsize$\pm$0.105} & 96.1\,{\scriptsize$\pm$3.9} & $-19.04$\,{\scriptsize$\pm$5.57} & $-0.11$\,{\scriptsize$\pm$0.17} \\
\quad $+\ \mathcal{L}_{\mathrm{nbr}}$ ($\lambda_n=0.01$) & 0.038\,{\scriptsize$\pm$0.118} & 0.103 & 3.615\,{\scriptsize$\pm$0.111} & 96.1\,{\scriptsize$\pm$3.8} & $-1.74$\,{\scriptsize$\pm$3.23}\textsuperscript{\dag} & $+0.36$\,{\scriptsize$\pm$0.37}\textsuperscript{\dag} \\
\midrule
NAU ($\lambda_s=0.01$, $\lambda_n=0.01$) & 0.010\,{\scriptsize$\pm$0.034} & 0.081 & 3.621\,{\scriptsize$\pm$0.107} & 96.2\,{\scriptsize$\pm$3.8} & $-4.32$\,{\scriptsize$\pm$3.39} & $+0.33$\,{\scriptsize$\pm$0.34} \\
\quad NAU at $\lambda_s=0.1$ & 0.000\,{\scriptsize$\pm$0.000} & 0.072 & 3.617\,{\scriptsize$\pm$0.108} & 96.1\,{\scriptsize$\pm$3.9} & $-18.53$\,{\scriptsize$\pm$4.08} & $+0.18$\,{\scriptsize$\pm$0.30} \\
\quad NAU at $\lambda_s=0.1$, $\lambda_n=0.1$ (over-strong dose) & 0.000\,{\scriptsize$\pm$0.000} & 0.072 & 3.519\,{\scriptsize$\pm$0.107} & 93.5\,{\scriptsize$\pm$3.5} & $-18.44$\,{\scriptsize$\pm$9.07} & $+1.55$\,{\scriptsize$\pm$0.63} \\
\bottomrule
\end{tabular}}
\end{table}

\begin{table*}[t]
\centering
\small
\setlength{\tabcolsep}{4pt}
\renewcommand{\arraystretch}{1.05}
\caption{Per-term ablation of the NAU objective on Sports with TIGER + RQ-KMeans (bandwagon, 1\% poison, $n_{\text{target}}=1$), at $\lambda_f=0.1$, $\lambda_s=0.1$, $\lambda_n=0.01$ and $\bm{w}_r=\bm{w}_f=\bm{1}$, with the generative separation score $\phi_{\mathrm{gen}}$ and embedding/mass neighbours in every row that activates the corresponding term, so only the WEIGHTS vary. The one exception is the over-strong dose row, which is only available with the cosine score and is marked accordingly. Each data seed is scored against its own clean and poisoned reference, with unlearning replicates averaged within a seed and the spread taken across seeds; $\pm$ is therefore absent wherever the column rests on a single data seed. $\mathrm{SH}@10$, $\mathrm{NDCG}@10$, $\tau_P$ and $\Delta\tau_P$ are reported as percentages, matching the main results tables. $\Delta\tau_P$ is paired against the $+\,\mathcal{L}_{\mathrm{forget}}$ row on (stratum, seed), which removes the seed intercept.}
\label{tab:ablation-unified-per-term-sports}
\resizebox{\textwidth}{!}{%
\begin{tabular}{llrrrr}
\toprule
Target & Loss terms & $\mathrm{SH}@10$ (\%) & $\mathrm{NDCG}@10$ (\%) & $\tau_P$ (\%) & $\Delta\tau_P$ (\%) \\
\midrule
\multirow{6}{*}{\rotatebox[origin=c]{90}{Unpopular}}
 & $\mathcal{L}_{\mathrm{retain}}$ only & 0.184$\pm$0.231 & 2.082$\pm$0.024 & 102.5$\pm$3.8 & +0.85$\pm$0.50 \\
 & $+\ \mathcal{L}_{\mathrm{forget}}$ & 0.000$\pm$0.000 & 2.065$\pm$0.029 & 101.7$\pm$4.0 & -- \\
 & \quad $+\ \mathcal{L}_{\mathrm{sep}}$ ($\lambda_s=0.1$) & 0.000$\pm$0.000 & 2.059$\pm$0.026 & 101.4$\pm$3.5 & -0.29$\pm$0.61 \\
 & \quad $+\ \mathcal{L}_{\mathrm{neighborhood}}$ ($\lambda_n=0.01$) & 0.000$\pm$0.000 & 2.069$\pm$0.041 & 101.9$\pm$4.0 & +0.18$\pm$1.14 \\
 & \quad $+\ \mathcal{L}_{\mathrm{sep}}+\mathcal{L}_{\mathrm{neighborhood}}$ $=$ NAU ($\lambda_s=0.1$, $\lambda_n=0.01$) & 0.000$\pm$0.000 & 2.063$\pm$0.031 & 101.6$\pm$3.3 & -0.11$\pm$0.83 \\
 & \quad \quad NAU at $\lambda_n=0.1$ (over-strong dose, cosine sep) & 0.000$\pm$0.000 & 2.007$\pm$0.049 & 98.8$\pm$3.7 & -2.88$\pm$1.75 \\
\midrule
\multirow{6}{*}{\rotatebox[origin=c]{90}{Mid}}
 & $\mathcal{L}_{\mathrm{retain}}$ only & 0.353$\pm$0.444 & 2.094$\pm$0.079 & 103.0$\pm$2.2 & +0.33$\pm$1.43 \\
 & $+\ \mathcal{L}_{\mathrm{forget}}$ & 0.004$\pm$0.007 & 2.087$\pm$0.070 & 102.7$\pm$3.0 & -- \\
 & \quad $+\ \mathcal{L}_{\mathrm{sep}}$ ($\lambda_s=0.1$) & 0.000$\pm$0.000 & 2.086$\pm$0.071 & 102.7$\pm$2.7 & -0.05$\pm$0.52 \\
 & \quad $+\ \mathcal{L}_{\mathrm{neighborhood}}$ ($\lambda_n=0.01$) & 0.001$\pm$0.001 & 2.081$\pm$0.085 & 102.4$\pm$3.4 & -0.32$\pm$0.86 \\
 & \quad $+\ \mathcal{L}_{\mathrm{sep}}+\mathcal{L}_{\mathrm{neighborhood}}$ $=$ NAU ($\lambda_s=0.1$, $\lambda_n=0.01$) & 0.000$\pm$0.000 & 2.084$\pm$0.073 & 102.5$\pm$3.4 & -0.16$\pm$0.61 \\
 & \quad \quad NAU at $\lambda_n=0.1$ (over-strong dose, cosine sep) & 0.000$\pm$0.000 & 2.001$\pm$0.103 & 98.5$\pm$5.6 & -4.22$\pm$3.02 \\
\midrule
\multirow{6}{*}{\rotatebox[origin=c]{90}{Popular}}
 & $\mathcal{L}_{\mathrm{retain}}$ only & 0.642$\pm$0.698 & 2.070$\pm$0.060 & 102.0$\pm$5.6 & +0.84$\pm$0.75 \\
 & $+\ \mathcal{L}_{\mathrm{forget}}$ & 0.017$\pm$0.014 & 2.053$\pm$0.049 & 101.1$\pm$5.2 & -- \\
 & \quad $+\ \mathcal{L}_{\mathrm{sep}}$ ($\lambda_s=0.1$) & 0.000$\pm$0.000 & 2.048$\pm$0.052 & 100.9$\pm$5.4 & -0.24$\pm$0.49 \\
 & \quad $+\ \mathcal{L}_{\mathrm{neighborhood}}$ ($\lambda_n=0.01$) & 0.068$\pm$0.118 & 2.061$\pm$0.040 & 101.5$\pm$4.6 & +0.37$\pm$0.66 \\
 & \quad $+\ \mathcal{L}_{\mathrm{sep}}+\mathcal{L}_{\mathrm{neighborhood}}$ $=$ NAU ($\lambda_s=0.1$, $\lambda_n=0.01$) & 0.000$\pm$0.000 & 2.054$\pm$0.057 & 101.2$\pm$5.5 & +0.06$\pm$0.77 \\
 & \quad \quad NAU at $\lambda_n=0.1$ (over-strong dose, cosine sep) & 0.000$\pm$0.000 & 1.921$\pm$0.115 & 94.6$\pm$7.3 & -6.50$\pm$3.76 \\
\bottomrule
\end{tabular}}
\end{table*}

\begin{table*}[t]
\centering
\small
\setlength{\tabcolsep}{4pt}
\renewcommand{\arraystretch}{1.05}
\caption{Per-term ablation of the NAU objective on Toys with TIGER + RQ-KMeans (bandwagon, 1\% poison, $n_{\text{target}}=1$), at $\lambda_f=0.1$, $\lambda_s=0.1$, $\lambda_n=0.01$ and $\bm{w}_r=\bm{w}_f=\bm{1}$, with the generative separation score $\phi_{\mathrm{gen}}$ and embedding/mass neighbours in every row that activates the corresponding term, so only the WEIGHTS vary. The one exception is the over-strong dose row, which is only available with the cosine score and is marked accordingly. Each data seed is scored against its own clean and poisoned reference, with unlearning replicates averaged within a seed and the spread taken across seeds; $\pm$ is therefore absent wherever the column rests on a single data seed. $\mathrm{SH}@10$, $\mathrm{NDCG}@10$, $\tau_P$ and $\Delta\tau_P$ are reported as percentages, matching the main results tables. $\Delta\tau_P$ is paired against the $+\,\mathcal{L}_{\mathrm{forget}}$ row on (stratum, seed), which removes the seed intercept.}
\label{tab:ablation-unified-per-term-toys}
\resizebox{\textwidth}{!}{%
\begin{tabular}{llrrrr}
\toprule
Target & Loss terms & $\mathrm{SH}@10$ (\%) & $\mathrm{NDCG}@10$ (\%) & $\tau_P$ (\%) & $\Delta\tau_P$ (\%) \\
\midrule
\multirow{6}{*}{\rotatebox[origin=c]{90}{Unpopular}}
 & $\mathcal{L}_{\mathrm{retain}}$ only & 1.092$\pm$1.290 & 3.342$\pm$0.058 & 96.5$\pm$3.8 & -0.35$\pm$0.41 \\
 & $+\ \mathcal{L}_{\mathrm{forget}}$ & 0.240$\pm$0.255 & 3.354$\pm$0.070 & 96.9$\pm$4.0 & -- \\
 & \quad $+\ \mathcal{L}_{\mathrm{sep}}$ ($\lambda_s=0.1$) & 0.000$\pm$0.000 & 3.337$\pm$0.056 & 96.4$\pm$3.2 & -0.50$\pm$0.99 \\
 & \quad $+\ \mathcal{L}_{\mathrm{neighborhood}}$ ($\lambda_n=0.01$) & 0.013$\pm$0.030 & 3.349$\pm$0.065 & 96.7$\pm$3.5 & -0.16$\pm$0.45 \\
 & \quad $+\ \mathcal{L}_{\mathrm{sep}}+\mathcal{L}_{\mathrm{neighborhood}}$ $=$ NAU ($\lambda_s=0.1$, $\lambda_n=0.01$) & 0.000$\pm$0.000 & 3.347$\pm$0.080 & 96.7$\pm$4.6 & -0.18$\pm$1.02 \\
 & \quad \quad NAU at $\lambda_n=0.1$ (over-strong dose, cosine sep) & 0.000$\pm$0.000 & 3.282$\pm$0.053 & 94.8$\pm$3.5 & -2.10$\pm$1.29 \\
\midrule
\multirow{6}{*}{\rotatebox[origin=c]{90}{Mid}}
 & $\mathcal{L}_{\mathrm{retain}}$ only & 0.660$\pm$0.572 & 3.344$\pm$0.038 & 96.6$\pm$3.6 & +0.04$\pm$1.28 \\
 & $+\ \mathcal{L}_{\mathrm{forget}}$ & 0.172$\pm$0.236 & 3.343$\pm$0.019 & 96.5$\pm$2.8 & -- \\
 & \quad $+\ \mathcal{L}_{\mathrm{sep}}$ ($\lambda_s=0.1$) & 0.016$\pm$0.037 & 3.362$\pm$0.041 & 97.1$\pm$4.0 & +0.58$\pm$1.45 \\
 & \quad $+\ \mathcal{L}_{\mathrm{neighborhood}}$ ($\lambda_n=0.01$) & 0.006$\pm$0.011 & 3.330$\pm$0.028 & 96.2$\pm$3.1 & -0.36$\pm$0.54 \\
 & \quad $+\ \mathcal{L}_{\mathrm{sep}}+\mathcal{L}_{\mathrm{neighborhood}}$ $=$ NAU ($\lambda_s=0.1$, $\lambda_n=0.01$) & 0.000$\pm$0.000 & 3.355$\pm$0.035 & 96.9$\pm$3.5 & +0.36$\pm$1.05 \\
 & \quad \quad NAU at $\lambda_n=0.1$ (over-strong dose, cosine sep) & 0.000$\pm$0.000 & 3.250$\pm$0.036 & 93.9$\pm$3.5 & -2.68$\pm$1.30 \\
\midrule
\multirow{6}{*}{\rotatebox[origin=c]{90}{Popular}}
 & $\mathcal{L}_{\mathrm{retain}}$ only & 0.848$\pm$0.956 & 3.365$\pm$0.113 & 97.2$\pm$5.1 & -0.12$\pm$0.81 \\
 & $+\ \mathcal{L}_{\mathrm{forget}}$ & 0.243$\pm$0.392 & 3.369$\pm$0.086 & 97.3$\pm$4.6 & -- \\
 & \quad $+\ \mathcal{L}_{\mathrm{sep}}$ ($\lambda_s=0.1$) & 0.007$\pm$0.010 & 3.326$\pm$0.078 & 96.1$\pm$4.5 & -1.23$\pm$0.89 \\
 & \quad $+\ \mathcal{L}_{\mathrm{neighborhood}}$ ($\lambda_n=0.01$) & 0.159$\pm$0.341 & 3.333$\pm$0.102 & 96.3$\pm$5.0 & -1.03$\pm$0.60 \\
 & \quad $+\ \mathcal{L}_{\mathrm{sep}}+\mathcal{L}_{\mathrm{neighborhood}}$ $=$ NAU ($\lambda_s=0.1$, $\lambda_n=0.01$) & 0.005$\pm$0.012 & 3.348$\pm$0.098 & 96.7$\pm$4.7 & -0.59$\pm$0.83 \\
 & \quad \quad NAU at $\lambda_n=0.1$ (over-strong dose, cosine sep) & 0.006$\pm$0.014 & 3.256$\pm$0.087 & 94.1$\pm$4.5 & -3.24$\pm$1.01 \\
\bottomrule
\end{tabular}}
\end{table*}

\begin{figure}[t]
    \centering
    \includegraphics[width=\textwidth]{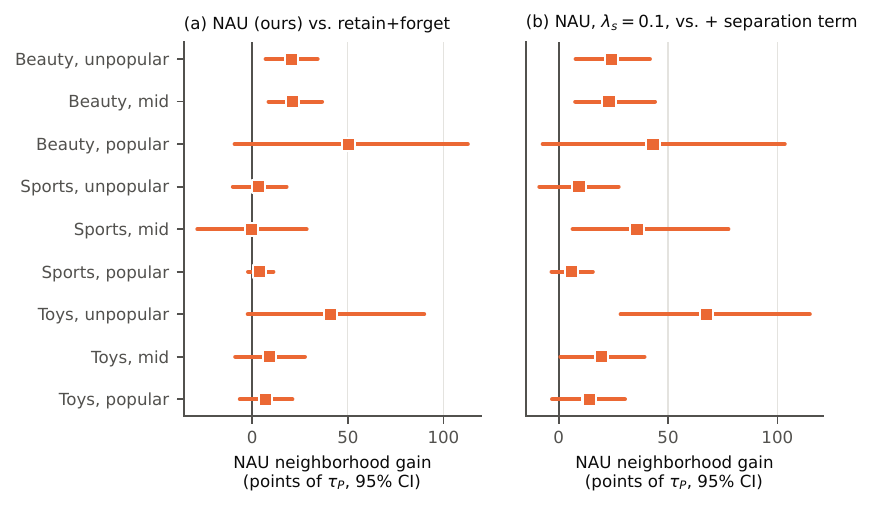}
    \caption{Neighborhood utility gained on each dataset and target popularity (TIGER + RQ-KMeans, five seeds per setting), in points of the retrained model's neighborhood utility: (a) NAU over fine-tuning on the retain and forget sets; (b) NAU with the stronger separation weight ($\lambda_s=0.1$) over fine-tuning on the retain and forget sets with the separation term ($\lambda_s=0.1$). Bars are 95\% paired bootstrap intervals over the shared test cases (\Cref{sec:appendix-bootstrap}); the neighborhood excludes the target's own test cases.}
    \label{fig:scope-grid}
\end{figure}

\header{Neighborhood effect across datasets and targets}
\Cref{fig:scope-grid} repeats the comparison of \Cref{fig:scope-locality} for all three datasets and target popularities. NAU's neighborhood gain over Forget+Repair is positive in eight of nine settings and resolved for the unpopular and mid-popularity Beauty targets; with the stronger separation weight ($\lambda_s=0.1$), the neighborhood term's gain over the separation term alone is positive in all nine settings and resolved in five.

\section{Additional Unwanted-Item Results}
\label[appendix]{sec:appendix-unwanted}

This section reports full unwanted-item deletion results against existing baselines.
They complement \Cref{sec:results-unwanted}, where \Cref{tab:paper-sensitive-beauty-tiger-rkmeans} compares methods on Beauty and \Cref{tab:ablation-unified-sensitive-lambda-n-all} isolates the neighborhood term on all three datasets.
Each method is at the configuration selected on the validation split (\Cref{tab:appendix-valsens}).
$\mathrm{UHF}@10$ measures unwanted-item exposure for users who requested deletion (lower is better), while $\mathrm{UHR}@10$ measures it for the remaining users, and $\Delta^{\mathrm{retr}}_{\mathrm{UHR}}$ how far the update moves them from retraining.
On Beauty and Toys the retrained model does not lower $\mathrm{UHF}@10$ below the original model, since only 12 to 14 users request deletion, so we report $\mathrm{UHF}@10$ itself rather than its distance to retraining.

\begin{table}[t]
\centering
\small
\setlength{\tabcolsep}{4pt}
\renewcommand{\arraystretch}{1.05}
\caption{
Unwanted item removal on Sports (Guns \& Rifles) with TIGER + RQ-KMeans, five data seeds.
$\mathrm{UHF}@10$ and $\mathrm{UHR}@10$ are the unwanted item hit rates of the requesting and the remaining users; $\Delta^{\mathrm{retr}}_{\mathrm{UHR}}$ is the mean over seeds of $|\mathrm{UHR}_U@10-\mathrm{UHR}_R@10|$, the distance to the retrained model of the same seed.
Each method uses the configuration selected on the validation split (\Cref{tab:appendix-valsens}; $^{\dagger}$: $\tau_P<95\%$, never ranked); NAU uses $\lambda_n=-1$, $\lambda_s=0$.
}
\label{tab:sensitive-sports-tiger-rkmeans}
\resizebox{\textwidth}{!}{%
\begin{tabular}{lcccccc}
\toprule
\textbf{Method} & $\mathrm{UHF}@10{\scriptstyle\,(\%)}\downarrow$ & $\mathrm{UHR}@10{\scriptstyle\,(\%)}$ & $\Delta^{\mathrm{retr}}_{\mathrm{UHR}}{\scriptstyle\,(\%)}\downarrow$ & $\mathrm{NDCG}@10\uparrow$ & $\tau_P$ (\%) & $\mathcal{E}{\scriptstyle\,(s)}\downarrow$ \\
\midrule
Finetune$^{\dagger}$ & 29.651\,{\scriptsize$\pm$14.035} & 1.602\,{\scriptsize$\pm$0.252} & 0.180\,{\scriptsize$\pm$0.126} & 1.894\,{\scriptsize$\pm$0.066} & 93.0\,{\scriptsize$\pm$6.4} & 0.3 \\
Forget only$^{\dagger}$ & 16.606\,{\scriptsize$\pm$12.736} & 0.694\,{\scriptsize$\pm$0.327} & 0.727\,{\scriptsize$\pm$0.358} & 1.915\,{\scriptsize$\pm$0.094} & 94.1\,{\scriptsize$\pm$8.6} & 0.3 \\
Forget+Repair$^{\dagger}$ & 30.718\,{\scriptsize$\pm$10.996} & 1.475\,{\scriptsize$\pm$0.338} & 0.221\,{\scriptsize$\pm$0.121} & 1.899\,{\scriptsize$\pm$0.071} & 93.2\,{\scriptsize$\pm$6.7} & 0.5 \\
SCIF & \textbf{32.870\,{\scriptsize$\pm$16.770}} & 1.707\,{\scriptsize$\pm$0.385} & \underline{0.334\,{\scriptsize$\pm$0.231}} & \textbf{2.045\,{\scriptsize$\pm$0.086}} & \textbf{100.5\,{\scriptsize$\pm$9.4}} & 34.5 \\
SEIF$^{\dagger}$ & 0.000\,{\scriptsize$\pm$0.000} & 0.000\,{\scriptsize$\pm$0.000} & 1.422\,{\scriptsize$\pm$0.197} & 0.003\,{\scriptsize$\pm$0.005} & 0.1\,{\scriptsize$\pm$0.2} & 0.3 \\
Kookmin$^{\dagger}$ & 31.332\,{\scriptsize$\pm$14.574} & 1.408\,{\scriptsize$\pm$0.295} & 0.126\,{\scriptsize$\pm$0.089} & 1.374\,{\scriptsize$\pm$0.094} & 67.4\,{\scriptsize$\pm$5.4} & 0.3 \\
Fanchuan$^{\dagger}$ & 0.000\,{\scriptsize$\pm$0.000} & 0.001\,{\scriptsize$\pm$0.003} & 1.420\,{\scriptsize$\pm$0.198} & 0.165\,{\scriptsize$\pm$0.085} & 8.1\,{\scriptsize$\pm$4.4} & 1.1 \\
TRACER & \textbf{32.870\,{\scriptsize$\pm$16.770}} & 1.662\,{\scriptsize$\pm$0.396} & \textbf{0.310\,{\scriptsize$\pm$0.226}} & \underline{2.043\,{\scriptsize$\pm$0.093}} & \underline{100.5\,{\scriptsize$\pm$9.7}} & 4.2 \\
\cmidrule(lr){1-7}
NAU \textit{(Ours)}$^{\dagger}$ & 11.144\,{\scriptsize$\pm$3.881} & 0.543\,{\scriptsize$\pm$0.132} & 0.878\,{\scriptsize$\pm$0.241} & 1.880\,{\scriptsize$\pm$0.086} & 92.4\,{\scriptsize$\pm$8.0} & 2.6 \\
\midrule
\textit{Original (before deletion)} & 49.838\,{\scriptsize$\pm$17.318} & 2.234\,{\scriptsize$\pm$0.378} & 0.812\,{\scriptsize$\pm$0.362} & 2.033\,{\scriptsize$\pm$0.071} & 99.8\,{\scriptsize$\pm$6.9} & -- \\
\textit{Retrain (reference)} & 29.289\,{\scriptsize$\pm$9.368} & 1.422\,{\scriptsize$\pm$0.197} & 0.000 & 2.043\,{\scriptsize$\pm$0.111} & 100.0 & \textit{20,808} \\
\bottomrule
\end{tabular}}
\end{table}

\begin{table}[t]
\centering
\small
\setlength{\tabcolsep}{4pt}
\renewcommand{\arraystretch}{1.05}
\caption{
Unwanted item removal on Toys (toy weapons) with TIGER + RQ-KMeans, five data seeds.
$\mathrm{UHF}@10$ and $\mathrm{UHR}@10$ are the unwanted item hit rates of the requesting and the remaining users; $\Delta^{\mathrm{retr}}_{\mathrm{UHR}}$ is the mean over seeds of $|\mathrm{UHR}_U@10-\mathrm{UHR}_R@10|$, the distance to the retrained model of the same seed.
Each method uses the configuration selected on the validation split (\Cref{tab:appendix-valsens}; $^{\dagger}$: $\tau_P<95\%$, never ranked); NAU uses $\lambda_n=-1$, $\lambda_s=0$.
Retraining does not lower $\mathrm{UHF}@10$ below the original model.
}
\label{tab:sensitive-toys-tiger-rkmeans}
\resizebox{\textwidth}{!}{%
\begin{tabular}{lcccccc}
\toprule
\textbf{Method} & $\mathrm{UHF}@10{\scriptstyle\,(\%)}\downarrow$ & $\mathrm{UHR}@10{\scriptstyle\,(\%)}$ & $\Delta^{\mathrm{retr}}_{\mathrm{UHR}}{\scriptstyle\,(\%)}\downarrow$ & $\mathrm{NDCG}@10\uparrow$ & $\tau_P$ (\%) & $\mathcal{E}{\scriptstyle\,(s)}\downarrow$ \\
\midrule
Finetune & \textbf{6.538\,{\scriptsize$\pm$6.940}} & 2.854\,{\scriptsize$\pm$0.387} & 0.767\,{\scriptsize$\pm$0.463} & 3.118\,{\scriptsize$\pm$0.077} & 95.1\,{\scriptsize$\pm$2.9} & 0.3 \\
Forget only$^{\dagger}$ & 4.872\,{\scriptsize$\pm$4.455} & 1.879\,{\scriptsize$\pm$0.513} & 1.634\,{\scriptsize$\pm$0.404} & 2.945\,{\scriptsize$\pm$0.113} & 89.9\,{\scriptsize$\pm$5.3} & 0.3 \\
Forget+Repair$^{\dagger}$ & 6.667\,{\scriptsize$\pm$9.129} & 2.363\,{\scriptsize$\pm$0.289} & 1.149\,{\scriptsize$\pm$0.374} & 3.067\,{\scriptsize$\pm$0.074} & 93.6\,{\scriptsize$\pm$3.4} & 0.5 \\
SCIF & 11.154\,{\scriptsize$\pm$7.045} & 2.911\,{\scriptsize$\pm$0.195} & \textbf{0.602\,{\scriptsize$\pm$0.340}} & \underline{3.392\,{\scriptsize$\pm$0.070}} & \underline{103.5\,{\scriptsize$\pm$3.7}} & 32.9 \\
SEIF$^{\dagger}$ & 0.000\,{\scriptsize$\pm$0.000} & 0.021\,{\scriptsize$\pm$0.046} & 3.492\,{\scriptsize$\pm$0.266} & 0.031\,{\scriptsize$\pm$0.029} & 1.0\,{\scriptsize$\pm$0.9} & 0.3 \\
Kookmin$^{\dagger}$ & 8.205\,{\scriptsize$\pm$8.338} & 2.809\,{\scriptsize$\pm$0.612} & 0.774\,{\scriptsize$\pm$0.742} & 2.500\,{\scriptsize$\pm$0.161} & 76.3\,{\scriptsize$\pm$6.2} & 0.3 \\
Fanchuan$^{\dagger}$ & 0.000\,{\scriptsize$\pm$0.000} & 0.000\,{\scriptsize$\pm$0.000} & 3.513\,{\scriptsize$\pm$0.305} & 0.182\,{\scriptsize$\pm$0.034} & 5.5\,{\scriptsize$\pm$1.0} & 1.1 \\
TRACER & 11.154\,{\scriptsize$\pm$8.901} & 2.937\,{\scriptsize$\pm$0.312} & \underline{0.663\,{\scriptsize$\pm$0.281}} & \textbf{3.416\,{\scriptsize$\pm$0.086}} & \textbf{104.2\,{\scriptsize$\pm$3.9}} & 4.0 \\
\cmidrule(lr){1-7}
NAU \textit{(Ours)} & \underline{9.744\,{\scriptsize$\pm$7.057}} & 2.380\,{\scriptsize$\pm$0.278} & 1.133\,{\scriptsize$\pm$0.253} & 3.119\,{\scriptsize$\pm$0.133} & 95.1\,{\scriptsize$\pm$5.3} & 2.4 \\
\midrule
\textit{Original (before deletion)} & 16.154\,{\scriptsize$\pm$12.999} & 2.947\,{\scriptsize$\pm$0.328} & 0.613\,{\scriptsize$\pm$0.488} & 3.465\,{\scriptsize$\pm$0.110} & 105.7\,{\scriptsize$\pm$4.5} & -- \\
\textit{Retrain (reference)} & 18.124\,{\scriptsize$\pm$9.249} & 3.513\,{\scriptsize$\pm$0.305} & 0.000 & 3.280\,{\scriptsize$\pm$0.095} & 100.0 & \textit{20,536} \\
\bottomrule
\end{tabular}}
\end{table}

\section{Intervals for Neighborhood Comparisons}
\label[appendix]{sec:appendix-bootstrap}

Neighborhood utility is estimated from few test cases: on Beauty, 523 test cases have a next item in the forgotten target's \ac{SID} neighborhood, and only about 30 of them have that item in the retrained model's top 10. A handful of test users can therefore move a neighborhood result by several points, and a difference between two methods is meaningful only if it survives that variation.

We measure this variation with a paired bootstrap. For every neighborhood test case $q$ and a method $A$ compared against a reference method $B$, we compute the difference $d_q = \mathrm{NDCG}@10_A(q) - \mathrm{NDCG}@10_B(q)$. Both methods are scored on the same test cases, which is what makes the comparison paired: differences in how easy a test case is cancel out. The reported gain is the mean of $d_q$ over all cases, expressed in percentage points of the retrained model's neighborhood NDCG@10. To see how much this mean would change with a different set of test users, we draw as many cases as we have, at random and with replacement, recompute the mean on this resampled set, and repeat 3000 times. The 95\% interval spans the middle 95\% of these 3000 means, i.e., from the 2.5th to the 97.5th percentile.

An interval that lies entirely above 0 means that method $A$ gains neighborhood utility over $B$ no matter which of the plausible test sets we happen to have; an interval entirely below 0 means a reliable loss; an interval that contains 0 means that the data cannot tell the two methods apart. The interval reflects the sampling of test cases for the trained and unlearned models at hand, pooled over the five data seeds; it does not resample the seeds themselves. As a complementary check, the sign of the NAU gain over fine-tuning on the retain and forget sets holds on every seed that has more than three neighborhood test cases (4 of 5). Comparisons against the retrained model itself are not resolved at this sample size (intervals of about $\pm 40$ points), because the retrained model is a separately trained network whose rankings differ from the unlearned models' on most neighborhood test cases; we therefore only compare unlearning methods with each other.

\section{SID and Neighborhood Diagnostics}
\label[appendix]{sec:appendix-diagnostics}

This section provides additional diagnostics for the \ac{SID} structure and its role in unlearning.
\Cref{tab:neighborhood_statistics_static} analyzes how many items share \ac{SID} prefixes at different levels, while \Cref{tab:ablation-unified-positions-beauty} studies how restricting updates to different \ac{SID} levels affects forgetting and utility.

\header{SID prefix-sharing statistics}
\Cref{tab:neighborhood_statistics_static} reports neighborhood statistics across \ac{SID} levels and tokenizers.
The results show that early \ac{SID} levels induce broad neighborhoods, while later levels often become highly sparse or singleton-dominated.
This supports our choice to treat neighborhood construction as a configurable component rather than assuming that every \ac{SID} level provides a meaningful neighborhood.

\begin{table*}[t]
\centering
\caption{
Neighborhood statistics across \ac{SID} levels.
Early levels define broader prefix-sharing neighborhoods, while later levels become increasingly sparse or singleton-dominated.
A high singleton share indicates that the corresponding level is less suitable for prefix-based neighborhood construction.
}
\label{tab:neighborhood_statistics_static}
\resizebox{\textwidth}{!}{
\begin{tabular}{llrrrrrrr}
\toprule
Dataset & Tokenizer & Level & Distinct prefixes & Unique (\%) & Mean nbrs. & Median & P90 & Singletons (\%) \\
\midrule
Beauty & RKMeans & 0 & 256    & 2.1  & 82.90   & 74   & 141  & 0.2 \\
       &         & 1 & 6,150  & 50.8 & 3.16    & 1    & 8    & 30.6 \\
       &         & 2 & 10,483 & 86.6 & 0.50    & 0    & 2    & 78.3 \\
       &         & 3 (dedup) & 12,101 & 100.0 & 0.00 & 0 & 0 & 100.0 \\
       & RQVAE   & 0 & 103    & 0.9  & 171.90  & 151  & 297  & 0.0 \\
       &         & 1 & 6,370  & 52.6 & 3.94    & 1    & 10   & 32.7 \\
       &         & 2 & 11,498 & 95.0 & 0.15    & 0    & 0    & 91.3 \\
       &         & 3 (dedup) & 12,101 & 100.0 & 0.00 & 0 & 0 & 100.0 \\
\midrule
Toys & RKMeans & 0 & 256    & 2.1  & 83.28   & 83   & 142  & 0.2 \\
     &         & 1 & 5,531  & 46.4 & 4.57    & 2    & 13   & 27.8 \\
     &         & 2 & 10,025 & 84.1 & 0.65    & 0    & 2    & 74.5 \\
     &         & 3 (dedup) & 11,924 & 100.0 & 0.00 & 0 & 0 & 100.0 \\
     & RQVAE   & 0 & 116    & 1.0  & 166.65  & 141  & 318  & 0.0 \\
     &         & 1 & 5,538  & 46.4 & 4.90    & 2    & 14   & 28.1 \\
     &         & 2 & 10,918 & 91.6 & 0.29    & 0    & 1    & 85.6 \\
     &         & 3 (dedup) & 11,924 & 100.0 & 0.00 & 0 & 0 & 100.0 \\
\midrule
Sports & RKMeans & 0 & 256    & 1.4  & 111.86  & 102  & 191  & 0.0 \\
       &         & 1 & 7,772  & 42.3 & 4.94    & 2    & 13   & 22.5 \\
       &         & 2 & 14,964 & 81.5 & 0.76    & 0    & 2    & 70.4 \\
       &         & 3 (dedup) & 18,357 & 100.0 & 0.00 & 0 & 0 & 100.0 \\
       & RQVAE   & 0 & 114    & 0.6  & 243.06  & 213  & 459  & 0.0 \\
       &         & 1 & 8,367  & 45.6 & 3.99    & 2    & 10   & 25.1 \\
       &         & 2 & 17,109 & 93.2 & 0.19    & 0    & 1    & 87.9 \\
       &         & 3 (dedup) & 18,357 & 100.0 & 0.00 & 0 & 0 & 100.0 \\
\midrule
Food k5 & RKMeans & 0 & 256     & 0.2  & 1034.07 & 1007 & 1684 & 0.0 \\
        &         & 1 & 19,600  & 11.7 & 31.69   & 18   & 74   & 2.8 \\
        &         & 2 & 90,371  & 54.0 & 5.06    & 1    & 13   & 37.6 \\
        &         & 3 (dedup) & 167,428 & 100.0 & 0.00 & 0 & 0 & 100.0 \\
        & RQVAE   & 0 & 148     & 0.1  & 1745.87 & 1537 & 3253 & 0.0 \\
        &         & 1 & 21,972  & 13.1 & 29.58   & 15   & 70   & 3.2 \\
        &         & 2 & 105,966 & 63.3 & 2.96    & 1    & 7    & 47.6 \\
        &         & 3 (dedup) & 167,428 & 100.0 & 0.00 & 0 & 0 & 100.0 \\
\bottomrule
\end{tabular}%
}
\end{table*}

\header{Level-restricted unlearning updates}
\Cref{tab:ablation-unified-positions-beauty} reports a diagnostic experiment that restricts the unlearning update to different \ac{SID} levels.
This analysis complements the main objective ablation by showing how different parts of the \ac{SID} representation contribute to forgetting and utility preservation.
The results indicate that effective unlearning is not only determined by the target item itself, but also by which identifier levels and model parameters are allowed to change.

\begin{table}[t]
\centering
\small
\caption{Restricting the unlearning update to individual semantic ID levels, on beauty, TIGER + RQ-KMeans, at the full objective's operating point. Levels are numbered $1$ to $L=4$, level $4$ being the deduplication digit. ``codes only'' freezes the backbone, so the update reaches the identifier embeddings and nothing else; those are the discriminating arms, because with the backbone free every restriction removes the target and the levels cannot be told apart. Each data seed is scored against its own reference, unlearning replicates averaged within a seed and the spread taken across seeds.}
\label{tab:ablation-unified-positions-beauty}
\resizebox{\textwidth}{!}{%
\begin{tabular}{llrrr}
\toprule
target & update restricted to & SH@10 & NDCG@10 & $\tau_{\mathcal{P}}$ \\
\midrule
\multicolumn{5}{l}{\textit{unpopular}: attack landed on all seeds} \\
 & all parameters (unrestricted) & 0.00000$\pm$0.00000 & 0.03586$\pm$0.00112 & 0.954$\pm$0.059 \\
 & levels 3, 4 frozen & 0.00000 & 0.03666 & 0.974 \\
 & levels 1, 2 + backbone & 0.00155$\pm$0.00326 & 0.03605$\pm$0.00127 & 0.959$\pm$0.065 \\
 & levels 1, 2, 3 + backbone & 0.00140$\pm$0.00292 & 0.03604$\pm$0.00130 & 0.959$\pm$0.065 \\
 & levels 1, 2, codes only & 0.00934$\pm$0.01058 & 0.03604$\pm$0.00103 & 0.958$\pm$0.055 \\
 & levels 3, 4, codes only & 0.01207 & 0.03684 & 0.979 \\
 & level 1 only, codes & 0.01283$\pm$0.01285 & 0.03591$\pm$0.00093 & 0.955$\pm$0.052 \\
 & level 2 only, codes & 0.01110$\pm$0.01211 & 0.03596$\pm$0.00094 & 0.956$\pm$0.053 \\
 & level 3 only, codes & 0.01426$\pm$0.01449 & 0.03595$\pm$0.00102 & 0.956$\pm$0.054 \\
 & level 4 only, codes & 0.01397 & 0.03667 & 0.975 \\
\midrule
\multicolumn{5}{l}{\textit{mid}: attack landed on all seeds} \\
 & all parameters (unrestricted) & 0.00000$\pm$0.00000 & 0.03656$\pm$0.00044 & 0.972$\pm$0.031 \\
 & levels 3, 4 frozen & 0.00000 & 0.03643 & 0.968 \\
 & levels 1, 2 + backbone & 0.00025$\pm$0.00041 & 0.03645$\pm$0.00059 & 0.969$\pm$0.027 \\
 & levels 1, 2, 3 + backbone & 0.00026$\pm$0.00043 & 0.03652$\pm$0.00066 & 0.970$\pm$0.027 \\
 & levels 1, 2, codes only & 0.00504$\pm$0.00641 & 0.03613$\pm$0.00077 & 0.960$\pm$0.017 \\
 & levels 3, 4, codes only & 0.01285 & 0.03631 & 0.965 \\
 & level 1 only, codes & 0.00747$\pm$0.00842 & 0.03606$\pm$0.00079 & 0.958$\pm$0.016 \\
 & level 2 only, codes & 0.00637$\pm$0.00799 & 0.03603$\pm$0.00077 & 0.957$\pm$0.017 \\
 & level 3 only, codes & 0.00936$\pm$0.01055 & 0.03609$\pm$0.00084 & 0.959$\pm$0.016 \\
 & level 4 only, codes & 0.01280 & 0.03630 & 0.965 \\
\midrule
\multicolumn{5}{l}{\textit{popular}: attack landed on 4/5 seeds} \\
 & all parameters (unrestricted) & 0.00000$\pm$0.00000 & 0.03603$\pm$0.00145 & 0.957$\pm$0.026 \\
 & levels 3, 4 frozen & 0.00000 & 0.03560 & 0.946 \\
 & levels 1, 2 + backbone & 0.00166$\pm$0.00181 & 0.03611$\pm$0.00137 & 0.959$\pm$0.019 \\
 & levels 1, 2, 3 + backbone & 0.00167$\pm$0.00190 & 0.03606$\pm$0.00138 & 0.958$\pm$0.022 \\
 & levels 1, 2, codes only & 0.01020$\pm$0.01180 & 0.03549$\pm$0.00133 & 0.943$\pm$0.021 \\
 & levels 3, 4, codes only & 0.00136 & 0.03541 & 0.941 \\
 & level 1 only, codes & 0.01319$\pm$0.01558 & 0.03547$\pm$0.00136 & 0.942$\pm$0.023 \\
 & level 2 only, codes & 0.01299$\pm$0.01353 & 0.03540$\pm$0.00105 & 0.940$\pm$0.020 \\
 & level 3 only, codes & 0.01597$\pm$0.01711 & 0.03542$\pm$0.00124 & 0.941$\pm$0.021 \\
 & level 4 only, codes & 0.00192 & 0.03537 & 0.940 \\
\bottomrule
\end{tabular}}
\end{table}

\end{document}